\documentclass[twocolumn,trackchanges]{aastex631}

\usepackage{amsmath}
\usepackage{lipsum}
\usepackage{graphics}
\usepackage{CJK}
\usepackage{tabularx}

\newcommand{\OIII}{\mbox{O\,\textsc{iii}}} 
 
\newcommand{\NII}{\mbox{N\,\textsc{ii}}} 
\newcommand{\SII}{\mbox{S\,\textsc{ii}}} 
\newcommand{\OI}{\mbox{O\,\textsc{i}}}

\begin{document}

\title{Extreme Neutral Outflow in a non-AGN Quiescent Galaxy at z$\sim$1.3}


\correspondingauthor{Yang Sun}
\email{sunyang@arizona.edu}

\author[0000-0001-6561-9443]{Yang Sun}
\affiliation{Steward Observatory, University of Arizona,
933 North Cherry Avenue, Tucson, AZ 85719, USA}

\author[0000-0001-7673-2257]{Zhiyuan Ji}
\affiliation{Steward Observatory, University of Arizona,
933 North Cherry Avenue, Tucson, AZ 85719, USA}

\author[0000-0003-2303-6519]{George H. Rieke}
\affiliation{Steward Observatory, University of Arizona,
933 North Cherry Avenue, Tucson, AZ 85719, USA}

\author[0000-0003-2388-8172]{Francesco D'Eugenio}
\affiliation{Kavli Institute for Cosmology, University of Cambridge, Madingley Road, Cambridge, CB3 0HA, UK}
\affiliation{Cavendish Laboratory, University of Cambridge, 19 JJ Thomson Avenue, Cambridge, CB3 0HE, UK}

\author[0000-0003-3307-7525]{Yongda Zhu}
\affiliation{Steward Observatory, University of Arizona,
933 North Cherry Avenue, Tucson, AZ 85719, USA}

\author[0000-0002-4622-6617]{Fengwu Sun}
\affiliation{Center for Astrophysics $|$ Harvard \& Smithsonian, 60 Garden St., Cambridge MA 02138 USA}

\author[0000-0001-6052-4234]{Xiaojing Lin}
\affiliation{Department of Astronomy, Tsinghua University, Beijing 100084, China}
\affil{Steward Observatory, University of Arizona,
933 North Cherry Avenue, Tucson, AZ 85719, USA}

\author[0000-0002-8651-9879]{Andrew J. Bunker}
\affiliation{Department of Physics, University of Oxford, Denys Wilkinson Building, Keble Road, Oxford OX1 3RH, UK}

\author[0000-0002-6221-1829]{Jianwei Lyu (\begin{CJK}{UTF8}{gbsn}吕建伟\end{CJK})}
\affiliation{Steward Observatory, University of Arizona,
933 North Cherry Avenue, Tucson, AZ 85719, USA}

\author[0000-0002-5104-8245]{Pierluigi Rinaldi}
\affiliation{Steward Observatory, University of Arizona, 933 North Cherry Avenue, Tucson, AZ 85719, USA}

\author[0000-0001-9262-9997]{Christopher N. A. Willmer}
\affiliation{Steward Observatory, University of Arizona, 933 North Cherry Avenue, Tucson, AZ 85719, USA}

\begin{abstract}

We report the discovery of a substantial sodium doublet (Na D $\lambda\lambda$5890, 5896\AA)–traced neutral outflow in the quiescent galaxy JADES-GS-206183 at $z=1.317$. Its JWST/NIRSpec-MSA spectrum shows a deep, blueshifted Na D absorption, revealing a neutral outflow with $v_{\rm out}=828^{+79}_{-49}\,\mathrm{km\,s^{-1}}$ and a mass outflow rate of $\log(\dot{M}_{\rm out}/\mathrm{M_{\odot}\,yr^{-1}})=2.40^{+0.11}_{-0.16}$. This outflow rate exceeds that of any neutral outflows identified beyond $z\sim1$ by the same line and is comparable with those in local galaxies with intensive star formation or luminous AGN. JADES-GS-206183 is also a peculiar quiescent galaxy with a spiral$+$bar morphology, high dust attenuation ($A_V=2.27\pm0.23$ mag). Paschen $\alpha$ (Pa$\alpha$) emission from the FRESCO NIRCam grism confirms its low star formation rate ($\mathrm{SFR_{Pa\alpha}}=10.78\pm 0.55\,\mathrm{M_{\odot}\,yr^{-1}}$), placing it 0.5 dex below the main-sequence ($\log(\mathrm{sSFR/yr^{-1}})=-10.2$). Despite the systematics introduced by different star formation history (SFH) priors, the SED modeling, combining HST-to-NIRCam photometry with the VLT/MUSE spectrum, suggests that JADES-GS-206183 experienced an older episode of star formation 0.5-2 Gyr ago and a possible rejuvenation within recent $\sim$10 Myr.
Moreover, rest-frame optical lines indicate that the current AGN activity of JADES-GS-206183, if present, is also weak. Even though we tentatively detect a broad component of the H$\alpha$ line, it likely traces an ionized outflow rather than an AGN. The results demonstrate that the Na D outflow in JADES-GS-206183 is highly unlikely to be driven by current star formation or nuclear activity. Instead, it may represent a long-lasting fossil outflow from past AGN activity, potentially co-triggered with the early phase of rejuvenation.

\end{abstract}

\keywords{
\href{http://astrothesaurus.org/uat/572}{Galactic Winds (572)},
\href{http://astrothesaurus.org/uat/594}{Galaxy Evolution (594)},
\href{http://astrothesaurus.org/uat/2040}{Galaxy Quenching (2040)},
\href{http://astrothesaurus.org/uat/1569}{Star Formation (1569)}
}

\section{Introduction} \label{sec:intro}

Galactic outflows, as an essential step of the galactic baryon cycle, play a fundamental role in driving galaxy evolution, regulating star formation and central nuclear activities within a galaxy. Current cosmological simulations of galaxy formation and evolution invoke feedback to reproduce observed galaxy baryonic mass functions and star formation efficiency, and also to quench galaxies \citep[][and references therein]{Keres2009,Hopkins2014,Somerville2015,Naab2017}. Massive stars, supernovae, and active galactic nuclei (AGN), as the sources of outflows and thus feedback, transfer their own thermal or kinetic energy to the interstellar medium (ISM) in galaxies, heating or directly expelling ISM components outwards to suppress star formation and central nuclear activity.

From the observational perspective, many works have reported galactic outflows in both AGN hosts \citep[e.g.,][]{Cicone2014,Feruglio2015,Baron2019,Davies2020} and inactive galaxies (without current AGN activity) \citep[e.g.,][]{Heckman2000,Rupke2005a,Chen2010,Perrotta2021} in the local Universe over the past few decades. Specifically,  many local post-starburst galaxies (PSB), whose star formation has shut down in the past$~1$ Gyr \citep{Dressler1983,Couch1987}, commonly show outflow signatures \citep[e.g.,][]{Tremonti2007,Coil2011,Luo2022,Baron2022,Sun2024}. This behavior indicates that  outflows should somehow contribute to galaxy quenching. Also, the outflows in PSBs are often multi-phase. Most of the PSB outflows are detected in a cool neutral phase through Na D absorption \citep{Baron2022, Sun2024}, but a large fraction of them also exhibit in the ionized phase traced by $\OIII$ $\lambda$5007, especially for those hosted by AGN-host PSBs. Additionally, \citet{Luo2022} reported a case of a quenching galaxy hosting outflows simultaneously in the ionized, neutral, and molecular phases.

Beyond the local Universe ($z>1$),  observations of outflows in quiescent galaxies using the same diagnostics as the local studies are challenging due to the limited wavelength coverage and sensitivities of pre-JWST facilities.
Historically, the neutral phase outflows at higher redshift could only be studied using a UV absorption line (such as Si II$\lambda\lambda$1304, 1636, Mg II$\lambda\lambda$2796, 2804) because the more common optical tracer, Na D, is shifted to the infrared range. Also, the stacking approach was often applied to enhance the outflow signatures \citep[e.g., ][]{Weiner2009,Steidel2010, Maltby2019}, which prevents us from studying the outflow and its impact on host galaxy properties in single galaxies. Since JWST launched in 2021, many more high-z quiescent galaxies have been spectroscopically discovered \citep[e.g., ][]{Carnall2023,deGraaff2024}. Also, an increasing number of outflows in the post-starburst and quiescent galaxies have been detected at cosmic noon and beyond \citep{Belli2024,Davies2024,D'Eugenio2024,Carniani2024, Zhu2024, Wu2025}, which enables us to better understand the outflow and quenching mechanism in the early Universe. Among them, \citet{Belli2024} and \citet{D'Eugenio2024} detected the neutral outflows in two AGN-host PSBs through blueshifted Na D absorption at z$\sim2.5$ and 3, respectively, for which the outflow mass rate is an order of magnitude higher than the current star formation rate of the hosts, indicating AGN-driven outflows can efficiently quench galaxies at high-z. In addition, a census of Na D outflows at cosmic noon from the JWST Cycle 1 Blue Jay survey \citep{Davies2024,Park2024} shows that powerful neutral outflows are commonly detected in massive quenching systems, which seem to be powered by AGN activity. In short, the recent studies of Na D outflows in high-z quiescents all focus on galaxies with ongoing AGN activity, leading us to naturally link the nuclear activity to those outflows. However, since the lifetimes of AGN and outflows are not necessarily the same, we cannot draw causality conclusions from comparing directly the relative incidence of the two. In particular, whether and how outflows are launched in the PSBs with quiet nuclear regions and how they drive or maintain galaxy quenching are still poorly understood at high-z.

In this work, we report the discovery of an extremely powerful Na D outflow in the z = 1.317 quiescent galaxy JADES-GS-206183 (RA$=53.176589$ deg, DEC$=-27.785519$ deg). The specific star formation rate of this galaxy is 0.5 dex below the main sequence \citep{2020ApJ...902..109B} and it has only weak signatures of an AGN. Surprisingly its mass outflow rate is the highest detected so far beyond the local Universe; the outflow energy and mass rate are even comparable to the energetic outflows driven by local luminous starbursts or AGNs. Our SED modelling of this galaxy suggests it might have started rejuvenation only in the recent 10 Myr and is still much lower than expected to power the outflow. Therefore, this galaxy gives crucial insight into the outflow launching mechanism and galaxy quenching around cosmic noon.

This paper is organized as follows:
We present the data used in this study in Section~\ref{sec:data}, and then describe the spectral analysis and the SED modelling in Section~\ref{sec:analys} and \ref{sec:sed_model}, respectively. The results for outflows in JADES-GS-206183 and its host galaxy properties are shown in Section~\ref{sec:result}, and we discuss them in Section~\ref{sec:disc}. Finally, we summarize our work in Section~\ref{sec:concl}. Throughout this paper, we assume a standard $\Lambda$CDM universe whose cosmological parameters are $\mathrm{H_0} =
70~\mathrm{km~s^{-1}~Mpc^{-1}}$, $\Omega_{\Lambda} = 0.7$, and $\Omega_{\mathrm{m}} = 0.3$.

\section{Data} \label{sec:data}

\subsection{NIRSpec} \label{sec:nirspec}
We used JWST/NIRSpec Micro Shutter Assembly (MSA; \citealt{Jakobsen2022,Ferruit2022}) spectra from the SMILES survey (PID 1207, PI: Rieke; \citealt{Alberts2024,Rieke2024}) to measure the properties of the Na D doublet ($\lambda\lambda$5890, 5896) in the spectrum of JADES-GS-206183. We observed with the G140M/F100LP and G235M/F170LP gratings and blocking filters to cover 0.97-3.07 $\mu$m in the observed frame (0.4-1.3 $\mu$m in the rest frame) at a spectral resolution of $R\sim 1000$. The effective exposure time with each disperser/filter combination was 7,000 seconds ($\sim 1.94$ hours).

We processed and calibrated the data using the {\tt JWST Calibration Pipeline} \citep{Bushouse2022}, version 1.14.0, with the CRDS\footnote{Calibration Reference Data System: \url{https://jwst-crds.stsci.edu/}} context {\tt jwst\_1236.pmap}. In brief, the {\tt calwebb\_detector1} stage converted ``uncal.fits'' files into uncalibrated rate images. Next, the {\tt calwebb\_spec2} stage applied flat-field corrections, flux calibration, and local background subtraction, yielding resampled spectra for each nodding position. The final spectral combination and extraction were carried out using {\tt calwebb\_spec3}. Additionally, we applied custom scripts to further reject hot pixels and remove $1/f$ noise. A detailed description of the reduction process, along with reduced spectra and the redshift catalog, will be provided in a forthcoming data release (\citealt{Zhu2025}; see also \citealt{Zhu2024arXiv}). Given the extended morphology of JADES-GS-206183, we used only the two nodding positions, namely, the two shutters far from the central shutter placed in the galaxy center for local background subtraction to avoid self-subtraction by the galaxy.

We also applied slit-loss corrections to the NIRSpec data, since the shutter covers only a fraction of the galaxy. The underlying assumption here, which might not be the case, is that the emission line-to-continuum ratio is invariant across the entire galaxy. However, the slit-loss correction does not affect our analysis of Na D absorption modelling (Section~\ref{sec:Na D}) and the subsequent estimation of outflow properties (Section~\ref{sec:nad_result}) because the scaling factor will be naturally normalized or removed along with the stellar continuum component and not change the relative strength of Na D absorption to the continuum (see Equation~\ref{eq1}). Keeping those in mind, we calculated  F115W and F210M synthetic photometry from the spectrum (the two filters avoid strong emission lines) and then determined the correction factors for the two bands by comparing them with the corresponding KRON convolved NIRCam photometry from the JADES v1.0 photometric catalog. We then did a linear interpolation between the two band correction factors to globally correct the whole spectrum. It reveals a wavelength-dependent trend in the correction factor, decreasing from approximately six at 1 $\mu m$ to two at 3 $\mu m$.
We confirm the slit-loss corrected NIRSpec spectrum matches the best-fit SED derived from multi-band photometry (Section~\ref{sec:sed_model}).

\subsection{Photometry}

\begin{figure*}
\centering
\includegraphics[width=1\textwidth]{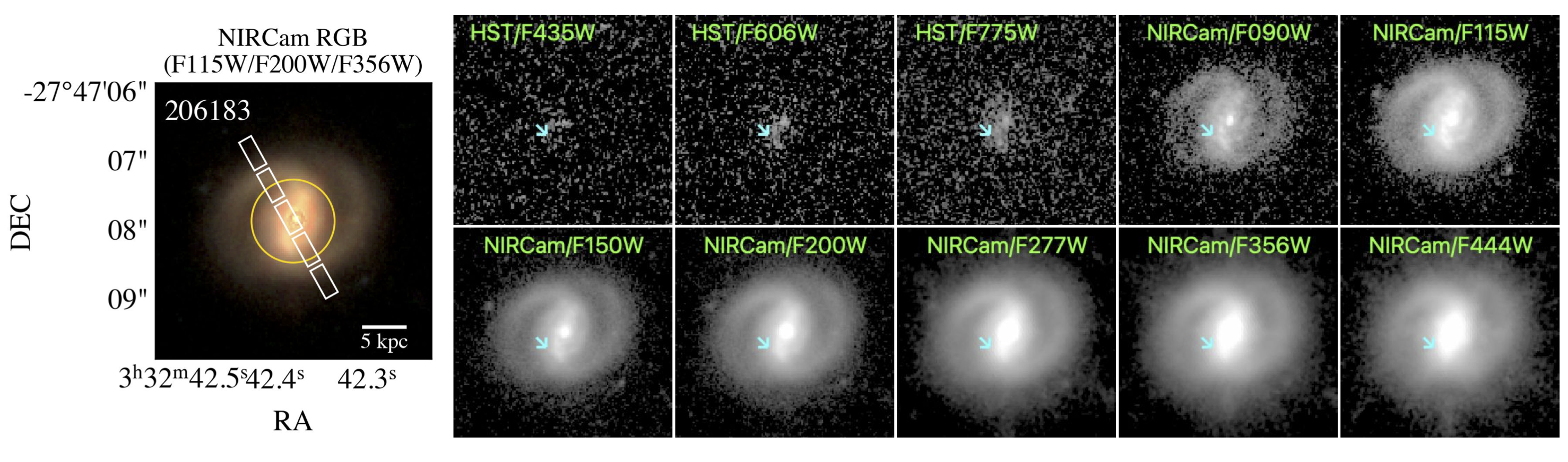}
\caption{HST and JWST images of JADES-GS-206183. 
{\bf Left:} NIRCam false-color RGB image generated from the F115W/F200W/F356W images overlaid with the position of the NIRSpec MSA shutters and a yellow circle representing the galaxy's effective radius (0.6 arcsec). {\bf Right:} Single-band HST/F435W, F606W, F775W, and JWST/NIRCam wide-band images. The cyan arrow indicate a southeast subregion close to the center in where the images shows an excess of UV light and thus where current low-level star formation is happening.}
\label{fig:images}
\end{figure*}

We obtained photometry in 13 wide- and medium-band NIRCam filters (F090W,
F115W, F150W, F182M, F200W, F210M, F277W, F356W, F410M, F430M, F444W,
F460M and F480M) of JADES-GS-206183, as well as legacy HST photometric data in the ACS/WFC filters (F435W, F606W, F775W, F814W and F850LP) and in the  WFC3 filters (F105W, F125W, F140W, and F160W) from the JADES catalog \citetext{\citealp{Rieke2023}, with updates to the photometry in \citealp{Eisenstein2023,Eisenstein2023b,DEugenio2025} and JADES Collaboration, in~prep.}. HST/ACS, HST/WFC3 and JWST/NIRCam filters together nicely cover the rest-frame optical to near-IR spectral energy distributions (SEDs) of the galaxy, allowing reasonable constraints on its stellar properties. In addition, to better reveal the morphological structure, we exhibit the single-band images and false-color RGB image made by F115W/F200W/F356W in Figure~\ref{fig:images}. The RGB image of JADES-GS-206183 shows that the light distribution is overall smooth, with structures that appear to be morphologically similar to a bar, with spiral arms emanating from both ends. The bar length is about 0.6 arcsec ($\sim$10 kpc), comparable with the effective radius of the galaxy.  Given this large bar size, observed in a quiescent galaxy at $z>1$, we remark that kinematic confirmation of this interpretation would be needed. We also notice that the southeast region close to the center(marked by the arrow in Figure~\ref{fig:images}) show an excess of UV light in the shorter-wavelength images (HST bands and JWST F090W and F115W), relative to the rest of the galaxy. However, based on the multi-band morphology, this excess is more likely associated with a UV-bright star-forming clump in one of the galaxy's spiral arms, although with the current data, we cannot fully rule out the possibility of a minor merger.

JADES-GS-206183 also has spectroscopic data obtained with JWST/NIRCam Wide Field Slitless Spectroscopy (WFSS) from the FRESCO survey \citep{Oesch2023} in the F444W band ($\lambda\sim$ 3.9--5.0~$\mu m$). The FRESCO survey covers a 7.4\arcmin $\times$8.4\arcmin\ area in both GOODS fields with the row-direction grisms on both modules of JWST/NIRCam providing a spectral resolution of R$\sim$1590--1680 from 3.9 to 5.0~$\mu m$. The 5$\sigma$ line sensitivity of FRESCO is $2\times10^{-18}~{\rm erg~s^{-1}cm^{-2}}$.

The NIRCam/WFSS data was processed by the publicly available reduction routine presented in \citet{Sun2023}\footnote{\url{https://zenodo.org/records/14052875} \citep{https://doi.org/10.5281/zenodo.14052875}}. 
We first processed the NIRCam data through the standard JWST stage-1 calibration pipeline\footnote{\url{https://zenodo.org/records/8140011} \citep{https://doi.org/10.5281/zenodo.6984365}} \texttt{v1.11.2}.
We assigned world coordinate system (WCS) information for the grism, performed flat-fielding, and removed the $\sigma$-clipped median sky background. 
The WCS of the grism exposure is registered to Gaia DR3 \citep{Gaia2023} using the NIRCam short-wavelength imaging data taken at the same time.
For JADES-GS-206183, we extracted the 2D spectrum from individual grism exposures, and coadded them in a common wavelength (1 nm/pixel) and spatial (0\farcs06/pixel) grid.
Prior to our scientific spectral extraction, we also extracted spectra of bright point sources ($\lesssim$\,21\,AB mag) to ensure the accuracy of the spectral tracing function and spectral flux calibration. 
We also extracted the spectrum of the galaxies with known ground-based spectroscopic redshifts, measuring the line center of detected Paschen $\alpha$ and $\beta$ lines to ensure the wavelength calibration error at $<$\,1\,nm.
We then optimally extracted the 1D spectrum of our target JADES-GS-206183 from coadded 2D spectrum using the surface brightness profile in the F444W band \citep{Horne1986}, for which the integrated flux perfectly matches the F444W photometry.

\section{Spectroscopic Analysis}\label{sec:analys}

\subsection{Stellar Continuum Modeling}
\label{sec:pPXF}

\begin{figure*}
\centering
\includegraphics[width=1\textwidth]{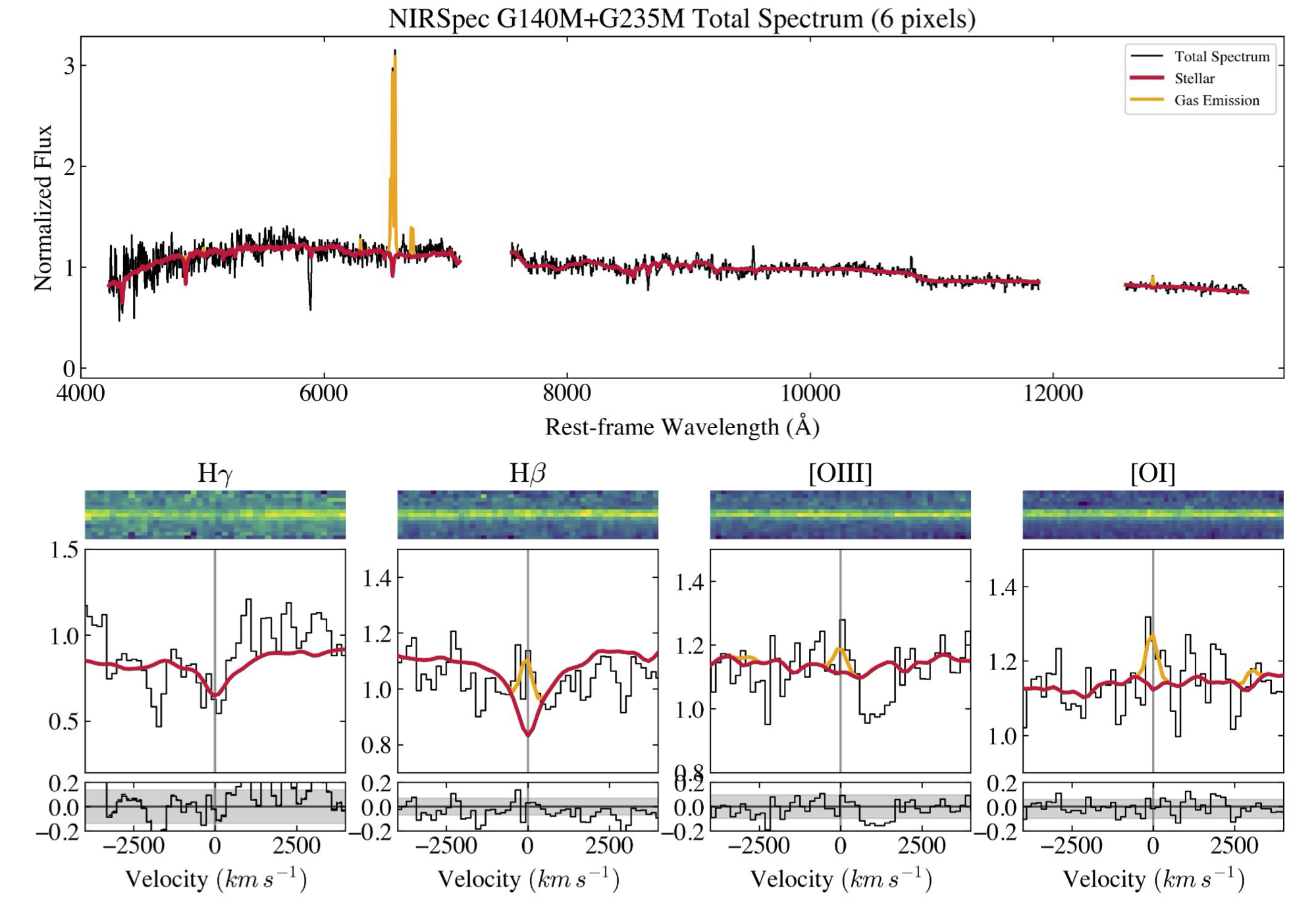}
\caption{JWST NIRSpec observations of JADES-GS-206183. 
{\bf Top:} NIRSpec MSA G140M and G235M spectrum (black),  the best-fit stellar continuum from pPXF (red),  and the nebular emission (orange). The two spectral gaps are caused by the NIRSpec detector gap. {\bf Bottom:} From left to right, the zoom-in original 2D spectra (top), best-fits (middle), residuals (bottom) of H$\gamma$, H$\beta$, $[\OIII]\lambda$5007, and $[\OI]\lambda$6300,  respectively.
The averaged $1\sigma$ noise levels are also plotted in the residual panels.}

\label{fig:SMILES_spec}
\end{figure*}

Both stellar absorption and the ISM can contribute to Na D absorptions; therefore, we first model the stellar continuum of the JADES-GS-206183 in its NIRSpec G140M and G235M spectra, using the penalized pixel fitting \citep[pPXF,][]{cappellari2004,cappellari2017,Cappellari2023} Python code. pPXF allows us to model the stellar continuum with a non-linear $\chi^2$ minimization in the galaxy spectrum, at the same time taking into account the instrument line spread function (LSF). In our modeling, we adopt the LSF from \citet{Jakobsen2022}.

We build the stellar templates using the Flexible Stellar Population Synthesis (FSPS; \citealt{Conroy2009,Conroy2010}) model built by \citet{Cappellari2023}, with the MILES stellar spectral library \citep{Sanchez-Blazquez2006,Falcon-Barroso2011}, the MIST isochrones \citep{Choi2016} and the Salpeter IMF \citep{Salpeter1955}\footnote{Even though we use the Salpeter IMF here,  which is inconsistent at the lower mass end compared to other widely-used IMFs, i.e. the Kroupa IMF that we use for SED fitting (see Section~\ref{sec:sed_model}), \citet{Cappellari2023} confirmed the pPXF spectral fitting result is insensitive to the slope of the IMF at low masses.}. The templates span a logarithmically spaced grid of ages and metallicities, covering the range of 1 Myr--15.85 Gyr with 0.1 dex sampling and [Z/H]$=$-1.75--0.25 with 0.25 dex sampling, respectively. We limit the age of the input templates to no longer than the Universe age at the redshift of JADES-GS-206183 (4.68 Gyr). 

In pPXF fitting, we mask the nebular emission lines, including Balmer lines, [$\OIII$] $\lambda\lambda4959, 5007$, [$\OI$] $\lambda6300$, [$\NII$] $\lambda\lambda6548, 6584$, and [$\SII$] $\lambda\lambda6717, 6731$. We will independently fit the profiles and calculate the fluxes of those emission lines after subtracting the stellar continuum derived by pPXF in the next section. We confirm the best-fit stellar continuum is not affected by whether the nebular emission lines are simultaneously fitted or not.
The Na D line is also masked in our modeling\footnote{We confirm masking the Na D region or not does not have impact on the best-fit stellar model around Na D and the further Na D-derived ISM properties}. 
Finally, a fourth-degree multiplicative polynomial is used to avoid mismatches between galaxy spectra and stellar templates and to ensure the accuracy of the fitted kinematic parameters. 

We first run pPXF by assuming the systemic redshift of JADES-GS-206183 to be the same as that derived from emission line measurements obtained with ALMA and VLT/MUSE \citep{Inami2017,Boogaard2019}. After obtaining the first pPXF results for stellar kinematics, we adjust the redshift of JADES-GS-206183 using the stellar velocity and re-ran the pPXF fitting. The pPXF fitting results are shown in Figure~\ref{fig:SMILES_spec}, and the best-fit stellar kinematics are shown in Table~\ref{tab:line}. The spectrum displays stellar absorption features (e.g., Balmer and Na D), in line with the typical spectral features observed in quiescent/post-starburst galaxies \citep[e.g., ][]{Park2024}.

\subsection{Optical Emission Line Fitting}\label{sec:linefitting}
\begin{figure*}[t!]
\centering
\includegraphics[width=\textwidth]{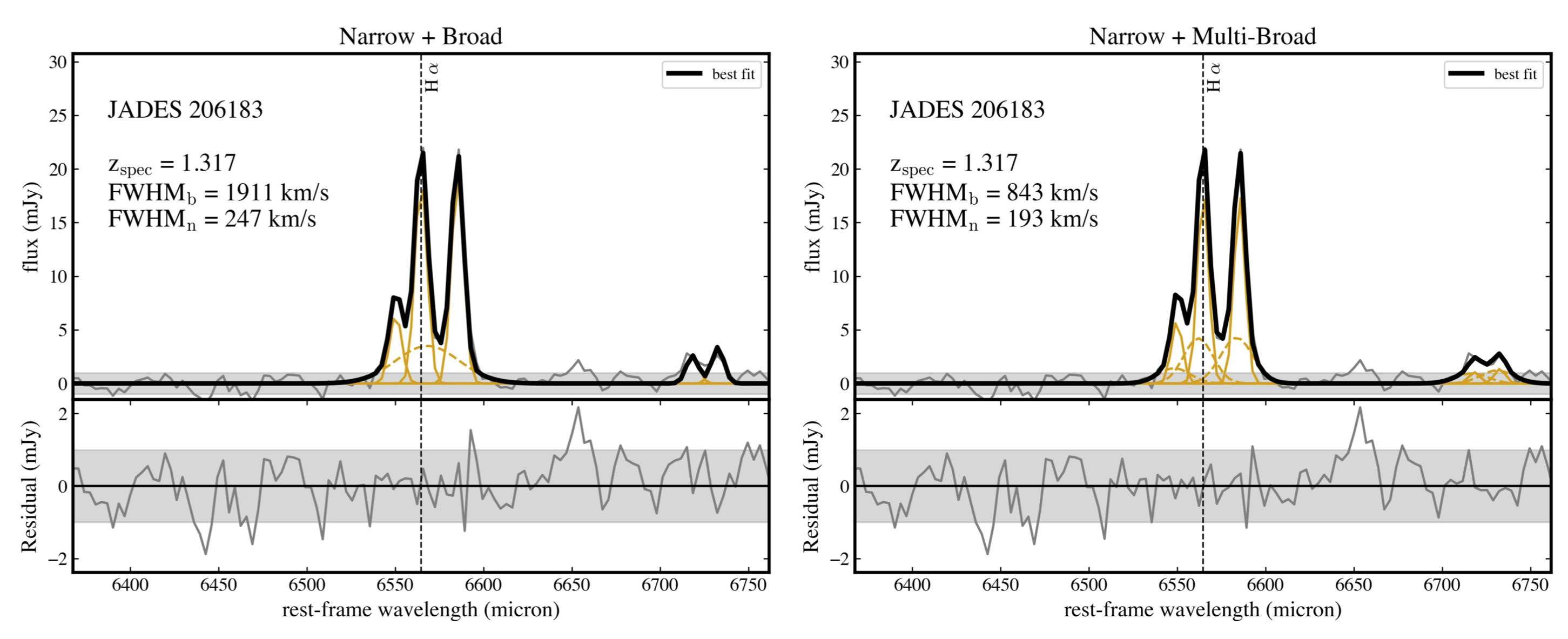}
\caption{Line profile fitting for the H $\alpha$, $[\NII]\lambda\lambda$6548, 6584, and $[\SII]\lambda\lambda$6717, 6731, assuming the single broad component illustrated in the bottom right panel of Figure~\ref{fig:SMILES_spec} is the superposition of three broad components originated from outflow. In the top panel, we show the stellar continuum subtracted 1D spectrum (gray line) and the best-fit line profile (black). The decomposed narrow components and broad components are plotted as the solid and dashed yellow lines, respectively. We show the residual and its 1$\sigma$ level in the bottom panel.}
\label{fig:smiles_Ha}
\end{figure*}
Next, we model the pure nebular emission features using the Python package \texttt{lmfit} on the stellar continuum-subtracted spectrum. A single Gaussian profile is assigned for Balmer lines, [$\OIII$] $\lambda\lambda4959, 5007$, [$\OI$] $\lambda6300$, [$\NII$] $\lambda\lambda6548, 6584$, and [$\SII$] $\lambda\lambda6717, 6731$, which shares the same velocity and velocity dispersion. The line width of the narrow component is limited within $300$ $\mathrm{km\,s^{-1}}$. For H$\alpha$, we add a broad Gaussian profile to the narrow component, which may be present due to outflows and/or AGN activity. For the broad components in our modeling, the width is allowed to vary between 300 -- 5000 $\mathrm{km\,s^{-1}}$. All the components are constrained to have velocity offsets within $\pm 200 \,\mathrm{km,s^{-1}}$ relative to the systemic redshift. We test the necessity of this broad component using the Bayesian information criterion (BIC) method. We calculate BIC parameters for the ``narrow-only'' and ``narrow+broad'' models, and define a broad component as significantly detected when the difference of BIC between the ``narrow-only'' and ``narrow+broad" model exceeds 10 ($\Delta \text{BIC} = \text{BIC}_n - \text{BIC}_{n+b}>10$). This threshold follows a commonly adopted criterion indicating strong evidence in favor of the more complex model \citep{Liddle2007}. Finally, we used a bootstrap approach to derive the uncertainties of the line profile parameters and fluxes.

The best-fit emission line profiles are shown in Figure~\ref{fig:SMILES_spec}. JADES-GS-206183 only exhibits low-ionization nebular lines (e.g., H$\alpha$, [$\NII$], [$\SII$]), lacking high-ionization emission lines, e.g. [$\OIII$] (see zoom-in line fittings in the bottom row of Figure~\ref{fig:SMILES_spec}). The line fluxes and kinematics properties are shown in Table~\ref{tab:line}. The gas velocity (relative to the stellar component) is $-27$ $\mathrm{km\,s^{-1}}$, much smaller than the instrumental velocity dispersion ($\sim100$ $\mathrm{km\,s^{-1}}$), suggesting that the narrow component of the emission lines is tracing gas well coupled with the stellar component. The gas velocity dispersion is in good agreement with the stellar one derived by pPXF, about $\sim100$ $\mathrm{km\,s^{-1}}$. 

Finally, the BIC difference between the ``narrow-only" and ``narrow+broad" model is 12, indicating that an additional broad component with a FWHM of $\sim2000$ $\mathrm{km\,s^{-1}}$ is needed to reproduce the spectral range of H$\alpha$$+[\NII]$ well (the bottom right panel of Figure~\ref{fig:SMILES_spec}), and thus the presence of additional gas component(s) with distinct kinematics in JADES-GS-206183. This single broad component, if entirely attributed to H$\alpha$, could suggest the presence of a broad-line AGN in JADES-GS-206183. However, as will be detailed in Section~\ref{sec:AGN}, multiple lines of evidence strongly suggest that JADES-GS-206183 does not exhibit AGN activity.

Another possibility is that the apparently required broad component(s) arise from the superposition of three broad components, namely H$\alpha$ and the $[\NII]$ doublet -- produced by an ionized outflow, rather than from a single broad H$\alpha$ line. To test this scenario, we re-model the $[\NII]$+H$\alpha$ line profile, along with the nearby $[\SII]$ emission, using the``narrow+multi-broad'' model on the stellar continuum-subtracted spectrum. The template contains a narrow and a broad component for $[\NII]$, H$\alpha$, and $[\SII]$. The resulting FWHM of the narrow component, broad H$\alpha$ component, and broad $[\NII]$(or $[\SII]$) component are $193\pm35$, $843\pm154$, and $ 940\pm169\,\mathrm{km\,s^{-1}}$, respectively, indicating that the broad components of H$\alpha$, $[\NII]$, and $[\SII]$ have almost the same width. Also, all of the broad components are consistent with systemic velocity considering the spectral resolution (see Table~\ref{tab:line}). 
Therefore, the best-fit  ``narrow+multi-broad'' model suggests the broad component of H$\alpha$, $[\NII]$, and $[\SII]$ is very likely to trace the same gas kinematics, which is consistent with the outflow-only scenario, without additional AGN BLR emission.
Also, we find that the BIC of the multiple broad $[\NII]+$H$\alpha$ model is comparable with that of the single broad H$\alpha$ model ($\Delta BIC<5$), suggesting two models can produce a similarly good fit for the data.

In Section~\ref{sec:AGN}, we will discuss the possible origin(s) of this broad H$\alpha$ component in more detail, concluding that the outflow scenario (i.e., the ``narrow+multi-broad'' model) is the more likely explanation. Therefore, in Table~\ref{tab:line}, the kinematics and flux measurements of H$\alpha$, $[\NII]$, and $[\SII]$ is from the ``narrow+multi-broad'' model.

\begin{table*}[]
\centering
\begin{tabular}{ccccc}
\hline\hline
Component      &   & $\Delta v$ & FWHM & Flux \\ 
      &   & ($\mathrm{km\,s^{-1}}$) & ($\mathrm{km\,s^{-1}}$) & ($\mathrm{10^{-17}\,erg\,s^{-1}\,cm^{-2}}$)\\ \hline
star           &   & $0\pm5$        &  $245\pm57$     &      \\ \hline
H$\beta$          &   &   $-31\pm19$       &  $223\pm28$     &   $0.54\pm0.10$   \\
$[\OIII]\lambda$5007      &   &   $-31\pm19$       &  $223\pm28$     &  $0.15\pm0.11$     \\
$[\OI]\lambda$6300        &   &   $-31\pm19$       &   $223\pm28$    &   $0.33\pm0.09$   \\
H$\alpha$         & narrow &   $-28\pm5$       &   $193\pm35$    &  $3.31\pm0.44$    \\
               & broad &   $-150\pm60$       &  $843\pm154$     &   $2.02\pm0.55$   \\
$[\NII]\lambda$6584       & narrow &        $-28\pm5$       &   $193\pm35$     & $3.19\pm0.45$     \\
               & broad &     $-100\pm43$       &  $940\pm169$    &  $2.25\pm0.40$    \\
$[\SII]\lambda\lambda$6717\&6731 & narrow &       $-28\pm5$       &   $193\pm35$       &  $0.44\pm0.21$    \\
               & broad &    $-100\pm43$       &  $940\pm169$   &   $1.14\pm0.30$   \\
Pa$\alpha$       &   &    $-4\pm6$      &  $213\pm17$&   $1.67\pm0.08$   \\ \hline
\end{tabular}
\caption{The kinematic properties (velocity and dispersion) of star and gas traced by multiple emission lines in JADES-GS-206183. The properties of $[\OIII]$, H$\beta$, $[\OI]$, and Pa$\alpha$ were derived using a single Gaussian component, while for the H$\alpha$, $[\NII]$ and $[\SII]$, the properties were estimated through multiple Gaussian components assuming the ionized outflow scenario (see Section \ref{sec:linefitting}).} 
\label{tab:line}
\end{table*}

\subsection{Na D Absorption Line Fitting}
\label{sec:Na D}

\begin{figure}
\centering
\includegraphics[width=\columnwidth]{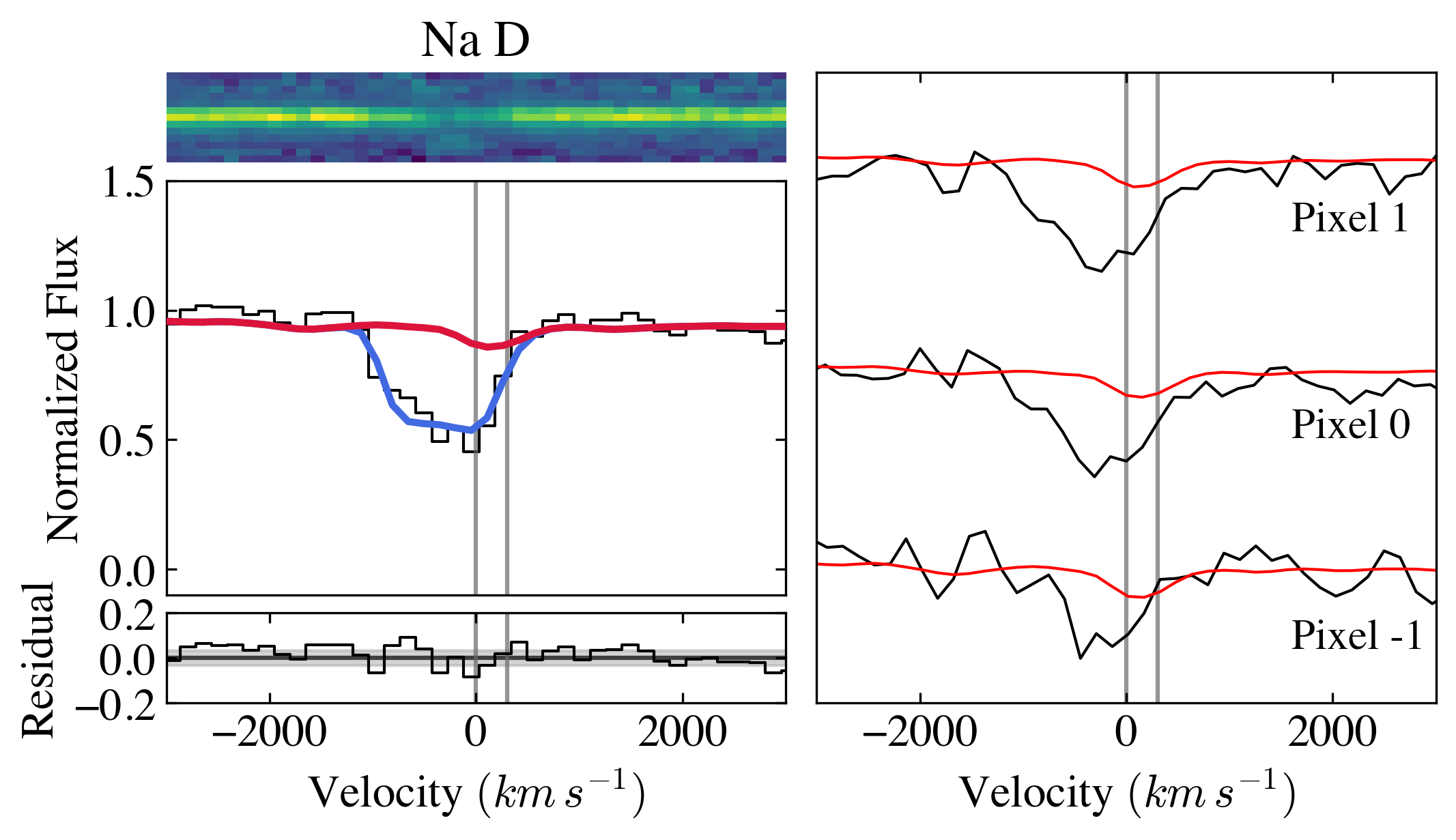}
\caption{{\bf Left:} Best-fit Na D profile of JADES-GS-206183. The top panel shows the original 2D MSA spectrum. In the middle panel, the black, red, and blue lines show the 1D MSA spectrum extracted from a 6-pixel-wide aperture, the best-fit stellar continuum by pPXF, and the best-fit ISM Na D absorption, respectively. The two gray vertical lines illustrate the systemic velocity of Na D$\lambda$ 5890, 5986. The bottom panel displays the fitting residual. {\bf Right:} Row-by-Row spectra of the central three rows of the JADES-GS-206183 spectrum. We label the rows with 1, 0, and -1 from top to bottom. }
\label{fig:NaD_pixel}
\end{figure}

\begin{figure}
\centering
\includegraphics[width=\columnwidth]{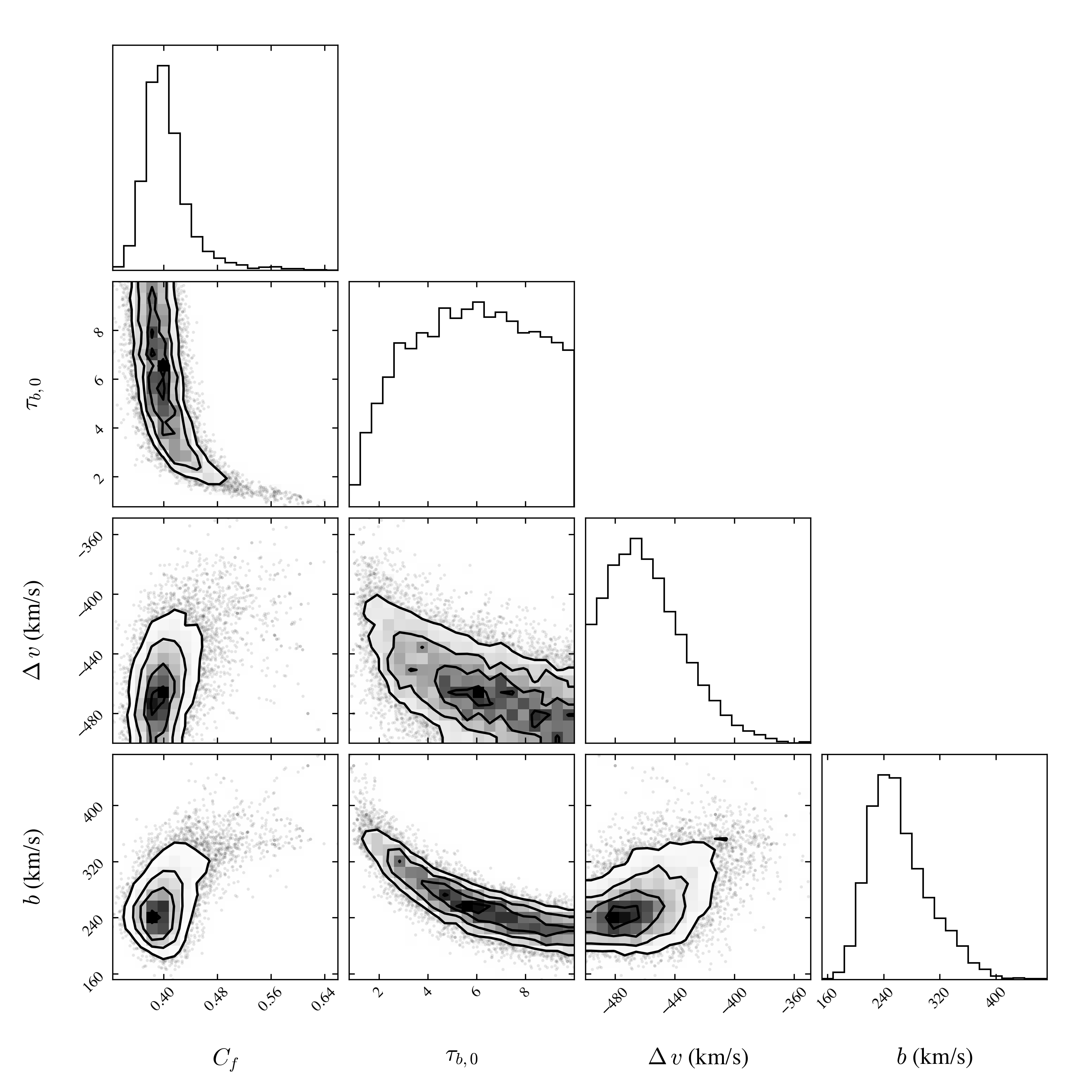}
\caption{MCMC posterior parameter distributions and best-fits of Na D profile parameters ($C_f$, $\tau_{b,0}$, $\Delta v$, and $b$).}
\label{fig:NaD_model_MCMC}
\end{figure}

After obtaining the stellar Na D contribution from pPXF, we are then able to model the part of the Na D absorption arising from the ISM.  

We fit the Na D line profile as:
\begin{align}
    F(\lambda) = F_{*} \times F_{\text{Na D, ISM}},
\label{eq1}
\end{align}
where $F_*$ is the stellar continuum derived from pPXF and $F_{\text{Na D, ISM}}$ is the ISM Na D absorption. 
$F_{\text{Na D, ISM}}$ is parametrized using the standard partial covering model from \citet{Rupke2005a}:  
\begin{align}
    F_{\text{Na D, ISM}}(\lambda) = 1 - C_f + C_f\,e^{(-\tau_{b}(\lambda) - \tau_{r}(\lambda))},
\end{align}
where $C_f$ is the covering fraction of the foreground absorbing gas against the background continuum source, $\tau_{b}$ and $\tau_{r}$ are the optical depths of the blue (5891\AA) and red (5897\AA) Na D lines, respectively. For Na D, $\tau_{b}/\tau_{r} = 2$. The optical depth is modeled by a Gaussian profile:
\begin{align}
    \tau(\lambda,b)=\tau_0\,e^{{-(\lambda-\lambda_0)^2/(\lambda_0\,b/c)^2}},
\end{align}
where $\lambda_0$ is the central wavelength of the line, $b$ is the Doppler width convolved with the instrumental dispersion, and $c$ is the speed of light. 
Therefore, in total, the four parameters of the Na D line model are $\lambda_0$, $\tau_b$, $b$, and $C_f$.

As many previous works cautioned, the absorption optical depth and covering fraction are degenerate when the Na D lines are blended \citep{Rupke2005a,Davies2024}. To properly account for this degeneracy and hence to better estimate the parameter uncertainties, we adopt the Markov Chain Monte Carlo (MCMC) method using \texttt{emcee} \citep{Foreman-Mackey2013} for the fitting procedure. 

The best-fit line parameters and their uncertainties are shown in Figure~\ref{fig:NaD_model_MCMC}, and the final best-fit ISM Na D profile is shown by the blue lines in the left panel of Figure~\ref{fig:NaD_pixel}. The ISM Na D component is significantly blueshifted ($\Delta v=-459^{+28}_{-24}\,\mathrm{km\,s^{-1}}$) from the stellar component (i.e. the systemic velocity), indicating that the Na D-traced neutral phase ISM is outflowing in JADES-GS-206183. 

The spectra of JADES-GS-206183 were extracted using a 6-pixel aperture (with a pixel size of 0.1"). However, as shown in the top left panel of Figure~\ref{fig:SMILES_spec}, JADES-GS-206183 exhibits an elongated substructure near its center in the NIRCam imaging -- presumably a stellar bar. This elongated structure is oriented at approximately $\sim$20 degrees from the north and is not well aligned with the orientation of the MSA shutter. As a result, the position of the flux centroid varies across the shutter. Since this effect is not accounted for in our spectral extraction procedure, it may artificially broaden the NaD absorption feature.

To investigate this, we extracted six spectra pixel by pixel. We find only the spectra of the central three pixels -- one from the center and two from positions one pixel above and below the center -- show a clear Na D feature. We then used pPXF to model the three spectra independently (right panel of Figure~\ref{fig:NaD_pixel}). We find that the Na D feature is significantly blueshifted in all three spectra. Moreover, the total strength of the ISM Na D absorption calculated by co-adding the fitting results of the three spectra is almost identical to that obtained from fitting the original combined spectrum extracted with a 6-pixel aperture. These results suggest that the blueshifted and strong Na D absorption measured using the single combined spectrum is robust.

Interestingly, we note that the strength of the Na D absorption is stronger in the spectrum extracted from one pixel above the center (the top spectrum in the right panel of Figure~\ref{fig:SMILES_bluejay_comp_Eout}) compared to the spectrum extracted from one pixel below the center (the bottom one). This suggests that the Na D outflow is preferentially originating from the upper-left (i.e., north-east) region of JADES-GS-206183. As we will present in Section~\ref{sec:OII_map}, we also observe possible evidence of an ionized outflow in the same direction based on the morphology of [OII] emission from MUSE observations of JADES-GS-206183.

\subsection{Pa $\alpha$}\label{sec:L_Paa}

\begin{figure}
\centering
\includegraphics[width=\columnwidth]{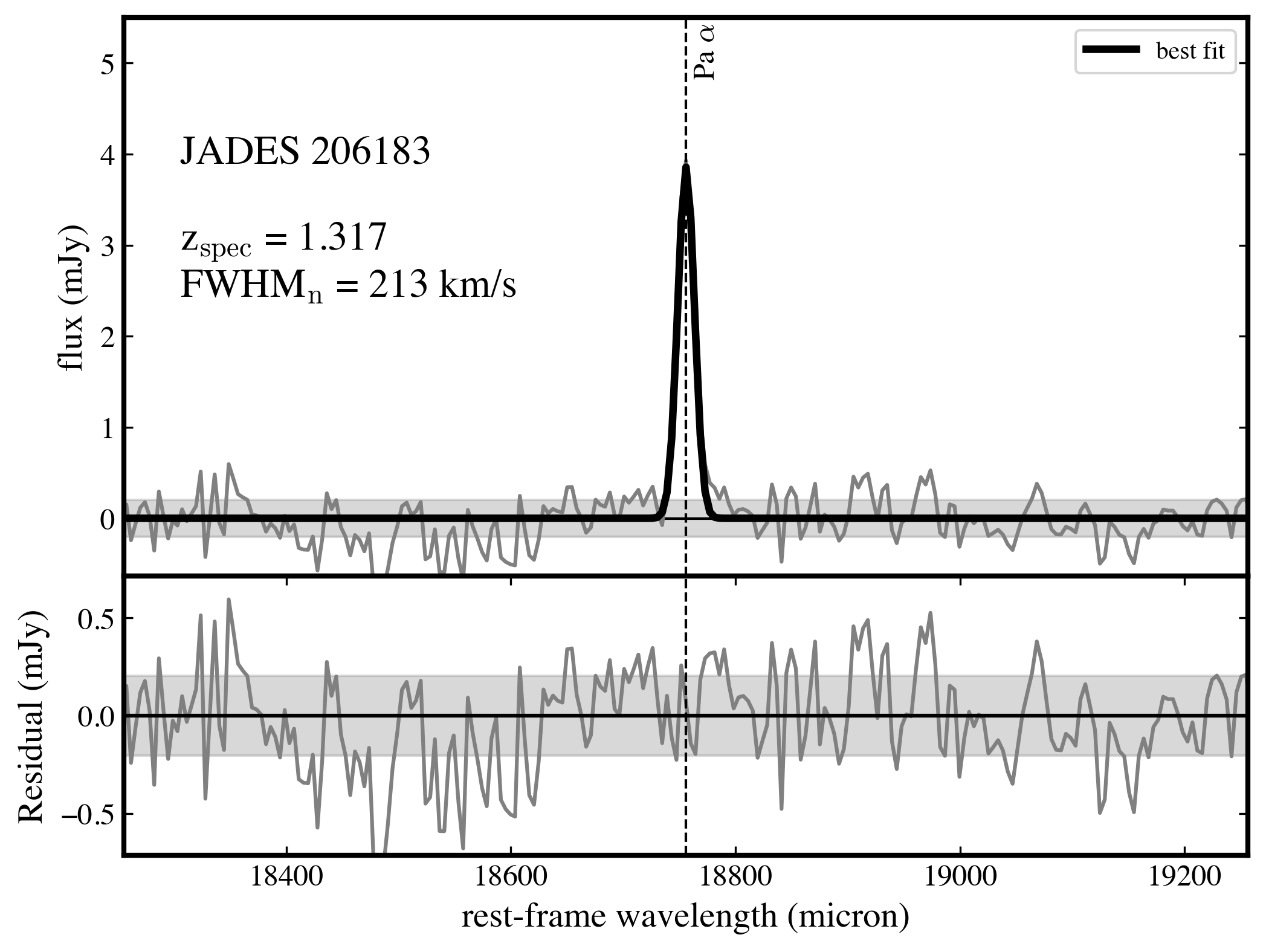}
\caption{Line profile fitting for the Pa $\alpha$. We show the extracted 1D spectrum (stellar continuum subtracted) and 1$\sigma$ error level (grey line and shading), and the best-fit line profile (black).}
\label{fig:fresco_paa}
\end{figure}

We estimated the SFR of JADES-GS-206183 using the Paschen $\alpha$ (Pa$\alpha$) emission in its FRESCO grism spectrum. Compared to other commonly-used SFR indicators, e.g., H $\alpha$, Pa$\alpha$ is a more accurate SFR tracer because it is substantially less affected by dust attenuation.

We first model the stellar continuum of the FRESCO spectrum using the same method as we introduced in Section~\ref{sec:pPXF} to estimate the contribution of the stellar Pa$\alpha$ absorption using pPXF. 
We mask Pa$\alpha$ during pPXF fitting. We fix the stellar kinematics to the pPXF results of the NIRSpec spectrum because the continuum is not well detected in the grism results.
We subtract the stellar continuum from the original spectrum and then apply single/double Gaussian profiles to model the Pa$\alpha$ emission using \texttt{lmfit}. Given that the H$\alpha$ from  JADES-GS-206183 shows a broad component (Section~\ref{sec:linefitting}), we also test the necessity of a broad component for Pa $\alpha$ using the BIC method. We find that the BIC of the ``narrow+broad" case is even larger than the narrow-only case ($\Delta \text{BIC}=-5$), indicating Pa $\alpha$ only needs a narrow component. The final best-fit of Pa $\alpha$ is shown in Figure~\ref{fig:fresco_paa}. The observed Pa $\alpha$ flux is $1.68\pm0.08\times10^{-17}\,\mathrm{erg\,s^{-1}\,cm^{-2}}$; and its intrinsic width is $213\pm17\,\mathrm{km\,s^{-1}}$, consistent with the width of the narrow component of the emission lines detected in the $R\sim1000$ NIRSpec spectrum.

With H$\alpha$ (details of the measurement process will be presented in Section~\ref{sec:AGN}) and Pa $\alpha$, we derive the gas reddening ($A_V$) using the Balmer decrement method\footnote{We do not adopt the traditional H$\beta$/H$\alpha$ method because the flux measurement of H$\beta$ is relatively uncertain since its apparent emission is reduced by the stellar absorption.} and the \citet{Calzetti2000} dust attenuation law. Given that the total H$\alpha$ profile contains a broad outflow component, we only use the narrow H$\alpha$ flux here to calculate the attenuation. Under Case B recombination (and assuming $T_e=10^4$ K and $n_e=100$~cm$^{-2}$), the intrinsic flux ratio between H$\alpha$ and Pa $\alpha$ is 8.13 \citet{Osterbrock2006}. The observed flux ratio of 1.98$\pm$0.28, corresponding to a gas reddening $A_V$ of 2.27 $\pm$ 0.23 magnitude. This spectroscopically-derived $A_V$ is generally consistent with the SED-derived value (see Section~\ref{sec:sed_model}).

It is important to note that the H$\alpha$ flux is derived from NIRSpec slit spectroscopy that primarily captures light within the MSA shutter, while the Pa $\alpha$ comes from the NIRCam grism spectrum of the entire galaxy. As a result, the $A_V$ inferred from the H$\alpha$/Pa$\alpha$ ratio is sensitive to the slit-loss correction applied to the NIRSpec data, which assumes the line-to-continuum flux ratio is invariant across galaxy radius. (see Section~\ref{sec:nirspec}).
To assess the impact of slit-loss correction on $A_V$, we extract the grism spectrum from an aperture approximately matching the NIRSpec MSA shutter size ($\sim0.4"$) and measure both the H$\alpha$ and Pa $\alpha$ fluxes from the central region of JADES-GS-206183 where the shutter is placed. In this configuration, the H$\alpha$ and Pa $\alpha$ fluxes would become $\sim$4 and 2 times fainter, respectively, resulting in a smaller H$\alpha$-to-Pa $\alpha$ ratio ($\sim0.9$) and thus a larger $A_V$ ($3.53$ mag). This indicates that the $A_V$ derived after slit-loss correction could underestimate the dust attenuation towards the central region of JADES-GS-206183 by about one magnitude. 
However, such a one magnitude difference of $A_V$ would not change any scientific results of this paper, specifically the AGN obscuration estimation and the broad-H$\alpha$ origin (Section~\ref{sec:BHa}).
Moreover, the $A_V$ derived from this alternative approach also carries significant uncertainties, as it neglects several important factors, including 1) differences in spectral dispersion directions and PSFs of the NIRSpec grating and NIRCam grism observations, and 2) the off-center placement of the NIRSpec shutter. 
Therefore, we still adopt the $A_V$ corrected for NIRSpec slit losses for the following analysis.

After correcting for the dust attenuation using the $A_V$,
we derive the intrinsic Pa$\alpha$ line luminosity, which is $2.34\pm0.11\times10^{41}\,\mathrm{erg\,s^{-1}}$.
Finally, using the Pa$\alpha$-SFR relation from \citet{Neufeld2024} which is based on the \citep{Chabrier2003} IMF: 
\begin{equation}
    \text{SFR}\,\mathrm{(M_{\odot}\,yr^{-1})}=4.6\times10^{-41}\times L(\text{Pa}\alpha)\,\mathrm{(erg\,s^{-1})},
\end{equation}
we obtain a $\text{SFR}_{\text{Pa}\alpha}$ of $10.78\pm 0.55\,\mathrm{M_{\odot}\,yr^{-1}}$. JADES-GS-206183 is $\sim0.5$ dex below the star forming galaxy main sequence at $z\sim1.5$ (with a scatter of 0.3-0.4 dex) \citet{Leja2022}, confirming its quiescent/PSB nature.

\subsection{[OII] $\lambda\lambda$ 3726,3729 map}\label{sec:OII_map}

\begin{figure*}
\centering
\includegraphics[width=1\textwidth]{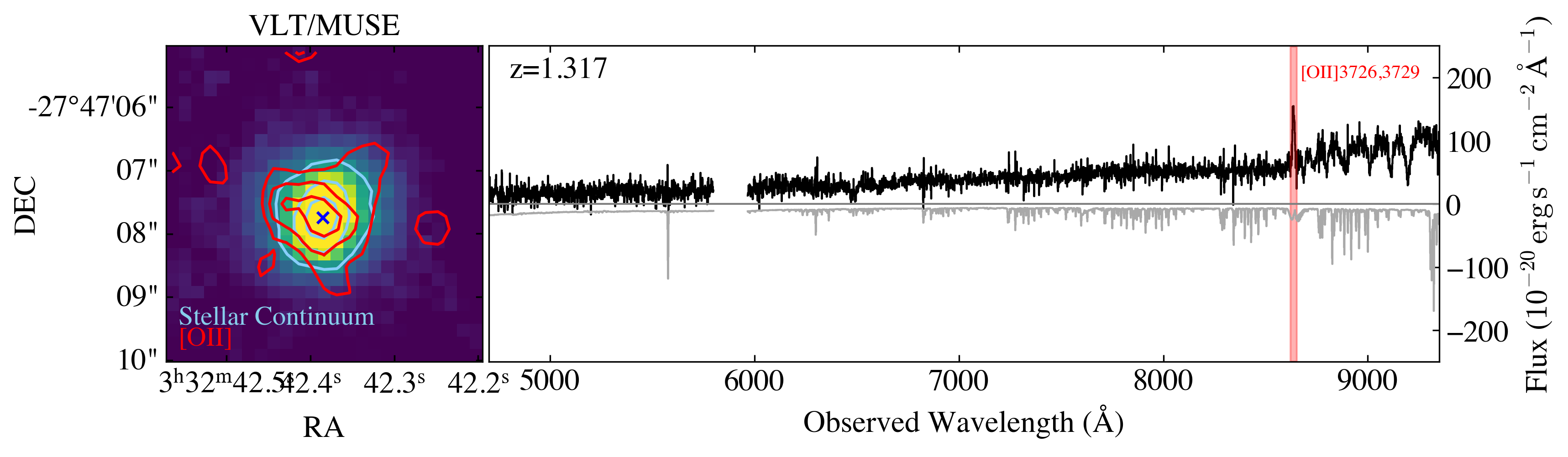}
\caption{MUSE observation of JADES-GS-206183. {\bf Left:} The AMUSED 5" $\times$ 5" white light map tracing stellar continuum light with the blue contour level of 10$\sigma$, 20$\sigma$, and 30$\sigma$. The blue cross marks the center of the stellar emission. The contour of [OII]$\lambda\lambda3726, 3729$ with levels of 1$\sigma$, 1.5$\sigma$ and 2$\sigma$ (red) is overlaid on the above.
{\bf Right:} The integrated 1D spectrum (black) extracted based on the Rafelski segmentation map \citep{Rafelski2015} and the corresponding flux uncertainties (gray). The red-shaded region represents the integrated spectral window for mapping the [OII] emission.}
\label{fig:MUSE_OII}
\end{figure*}

We use VLT/MUSE data for JADES-GS-206183 acquired by the MUSE Hubble Ultra-Deep Field survey from AMUSED\footnote{\url{https://amused.univ-lyon1.fr/project/UDF/}}.
Specifically, JADES-GS-206183 was covered by the MXDF -- a part of the MUSE HUDF survey. The MUSE spectrum covers the optical regime (4750--9300\AA) at a average spectral resolution of $R\sim3000$. The details of MUSE data reduction and AMUSED data products can be found in \citet{Bacon2023}.

We study the spatial distribution of [O II] $\lambda\lambda3726\&3729$ -- the only strong emission line detected in the MUSE data for  JADES-GS-206183. The full 1D spectrum is shown in Figure~\ref{fig:MUSE_OII}. To obtain the [O II] map, we derive the [O II] $\lambda3726\&3729$ flux in each spaxel by integrating the fluxes within the rest-frame wavelength range of [3721, 3735]\AA~ after subtracting the average of the adjacent continuum. 

The white light stellar continuum image along with the contours of [O II] flux derived from the MUSE data for JADES-GS-206183 is shown in Figure~\ref{fig:MUSE_OII}. Relative to the stellar continuum, the [O II] emission appears to lie toward the upper-left of the galaxy, suggesting a possible ionized outflow (traced by [O II]) in JADES-GS-206183. Notably, this direction also coincides with the observed Na D outflow (Section~\ref{sec:Na D}), indicating that, if real, the ionized outflow may be coupled with the neutral gas outflow. However, given that the MUSE observation of JADES-GS-206183 was conducted with a seeing limit of $\sim0.8''$, we emphasize that no definitive conclusions can be drawn at this stage from the MUSE observation regarding the presence of an ionized outflow in JADES-GS-206183.

\section{SED modelling}
\label{sec:sed_model}

\begin{figure*}
\centering
\includegraphics[width=\textwidth]{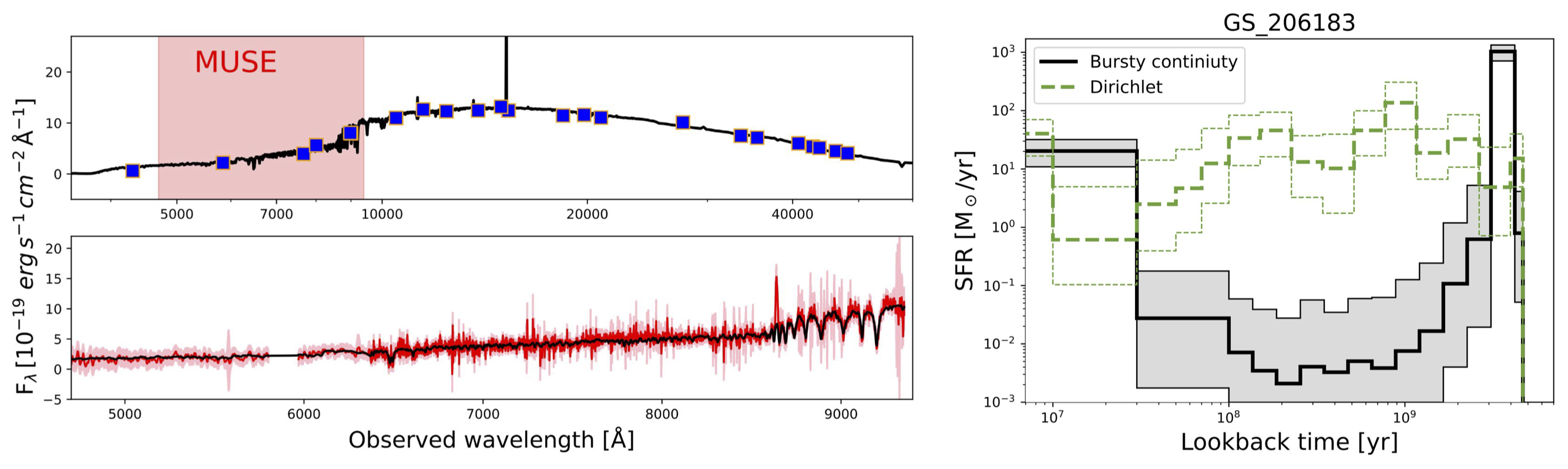}
\caption{{\bf Left: } Prospector SED fitting results for JADES-GS-206183. We simultaneously model the photometry from HST and JWST/NIRCam and the spectrum from VLT/MUSE (see Section~\ref{sec:sed_model}). The best-fit SED model, based on our default bursty continuity SFH, is shown as a black line. The blue squares in the top panel indicate the observed HST/ACS and JWST/NIRCam photometry. We zoom into the region where the MUSE spectrum covers in the bottom panel, where the observed MUSE spectrum (same as the right panel of Figure \ref{fig:MUSE_OII}) is shown in red, and again, the black line is the best-fit SED model. The SED model shows excellent agreement with the data, successfully reproducing not only the overall SED shape from rest-frame UV to NIR, but also the key stellar absorption features around rest-frame 4000 \AA\ in the MUSE spectrum. {\bf Right:} Reconstructed nonparametric SFHs of JADES-GS-206183. The black solid line with shaded region shows the best-fit and 1$\sigma$ uncertainty from the bursty continuity SFH fitting, while the green dashed line shows the result from the Dirichlet SFH fitting. Systematic uncertainties regarding the assumed SFH priors are evident in the reconstructed SFHs. Nonetheless, both reconstructions suggest that the SFR of JADES-GS-206183 has risen again over the past 10-30 Myr, following a decline since an earlier star-forming period.}
\label{fig:JADES-GS-206183_sfh}
\end{figure*}
We infer the stellar population properties of JADES-GS-206183 through SED fitting with Prospector \citep{Johnson2021}, using photometry from 22 filters spanning HST/ACS, HST/WFC3, and JWST/NIRCam, along with the integrated 1D spectrum from VLT/MUSE. The MUSE spectrum covers key stellar absorption features, including Balmer lines and the Ca H+K doublet, which provide strong constraints on stellar ages and star formation histories.

\subsection{Assumptions}
Our SED assumptions, including parameter priors, are similar to those used in \citet{Ji2024}. In brief,
we fix the redshift to the spectroscopic redshift by pPXF ($z=1.317$). We use the Kroupa stellar IMF \citep{Kroupa2001}, and adopt the FSPS stellar synthesis code \citep{Conroy2009,Conroy2010} with the stellar isochrone libraries MIST \citep{Choi2016}
and the MILES stellar spectral libraries \citep{Sanchez-Blazquez2006,Falcon-Barroso2011}. We use the Madau IGM transmission model \citep{Madau1995}. We include the nebular emission model from \citet{Byler2017}. We set the stellar metallicity ($Z_*$), the gas metallicity($Z_{\text{gas}}$), and the ionization parameter (U) as free parameters. We use a two-component dust attenuation model, where the attenuation toward nebular emission and young ($<10$ Myr) stellar
populations are modeled by a power law, and towards old ($>10$ Myr) stellar populations are treated following the
parameterization of \citet{Noll2009}, a modified \citet{Calzetti2000} law with the 2175\AA ~dust feature.

To properly model the MUSE spectrum, we include the LSF information from \citep{Bacon2023}. We include a multiplicative 12th-order Chebyshev polynomial calibration model to account for effects such as imperfect flux calibration in ground-based spectroscopy that can impact the continuum shape -- a strategy widely adopted in full-spectrum modeling of galaxies \citep[e.g.,][]{Kelson2000,Conroy2018,Tacchella2022}. In addition, we incorporate a noise model to account for outliers in the spectral data points, following the approach of \citet{Hogg2010}, which mitigates the influence of ``bad'' data points by reducing sensitivity to them. For this noise model, we adopt the same priors as described in \citet[][Section 4.2]{Tacchella2022}.

For the star formation history (SFH), our default model adopts a nonparametric, piecewise form with 15 lookback time bins and the bursty continuity prior \citep{Tacchella2022}. In the bursty continuity prior, the SFR change between two adjacent bins is drawn from a Student's T distribution with $\sigma=1$ and $\nu=2$ \footnote{In Student's T distribution, $\sigma$ controls the width of the distribution, and $\nu$ is the number of degrees of freedom that governs the heaviness of the distribution's tails.}. 
This is a modified version of the standard continuity prior of \citet{Leja2019} ($\sigma=0.3$), allowing for more flexible (i.e., burstier) change while retaining an overall smooth SFH. The bursty continuity prior has been commonly adopted for high-redshift galaxy SED modelling (e.g., \citealt{Tacchella2022,Park2024,Looser2024}), motivated by growing evidence that galaxy SFHs are significantly more bursty in the early Universe. The first five bins are fixed at 0-10, 10-30, 30-50, 50-70, and 70-100 Myr to capture recent episodes of star formation with relatively high time resolution. The final bin spans 
$0.9t_H - t_H$ where $t_H$ is the Hubble time at $z=1.317$. The remaining nine bins are evenly spaced in logarithmic time between 100 Myr and $0.9 t_H$. We also test our fitting results using the commonly adopted Dirichlet prior from \citet{Leja2017}. We find the derived SED properties -- stellar mass, SFR, or $A_V$ -- for JADES-GS-206183 between these two priors are generally consistent with each other within systematic uncertainties, although the SFR and $A_V$ based on the bursty continuity prior are closer to the emission line results. In addition, some differences arise in the reconstructed SFHs between two priors. We discuss these SED results and associated systematics in detail below.

\subsection{SED properties}\label{sec:sed_result}
The best-fit SED model and reconstructed SFHs are shown in Figure~\ref{fig:JADES-GS-206183_sfh}. Our modeling provides an excellent match to the data, successfully reproducing not only the overall panchromatic SED shape but also the key stellar absorption features and the continuum shape in the MUSE spectrum. The inferred stellar mass is $\log(M_*/M_{\odot}) = 11.25 \pm 0.02$ ($10.97 \pm 0.03$) for the bursty continuity (Dirichlet) SFH prior. The SED-inferred $A_V$ is $1.76 \pm 0.05$ ($1.09 \pm 0.05$), which is slightly lower but broadly consistent with the value of $2.27 \pm 0.23$ derived from H$\alpha$ and Pa$\alpha$ (Section~\ref{sec:L_Paa}).

Regarding the star formation rate, the burst continuity (Dirichlet) prior results in $\text{SFR}_\text{SED} = 20 \pm 10\,(40 \pm 27)\,\mathrm{M_{\odot}\,yr^{-1}}$, which would place this galaxy at $\sim0.3$ (0.1) dex below the main-sequence. The SED-inferred SFRs of JADES-GS-206183 are 2-4 times higher than its Pa$\alpha$ ($\sim 10.8\,\mathrm{M_{\odot} yr^{-1}}$; Section~\ref{sec:L_Paa}) or CO-inferred SFR ($\sim 11.0\,\mathrm{M_{\odot} yr^{-1}}$; \citealt{Boogaard2019,Aravena2020}), but such differences can be explained by the systematics between SED-modelled and observed H$\alpha$ found by \citet{Leja2017} (see their Fig. 9).
In addition, the higher SFR derived from the Dirichlet prior compared to the bursty continuity prior for JADES-GS-206183 ($\sim0.3$ dex) is also consistent with the result of \citep{Leja2019}, who found that, for the systems with quiescent-like SFH, the Dirichlet SFR is systemically higher than that obtained with the continuity prior with the latter being closer to the intrinsic SFR (their Fig. 5). Considering all the systematic effects, we think the SED results with the bursty continuity prior is in good agreement with the Pa$\alpha$-based estimates presented in Section~\ref{sec:L_Paa} and thus more reliable for our target. Hereafter, we adopt the Pa$\alpha$-derived SFR and H$\alpha/$Pa$\alpha$-derived $A_V$ for our analysis.

Finally, we note that our SED fitting does not include an AGN component. Nevertheless, the excellent agreement between the data and a stellar-only model suggests that JADES-GS-206183 does not host a strong ongoing AGN. Further evidence against the presence of an AGN will be presented in Section~\ref{sec:AGN}.

Although the two SFH fittings yield overall consistent results for the physical properties discussed above, the reconstructed SFHs reveal systematic uncertainties introduced by the choice of SFH prior (right panel of Figure~\ref{fig:JADES-GS-206183_sfh}). The bursty continuity prior suggests that the bulk of the stellar mass in JADES-GS-206183 was assembled very early, around 3 Gyr ago, whereas the Dirichlet prior places the main formation episode much later, at approximately 0.5-1 Gyr ago. This systematic effect -- where the continuity prior tends to return older stellar populations than the Dirichlet prior -- has also been reported in previous studies \citep[e.g.,][]{Tacchella2022,Ji2022}. Despite this difference, both fittings converge on a qualitatively similar evolutionary trend: a decline in star formation rate following the primary buildup of stellar mass. By the time of observation, both reconstructed SFHs suggest a modest rise in star formation activity within the past $\sim$10-30 Myr. Nevertheless, by the time of observation, the sSFR of JADES-S-206183 remains low, on the order of $10^{-10}$ yr$^{-1}$ (and 0.5 dex below the main-sequence), consistent with the galaxy's overall quiescent nature. These findings suggest that we may be witnessing the onset of star-formation rejuvenation in JADES-GS-206183, which could either bring it back toward the star-forming main sequence or represent only a brief, weak episode of star formation that may soon be fully quenched again (see Section~\ref{sec:nad_result}).

\section{Results} \label{sec:result}

\subsection{Physical properties of the Na D outflow}
\label{sec:nad_result}

Our SMILES NIRSpec spectroscopy reveals the presence of a substantial outflow of neutral gas in JADES-GS-206183, as traced by Na D absorption which is blueshifted by $-459\,\mathrm{km\,s^{-1}}$ (Section~\ref{sec:Na D}). With modeling results for the Na D line presented in this section, we further
measure the physical properties of this neutral outflow in JADES-GS-206183.

In particular, we derive the outflow velocity ($v_{\rm out}$), outflow mass and rate (${M}_{\rm out}$ and $\dot{M}_{\rm out}$), outflow momentum and its rate (${p}_{\rm out}$ and $\dot{p}_{\rm out}$), and total energy and its rate (${E}_{\rm out}
$ and $\dot{E}_{\rm out}$). We also calculate the equivalent width ($\text{EW}_{\text{NaD, ISM}}$), a line quantity that also indicates the amount of outflowing mass. The assumptions used to infer these physical properties of outflows are the same as those adopted by the Blue Jay survey \citep{Davies2024}, which has also conducted a JWST/NIRSpec study of Na D outflows for a sample at cosmic noon. Specifically, the outflow velocity is defined as $v_{\rm out} = \left|\Delta v\right| + 2\sigma$, where $\sigma=b/\sqrt{2}$ is the velocity dispersion of the ISM Na D line. For the outflow mass rate calculation, we use the relation:
\begin{align*}
\dot{M}_{\rm out} &= \frac{M_{\text{\rm out}}v_{\text{\rm out}}}{r_{\text{\rm out}}}\\
&= 11.45\left(C_{\Omega}\frac{C_f}{0.4}\right)\left(\frac{\text{N(H I)}}{10^{21}\,\mathrm{cm^{-2}}}\right)\\
&\times\left(\frac{r_{\text{\rm out}}}{1\,\mathrm{kpc}}\right)\left(\frac{v_{\text{\rm out}}}{200\,\mathrm{km\,s^{-1}}}\right)\,\mathrm{M_{\odot}\,yr^{-1}},
\end{align*}
where $r_{\text{\rm out}}$ is the outflow extent, $C_{\Omega}$ is the covering factor related to the wind opening angle, $\text{N(H I)}$ is the hydrogen column density. We assume $r_{\text{\rm out}}=1\,\mathrm{kpc}$ and $C_{\Omega}=0.5$ as \citet{Davies2024} did. Another reason for assuming, $r_{\text{\rm out}}=1\,\mathrm{kpc}$ is that it is comparable to the physical scale that the NIRSpec MSA shutter size ($r\sim$0.15 arcsec) corresponds to at z = 1.317, which ensures the mass outflow rate measurement is spatially self-consistent. $\text{N(H I)}$ is derived from $\text{N(Na I)}$ assuming Milky-Way-like Na abundance and dust depletion factors and a 10\% neutral fraction (see \citealt{Rupke2005a}), where $\text{N(Na I)}$ is measured from the best-fit optical depth at the central ISM red Na D line $\tau_{r, 0}$:
\begin{align*}
\text{N(Na I)} &= 10^{13}\left(\frac{\tau_{r, 0}}{0.758}\right)\left(\frac{0.4164}{f_{lu}}\right)\\
&\times\left(\frac{1215\mathrm{\AA}}{\lambda_{lu}}\right)\left(\frac{b}{10\,\mathrm{km\,s^{-1}}}\right)\,\mathrm{cm^{-2}},
\end{align*}
where $f_{lu}=0.32$ and $\lambda_{lu}=5897.55\,\mathrm{\AA}$ are the oscillator strength and rest-frame wavelength of the transition. 

The uncertainty of $\dot{M}_{\rm out}$ derived from the MCMC fitting, i.e., random error, is only about $\sim0.1$ dex, while its systematic errors coming from the assumptions of outflow structures and composition can be much larger. As \citet{Davies2024} discussed, $C_{\Omega}$ is unknown but cannot be smaller than 0.25 due to the outflow detection fraction in the Blue Jay survey, which can cause a factor of two difference in the inferred $\dot{M}_{\rm out}$. Also, the neutral fraction of the outflow can range from 5\% \citep{Baron2020} to 10\% (used by the Blue Jay survey), which could introduce another factor of two difference. Considering those two factors, the systematic uncertainty of the $\dot{M}_{\rm out}$ is about 0.6 dex. Furthermore, the radius of the outflow extent $r_{\text{\rm out}}$ is poorly constrained for Na D outflows at cosmic noon. However, given that the previous spatially-resolved outflow studies \citep[e.g.,][]{Rupke2015,Baron2020,D'Eugenio2024} suggest that 1 kpc is a lower limit of the size of Na D outflows which usually are even larger (a few to 15 kpc), our assumption would only result in the underestimation of $\dot{M}_{\rm out}$ that could even make JADES-GS-206183 more interesting than what we will state in Section~\ref{disc: current}. Moreover, since the NIRSpec observation of JADES-GS-206183 only points to its central region, the exact $\dot{M}_{\rm out}$ and $\dot{M}_{\rm out}$ of this galaxy are expect to be higher than what we report, which could only be derived from IFU data.

After measuring the outflowing mass and mass outflow rate, we are then able to calculate the momentum and energy outflow rates. The final measurements of the outflow properties are listed in Table~\ref{tbl:JADES-GS-206183_info}.

\begin{table*}[h!]
    \centering
    \begin{tabular}{cccccccccc}
    \hline
    \hline
        $v_{\text{\rm out}}$ & $\text{EW}_{\text{NaD, ISM}}$ & $\log(\text{N(Na I)})$ &          $\log(\text{N(H I)})$ & $\log({M}_{\text{\rm out}})$ &
        $\log(\dot{M}_{\text{\rm out}})$ &
        $\log({p}_{\text{\rm out}})$ &
        $\log(\dot{p}_{\text{\rm out}})$ &
        $\log({E}_{\text{\rm out}})$ &
        $\log(\dot{E}_{\text{\rm out}})$ \\
        $(\mathrm{km\,s^{-1}})$ & $(\mathrm{\AA})$ & $(\mathrm{cm^{-2}})$ & 
        $(\mathrm{cm^{-2}})$ & $(\mathrm{M_{\odot}})$ &
        $(\mathrm{M_{\odot}\,yr^{-1}})$ & $(\text{dyn\,s})$
        & $(\text{dyn})$ & $(\mathrm{erg})$  &
        $(\mathrm{erg\,s^{-1}})$ \\
         \hline
        $828^{+79}_{-49}$ & 
        $8.65^{+0.97}_{-0.96}$&
        $14.39^{+0.13}_{-0.23}$ & $22.03^{+0.13}_{-0.23}$ & $8.47^{+0.12}_{-0.19}$ & $2.40^{+0.11}_{-0.16}$ & $49.69^{+0.11}_{-0.16}$ & $36.11^{+0.11}_{-0.14}$ & $57.30^{+0.11}_{-0.14}$ & $43.73^{+0.11}_{-0.12}$ \\
    \hline
    \end{tabular}
    \caption{Neutral outflow properties of 206183.}
    \label{tbl:JADES-GS-206183_info}
\end{table*}

\subsection{Comparison with Na D outflows in the literature}\label{sec:comparison}

\begin{figure*}
\centering
\includegraphics[width=1\textwidth]{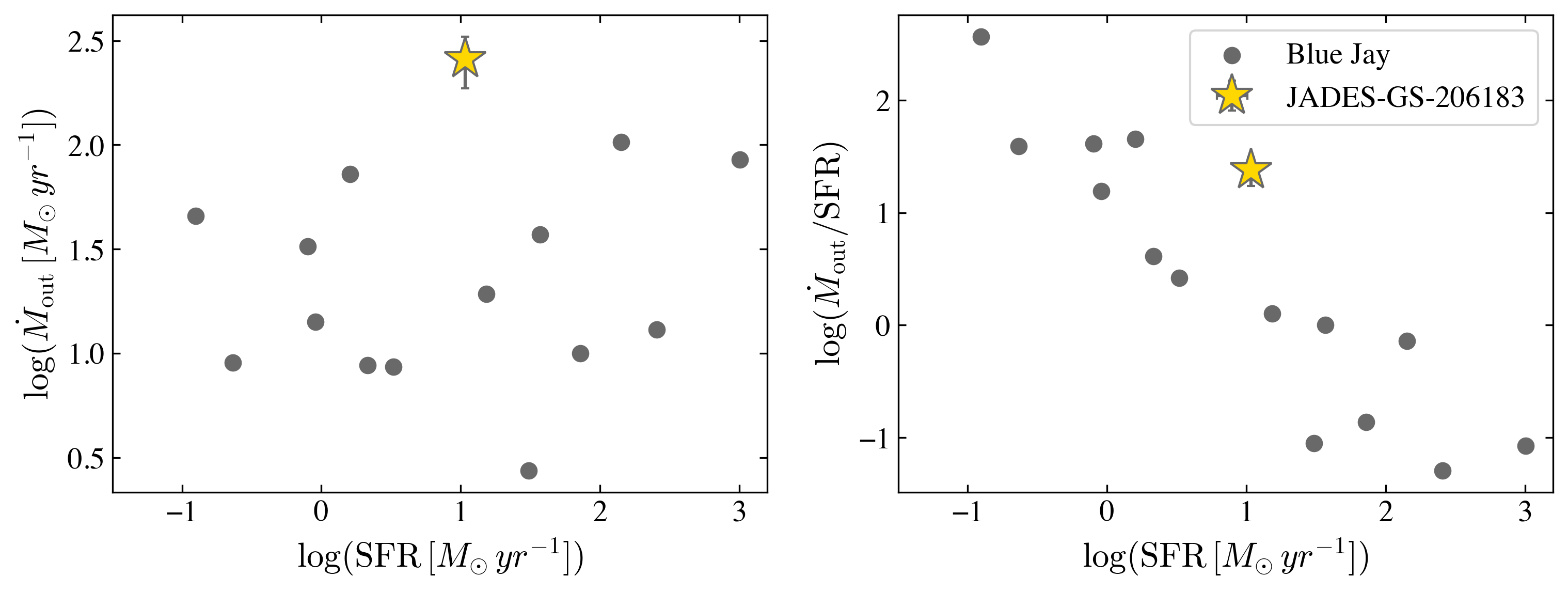}
\caption{Comparison between JADES-GS-206183 (yellow stars) and the 14 Blue Jay z$\sim$2 Na D outflows \citep{Davies2024} (gray dots) in the SFR-mass outflow rate (left) and SFR-mass loading factor (right) planes. JADES-GS-206183's Na D outflow exhibits a higher mass outflow rate and thus higher mass loading factor compared to the Na D outflows detected in the galaxies with similar SFR ($\log(\text{SFR}/{\rm M_{\odot}\,yr^{-1}})\sim1$) and redshift.}
\label{fig:SMILES_bluejay_comp_Mout}
\end{figure*}

\begin{figure*}
\centering
\includegraphics[width=1\textwidth]{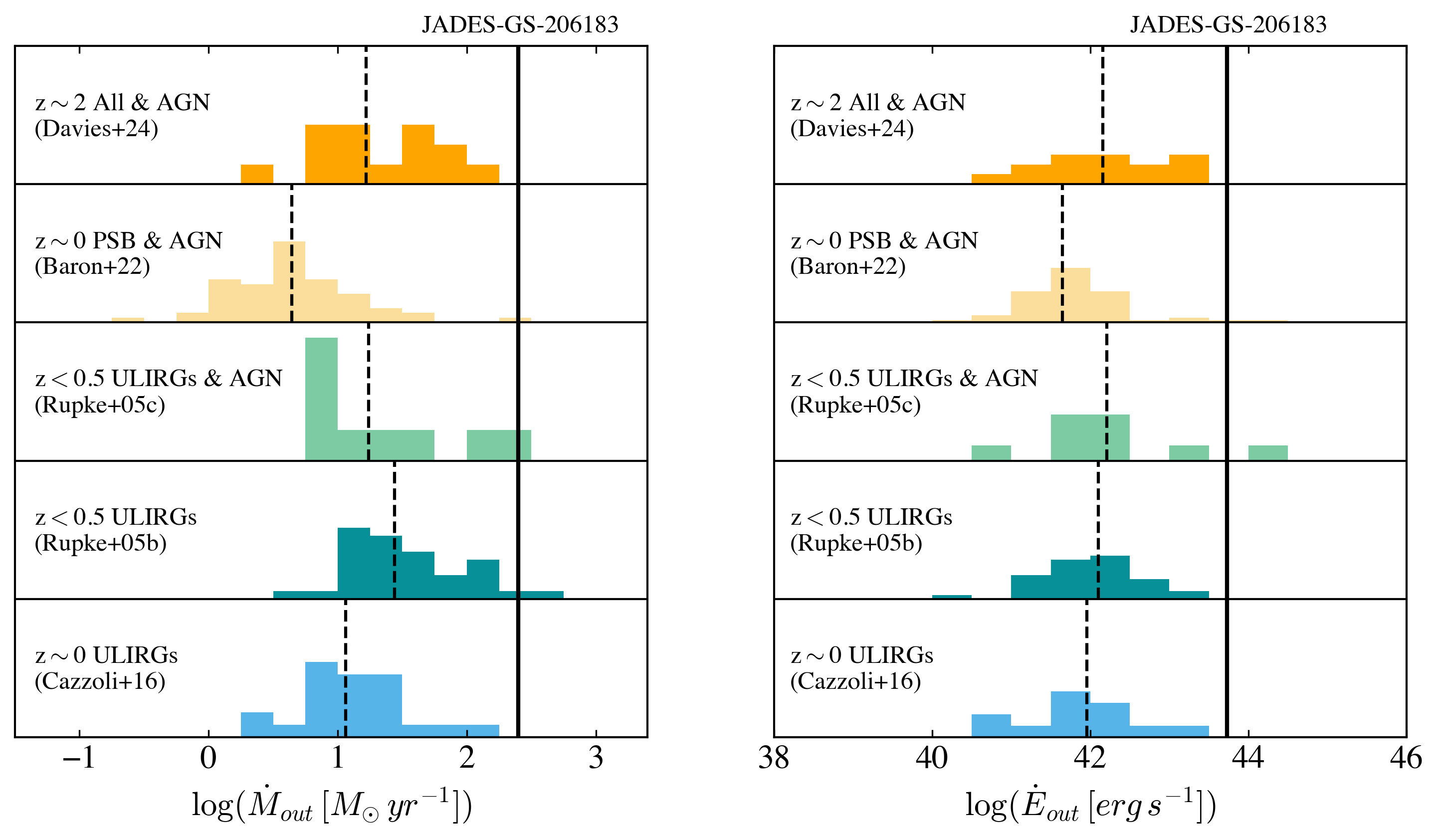}
\caption{Comparison of neutral Na D outflow properties ($\log(\dot{M}_{\rm out})$ on the left and $\log(\dot{E}_{\rm out})$ on the right) in different samples, including 1) the outflows detected in the  various types of $z\sim2$ galaxies from the Blue Jay survey \citep{Davies2024}; 2) the local PSB AGN hosts from \citet{Baron2022}; 3) the $z<0.5$ Seyfert-2 ULIRGs from \citet{Rupke2005c}; 4) the $z<0.5$ SF/LINER ULIRGs from \citet{Rupke2005b}; 5) the local pure-star forming ULIRGs from \citet{Cazzoli2016}. The median values of each sample are shown by a black dashed line. The ($\log(\dot{M}_{\rm out})$ and $\log(\dot{E}_{\rm out})$ measurements of JADES-GS-206183 are illustrated by a solid  line. Considering JADES-GS-206183 is an inactive PSB, the mass and energy outflow rate of its Na D outflow is surprisingly high compared to either the local or $z\sim2$ outflows that are driven by powerful AGN or starbursts.}
\label{fig:loal+z2_comp}
\end{figure*}

In this section, we compare the properties of the Na D outflow in JADES-GS-206183 with Na D outflows detected in other galaxies at cosmic noon and in the local Universe. 

At cosmic noon, the Blue Jay survey identified 14 Na D outflows (referred to as Blue Jay Na D outflows hereafter) at z$\sim$1.7--3.5 using the NIRSpec MSA R$\sim$1000 spectra \citep{Davies2024}. These outflows are found in galaxies spanning a range of types, from star-forming to quiescent, as well as both AGN-host and non-AGN host galaxies. The range of $\log(\dot{M}_{\rm out}/[M_{\odot}\,{\rm yr^{-1}}])$ found for those Blue Jay Na D outflows is about 0.5--2. Additionally, they found a decreasing trend in the mass loading factors ($\eta=\dot{M}_{\text{\rm out}}/{\rm SFR}$) of outflows with the SFRs of their host galaxies: the Na D outflows hosted by star-forming galaxies (${\rm SFR}>10\,M_{\odot}\,yr^{-1}$) have a low mass loading factor ($\eta<1$) while those in quiescent galaxies have $\eta>1$. Interestingly, as shown in Figure~\ref{fig:SMILES_bluejay_comp_Mout},
JADES-GS-206183 has a mass loading factor $\sim1$ dex higher than the values of the Blue Jay outflows at this SFR range (typically around $\eta=0$). The result indicates that the neutral outflow in JADES-GS-206183 is much more efficient in removing gas from the galaxy relative to the current formation of stars, compared to other observed Na D outflows launched in galaxies with comparable SFR.

We also compare the $\dot{M}_{\rm out}$ and $\dot{E}_{\rm out}$ of JADES-GS-206183 with Na D outflow measurements in the local Universe, including outflows found in Ultra Luminous Infrared Galaxies (ULIRGs) with AGN \citep{Rupke2005c} and without AGN (\citealt{Rupke2005b} and \citealt{Cazzoli2016}), and post-starburst galaxies with AGN \citep{Baron2022}. 
As shown in Figure~\ref{fig:loal+z2_comp}, the mass and energy loss through the Na D outflow in JADES-GS-206183 are comparable with the highest of the local Na D outflows, rivaling  those driven by intensive star forming or AGN activity. It is worth noting that, except for \citealt{Cazzoli2016}, in which the local Na D outflow properties are relatively precisely measured through spatially-resolved spectroscopic data, other literature derived the outflow properties using limited-aperture or long-slit spectra  probe only the central region of the galaxy, similar to our analysis on JADES-GS-206183. 
As we mentioned in Section~\ref{sec:nad_result}, such aperture-limited observations generally yield lower limits on the mass and energy outflow rates; thus, the comparison between JADES-GS-206183 and values reported in the local Na D outflow literature is fair, and the outflow in JADES-GS-206183 unquestionably stands out.

\subsection{AGN or not?}
\label{sec:AGN}

\begin{figure*}
\centering
\includegraphics[width=\textwidth]{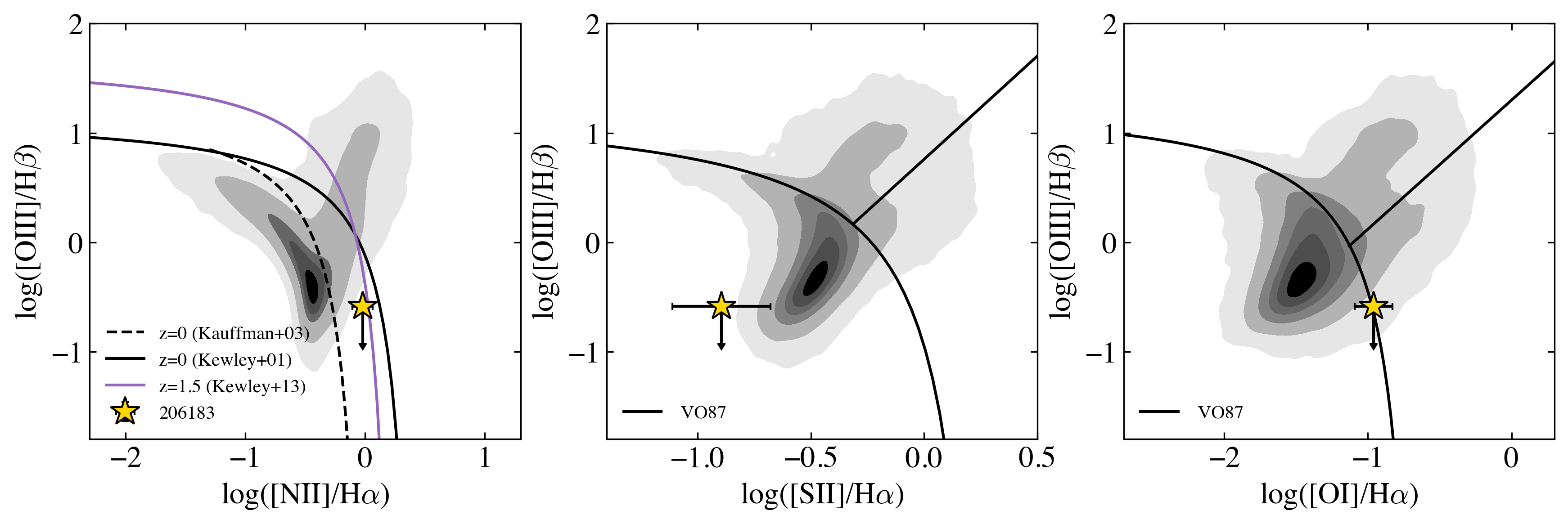}
\caption{Location of JADES-GS-206183 (yellow stars) in the $[\NII]$-BPT diagram (left), $[\SII]$-VO87 diagram (middle), $[\OI]$-VO87 diagram (right). The gray contours show the distribution of SDSS galaxies. The purple line in the left panel represents the redshift-dependent SF-AGN boundary at z$\sim$1.5 determined by \citet{Kewley2013}. }
\label{fig: BPT}
\end{figure*}

A natural explanation for exceptionally strong outflows is the presence of an AGN. In this section, we investigate the AGN activity in JADES-GS-206183 using different AGN diagnostics. 

\subsubsection{Archival multi-wavelength studies} 
First, we check whether JADES-GS-206183 has been classified as an AGN in the literature. In particular, we use the AGN catalogs from \citet{Lyu2022} and \citet{Lyu2024}, which conducted one of the most comprehensive searches -- using data from X-ray to radio, including the latest NIRCam and MIRI data -- for AGN populations at cosmic noon in GOODS-S. JADES-GS-206183 was not classified as an AGN host in these catalogs. In addition, JADES-GS-206183 has CO(2-1) and CO(5-4) observations from the ASPECS ALMA large program, which suggests that the CO excitation level of this galaxy is lower than that of AGN hosts, being more comparable to that of normal star-forming galaxies (Figure 5 in \citealt{2020ApJ...902..109B}), again, indicating JADES-GS-206183 does not host an on-going AGN.

\subsubsection{Optical line ratio diagnostics}

With our new NIRSpec/MSA observations, we can now investigate the AGN signatures (if any) in JADES-GS-206183 through its rest-optical spectral features. We first use the optical line ratio diagnostics, including the traditional [$\NII$]-BPT diagram \citep{Baldwin1981}, and also [$\SII$]- and [$\OI$]-VO87 diagrams \citep{Veilleux1987}. As we mentioned in Section~\ref{sec:linefitting}, the [$\OIII$] emission in JADES-GS-206183 is tentatively detected (1.5$\sigma$)(see Figure~\ref{fig:SMILES_spec}). Thus, we treat the [$\OIII$] flux as an upper limit. The BPT diagram is displayed in the left panel of Figure~\ref{fig: BPT}, where JADES-GS-206183 is located in the AGN-SF composite region between the boundaries defined by \citet{Kewley2001} and \citet{Kauffmann2003} at $z=0$. Additionally, JADES-GS-206183 falls below the SF-AGN boundary at z$\sim$1.5, as determined by \citet{Kewley2013}. Thus, based on the BPT diagram, the ionization of JADES-GS-206183 is not consistent with an AGN. 

The right two panels of Figure~\ref{fig: BPT} also display the [$\SII$]- and [$\OI$]-VO87 diagrams. On the [$\SII$]-VO87 diagram, JADES-GS-206183 is sitting in the SF region, with a remarkably low [$\SII$]/H$\alpha$ ratio. Given the low SNR of [$\SII$] in JADES-GS-206183, we caution that this abnormally low [$\SII$]/H$\alpha$ ratio may be due to the uncertainty in the measurement of the [$\SII$] flux. On the [$\OI$]-version VO87 diagram, JADES-GS-206183 is just crossing the SF-LINER boundary. 

To conclude, based on multiple AGN diagnostics using optical line ratios, there is no clear evidence for the presence of a strong AGN in JADES-GS-206183.

\subsubsection{Broad H$\alpha$: AGN BLR or outflow?}\label{sec:BHa}
As we mentioned in Section~\ref{sec:linefitting}, our nebular emission line fitting suggests the presence of a broad component(s) to better fit the R$\sim$1000 NIRSpec spectrum over the wavelength range covering H$\alpha$ and $[\NII]$. If we assume this to be a single broad H$\alpha$ emission component, we obtain a line width of $\sim2000\,\mathrm{km\,s^{-1}}$. Such a line width is broadly consistent with what is usually observed in broad-line AGN. The brightness of this broad H$\alpha$ component would indicate an AGN with a bolometric luminosity of $\log(L_{bol}/{\rm erg\,s^{-1}})\sim44.6$ using the bolometric corrections from \citet{Greene2005,Netzer2019}. This corresponds to an X-ray Luminosity of $\log(L_{X, 2-10 \mathrm{keV}}/{\rm erg\,s^{-1}})\sim43.5$ using the bolometric correction determined by \citet{Duras2020}.

JADES-GS-206183 is close ($d\sim4\,\text{arcmin}$) to the center of {\it Chandra} Deep Field South which has 7 Ms of deep X-ray imaging \citep{Luo2017}. With the broad-H$\alpha$-inferred intrinsic X-ray luminosity, i.e. $\log(L_{X, 2-10 \mathrm{keV}})\sim43.5$, JADES-GS-206183 would have been detected in the {\it Chandra} imaging, unless it hosts a highly obscured SMBH with obscuration of $\log(\text{N}_{\rm H}/{\rm cm^{-2}})\gtrsim24$ (Figure 25 of \citealt{Liu2017}). Unfortunately, it is impossible for us to directly estimate obscuration towards JADES-GS-206183's central SMBH. Nonetheless, if we simply use the $A_V=2.27$ from Hydrogen-line ratios (Section~\ref{sec:L_Paa}), we get a $\log(\text{N}_H/{\rm cm^{-2}})\sim21.7$ using the $A_V$-to-$\text{N}_H$ conversion of the Milky Way, which is orders of magnitude smaller than the obscuration inferred based on the non-detection of JADES-GS-206183 in X-ray. Although the levels of obscuration towards SMBH and averaged over the entire galaxy can be differ significantly, this suggests that the broad H$\alpha$ component suggested by our previous line fitting may not originate from an AGN BLR.

More evidence against the AGN interpretation for the single, $2000\,\mathrm{km\,s^{-1}}
$-wide H$\alpha$ line comes from the analysis of the Pa$\alpha$ line profile observed in the FRESCO grism spectrum of JADES-GS-206183. To be specific, assuming the case B recombination and the AGN obscuration is comparable with that of the galactic ISM ($A_V=2.27$), we manually add a mock broad Pa$\alpha$ with the same velocity and dispersion as the broad H$\alpha$ component but a lower amplitude scaled from H$\alpha$. We confirm that our line fitting procedure should identify this broad Pa$\alpha$ component with $\Delta BIC>10$. We also find, even in the most extreme case of $A_V=0$, namely no dust attenuation towards the SMBH of JADES-GS-206183, we should still be able to detect the broad component in the Pa$\alpha$ line with $\Delta$BIC$>$10. Therefore, the absence of clear evidence for a broad Pa$\alpha$ line in JADES-GS-206183's FRESCO grism spectrum  strongly argues against the scenario of a single, $2000 \,\mathrm{km\,s^{-1}}$-wide broad H$\alpha$ line.

After considering all the aforementioned commonly used AGN diagnostics, we conclude that JADES-GS-206183 does not appear to host a strong AGN. This thus raises the question: what is the origin of the broad emission line component at the H$\alpha +[\NII]$ wavelength range found in our spectral fitting? We note that the NIRSpec MSA spectroscopy for JADES-GS-206183 was conducted using medium-resolution gratings ($R\sim1000$). At this spectral resolution, multiple broad emission components with moderately large widths can blend together, mimicking a single broad component with a much larger width. Thus, the broad emission line component could result from the superposition of three broad components, namely H$\alpha$ and the $[\NII]$ doublet, generated by an ionized outflow. To this end, we used the ``narrow+multi-broad'' model to fit the $[\NII]$+H$\alpha$ line profile in Section~\ref{sec:linefitting} as well. It turns out that the best-fit  ``narrow+multi-broad'' model can produce 
a broad component for H$\alpha$, $[\NII]$, and $[\SII]$ that has the same width and the velocity consistent with the systemic redshift (see Table~\ref{tab:line}). 
This indicates that the broad component of H$\alpha$, $[\NII]$, and $[\SII]$ should trace the same gas kinematics, which is consistent with the outflow-only scenario, without additional AGN BLR emission.
Also, as mentioned in Section~\ref{sec:linefitting}, the BIC of the multiple broad $[\NII]+$H$\alpha$ model is comparable with that of the single broad H$\alpha$ model ($\Delta BIC<5$), suggesting that the model representing the outflow scenario can produce a similarly good fit for the data compared to the AGN BLR model.

Again, we also test whether we could detect a similarly outflow-broadened Pa$\alpha$ component, as we did for a single broad H$\alpha$ scenario. The result turns out that, if the outflowing materials in JADES-GS-206183 experience the same attenuation as the galaxy ISM ($A_V=2.27$), then we would also be able to detect a broad Pa$\alpha$ component. However, a systematic study of outflows in local ULIRGs \citep{Fluetsch2021} suggests that the $A_V$ of outflowing gas is typically one magnitude lower than the global attenuation. In this case, if assuming the $A_V$ for the outflow is 1.27 magnitude, we would not have a robust detection of the Pa$\alpha$ broad component resulting from the outflow ($\Delta BIC < 10$). While the attenuation level of the outflow in quiescent galaxies at z$>$1 may differ from those in local ULIRGs, this test could suggest that the ionized outflowing gas in JADES-GS-206183 has lower attenuation than the systemic ISM.

In conclusion, we do not find strong evidence for AGN presence in JADES-GS-206183. The broad ($\sim2000\,\mathrm{km\,s^{-1}}$) H$\alpha$ component suggested by our initial spectral fitting is most likely the result of three moderately broad ($\sim900\,\mathrm{km\,s^{-1}}$) lines--$[\NII]$ doublet and H$\alpha$--blended together due to the limited spectral resolution, which is consistent with expectations for an outflow. Thus, we argue that the origin of these broad lines is more consistent with ionized outflows, rather than AGN.

\section{Discussion}\label{sec:disc}

\subsection{Outflow Driving Mechanism}
\label{sec:outf_mech}
As highlighted in Section~\ref{sec:comparison}, the Na D outflow in JADES-GS-206183 is among the most powerful and energetic neutral outflows known at $z<3$. This naturally raises the questions: {\it Why is this neutral outflow so powerful? what is driving such a strong outflow in this galaxy?} 

\subsubsection{Current SF or AGN?}\label{disc: current}

\begin{figure*}
\centering
\includegraphics[width=1\textwidth]{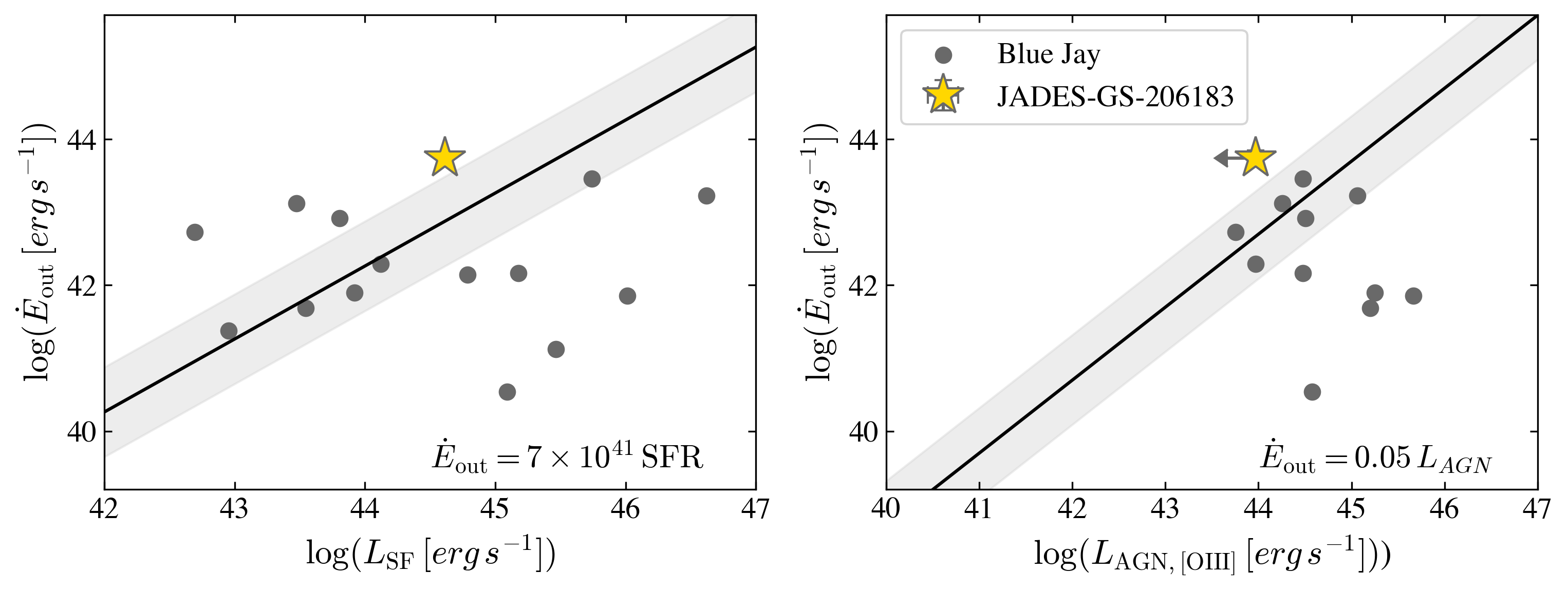}
\caption{Energy outflow rate of JADES-GS-206183 (yellow stars) compared to the luminosity from Pa$\alpha$-based SFR (left) and AGN (right). Given the non-detection of $[\OIII]\lambda5007$, we estimate an upper limit of $[\OIII]$-derived AGN luminosity based on the flux uncertainty. For comparison, the Blue Jay Na D outflows are also shown in the figures by gray dots. The black solid lines represent the expected outflow kinetic energy driven by star formation or AGN. The surrounding shaded regions are the typical uncertainties of outflow energy rate measurement brought by the assumptions of outflow opening angle and covering fraction (see Section~\ref{sec:nad_result}).}
\label{fig:SMILES_bluejay_comp_Eout}
\end{figure*}

Star formation and AGN are the two sources that can power strong galactic outflows \citep{King2015,Rupke2018}. In the local universe,  Na D outflow is commonly detected in star-forming systems \citep[e.g., ][]{Rupke2005b, Chen2010,Concas2019}, especially starburst galaxies and ULIRGs. Meanwhile, many works have found that the local Na D outflow strength and incidence are independent of the level of nuclear activity, suggesting that the physical mechanisms driving neutral gas outflows may be more closely related to galaxy star formation than to AGN activity in star-forming galaxies \citep[e.g., ][]{Sarzi2016, Concas2019}. However, Na D outflows are also detected in local PSBs \citep[e.g., ][]{Baron2022, Sun2024} -- whose star formation has been quickly shut down in the past 1 Gyr -- indicating that neutral outflows found in these systems cannot be powered by their {\it instantaneous} star formation. For PSBs hosting AGNs \citep[e.g.,][]{Baron2022}, the neutral outflows are therefore naturally attributed to AGN activity, particularly through AGN-driven galactic winds.

At cosmic noon, a consensus on neutral gas outflows, particularly regarding their driving mechanisms, has yet to be reached. The 14 Na D outflows recently identified by the Blue Jay survey \citep{Davies2024} are more likely driven by AGNs, given that the majority of these outflows are found in galaxies with low SFRs, which cannot account for the strengths of the outflows inferred from their Na D features. \citet{Belli2024} studied one of those Na D outflows detected in a PSB at z$\sim$2.5 in detail,  finding its host exhibits AGN signatures through the BPT line-ratio diagnostics. In addition, \citet{D'Eugenio2024} also detected a multiphase outflow (including Na D-traced neutral phase) at a z$\sim$3 PSB which also has ongoing AGN activities.

As we discussed above (Section~\ref{sec:sed_model} and \ref{sec:L_Paa}), JADES-GS-206183 is a quiescent galaxy that has kept its low star formation for about one Gyr, even though its SED analysis suggests it might be slightly rejuvenating recently. In addition, its central BH seems not to be actively accreting (Section~\ref{sec:AGN}). Those findings seem to question the scenario that current intense activity powers the outflow.

To quantitatively understand whether instantaneous star formation or AGN could drive the neutral outflow in JADES-GS-206183, we follow \citet{Davies2024} to estimate the energies that could be provided by these mechanisms and compare them with the kinetic energy of the Na D outflow. 
We first calculate the star formation luminosity of JADES-GS-206183 as $L_\text{SF}\approx \text{SFR}_{\text{Pa}\alpha}\times10^{10}L_{\odot}\,(\mathrm{erg\,s^{-1}})$. We also estimate the AGN luminosity using the $[\OIII]$ bolometric correction by \citet{Netzer2009} ($L_{bol}=600L_{[\OIII]}\,\mathrm{erg\,s^{-1}}$). However, we note that the $[\OIII]$-derived AGN bolometric luminosity is problematic because $L_{[\OIII]}$ of an AGN is sensitive to the physical state of the AGN BLR and obscuration \citep{Diamond-Stanic2009,Pennell2017,Netzer2019}. Moreover, as we mentioned in Section~\ref{sec:AGN}, the $[\OIII]$ emission is actually not detected for which we derive just a flux upper limit. Taking those caveats, we obtain the $[\OIII]$-derived AGN luminosity upper limit with $\log(L_{bol}/{\rm erg\,s^{-1}})<43.97$.

Figure~\ref{fig:SMILES_bluejay_comp_Eout} displays the comparison between the energy rate of the Na D outflows and the star formation and AGN luminosities, for both JADES-GS-206183 and the Blue Jay outflows. We overplot two reference lines -- based on theoretical models -- of the maximum energy rate that could be provided by star formation and AGN. If we assume that supernovae are the main energy source of a star formation-driven outflow, the mechanical energy rate they could provide is $\dot{E}_\text{SN} \mathrm{(erg\,s^{-1})} = 7\times10^{41} \text{SFR}\mathrm{(M_{\odot}\,yr^{-1})}$, the relation from \citet{Veilleux2005} based on solar metallicity Starburst99 models \citep{Leitherer1999}. For an AGN-driven outflow, \citet{King2015} found that 5\% of the AGN bolometric luminosity could be transferred to the outflow kinetic energy in the energy-conserving scenario. 

Unlike the strengths of the Blue Jay outflows which can all be explained by either SF or AGN, for JADES-GS-206183, we find that its observed outflow energy rate is about 1 dex higher than both the SN-powered outflow energy rate given its current SFR, and the expected AGN-driven outflow energy rate considering its $[\OIII]$-derived AGN luminosity. We note that, even when combining the SN$+$AGN energy budget, or adopting the highest (and likely overestimated) SFR value among all our estimates ($40\,\mathrm{M_\odot\,yr^{-1}}$ obtained from SED fitting with the Dirichlet prior, see Section~\ref{sec:sed_result}), the available energy is still insufficient to power the neutral outflow in JADES-GS-206183. That is, neither ongoing star formation nor nuclear activity could power the strong Na D outflow in JADES-GS-206183. This result will be even more robust if:
\begin{enumerate}
    \item the  extent  of the JADES-GS-206183 outflow is larger than 1 kpc in radius, the value we conservatively assume, a possibility supported by the observations of other Na D outflows that can extend to $\sim10$ kpc;
    \item the true AGN bolometric luminosity of JADES-GS-206183 is lower than the upper limit estimation derived from the $[\OIII]$ flux upper limit.
\end{enumerate}

\subsubsection{Fossil Outflow?}

Since neither instantaneous star formation nor AGN activity appears to be sufficiently powerful to produce the strength of the neutral outflow observed in JADES-GS-206183, we now explore another possible mechanism: a fossil outflow powered by previous intense starbursts and/or AGN activity. This process has been proposed to explain some neutral outflows found in nearby PSBs \citep{Sun2024}. 

The dynamical timescale of the observed outflow can be estimated as $t_{dyn}= r_{\rm out}/v_{\rm out}$. Considering $r_{\rm out}$ is 1 kpc, and its current outflow velocity is about 800 $\mathrm{km\,s^{-1}}$, the $t_{dyn}$ of the Na D outflow in JADES-GS-206183 is about 2 Myr, which can reach to $\sim$10 Myr if considering a much larger $r_{\rm out}$. Therefore, to test this fossil outflow scenario, from this point on, we will be focusing on estimating the star formation and AGN activities over the past 10 Myr in JADES-GS-206183.

Robustly reconstructing the SFH within the past 10 Myr requires observations such as rest-frame UV spectroscopy and/or X-ray data, which are  unavailable for JADES-GS-206183. Nonetheless, the Pa$\alpha$ emission provides a reliable estimate of the average SFR over the past 10 Myr \citep{Rieke2009}. Thus, it is reasonable to assume that JADES-GS-206183 has maintained the star formation rate derived in Section~\ref{sec:L_Paa} (SFR ${\mathrm{Pa}\alpha} \sim 10 ~\mathrm{M_\odot yr^{-1}}$). Therefore, the total energy injected by star formation over the past 10 Myr is unlikely to be sufficient to drive the neutral outflow we observe.

On the other hand, the dynamical timescale of the Na D outflow in JADES-GS-206183 is slightly shorter than but overall comparable to the model predicted long-term AGN activity phase duration, which is typically about $10\,\rm{Myr}$ \citep{Hopkins2005}. Within each phase, numerous individual episodes with durations $\sim0.1\,Myr$ may occur \citep[e.g.,][]{Hickox2014}. As a consequence, it is likely that the outflow could persist within the galaxy after an AGN has switched off. Moreover, we would expect multiple AGN energy injections to occur during the expansion of the outflow, which could continuously power the outflow \citep{Zubovas2020} and affect its dynamical and thermal equilibrium. Interestingly, \citet{Zubovas2022} adopted a neural network to analyze AGN luminosity histories in 59 observed AGN-driven molecular outflows and found that the average AGN duty cycle in those outflow systems is $\delta_{AGN}\sim0.6$. The result from \citet{Zubovas2022} suggests that more than half of the AGN-driven outflows may be detected in inactive galaxies whose AGN have shut down. The possible rejuvenation of JADES-GS-206183, as suggested by SED fitting (Section~\ref{sec:sed_model}), also fits into this picture. The apparent rise in star formation over the past 10-30 Myr, following a period of decline in SFR, may point to new cold-gas fueling -- whether through accretion (including minor mergers) from the cosmic web or re-cycled gas from a galactic fountain -- toward JADES-GS-206183. Such fueling can also trigger AGN activity, a phenomenon well established in the nearby Universe \citep{Sanders1988}, and the resulting AGN feedback could in turn have driven the strong outflow observed in JADES-GS-206183.

In addition, JADES-GS-206183 has a clear elongated central structure (presumably a bar) and two spiral arms shown in its NIRCam F115W/F200W/F356W RGB image (Figure~\ref{fig:images}). Bar-driven processes are often considered as an efficient funnel to drive gas to the center of the galaxy, feeding the BH accretion and thus lighting it up as an AGN. Many observational works have confirmed the bar-AGN connection through the correlation between the AGN luminosity and bar strength, as well as the higher AGN incidence in barred galaxies (\citealt{Silva-Lima2022} and references therein). In terms of this, the well-constructed bar shown in JADES-GS-206183 could also indicate that, maybe at some point before, its bar was able to transport a lot of gas into the nuclear region to trigger the AGN activity, but later on, the tunnel had been shut down, given that we do not observe any ongoing AGN nowadays. 

We estimate whether JADES-GS-206183 could have a powerful enough AGN to launch outflows in the past by estimating its central black hole mass ($M_{\bullet}$) and the inferred Eddington luminosity using the stellar mass derived through SED fitting (Section~\ref{sec:sed_model}). Adopting the $M_{\bullet}$-$M_{*}$ relation at $1<z<4$ ($\log(M_\bullet/M_*)=-2.5$) from \citet{Sun2025}, we obtain an  $M_{\bullet}$ of JADES-GS-206183 of  $\sim10^{8.5}\,M_{\odot}$, corresponding to an Eddington luminosity of $\sim5\times10^{46}\,\mathrm{erg\,s^{-1}}$. This means that past AGN activity in JADES-GS-206183 could easily drive the strong outflows we observe now: the observed outflow energy rate $\dot{E}_{\rm out}\sim0.05L_{bol}$ can be achieved once its Eddington ratio ($\lambda_{Edd}=L_{bol}/L_{Edd}$) is higher than $0.02$.

Our study can be put into a broader context through the work of \citet{Taylor2024}, who stacked the Mg II spectra of 264 galaxies from the UKIRT Ultra-Deep Survey (UDS) to search for the neutral outflows at z$>$1. They found that high-velocity outflows were not found in galaxies whose star formation has shut down $>$ 1 Gyr ago, but that outflow velocities similar to that in JADES-GS-206183 did occur in their two samples with star formation of age $<$ 0.6 Gyr and between 0.6 and 1 Gyr. This presented them with a similar dilemma to the one we face: (1) contrary to the high outflow velocities, \citet{Sun2024} have shown that outflow velocities significantly decrease with time from an energetic episode of star-formation; and (2) if driven by star formation, the very high mass loading factor for JADES-GS-206183 reflects a high past energy input to the outflow coupled with a small current SFR, but as a stellar population ages, the energy input into outflows (e.g., type II supernovae) decreases. \citet{Taylor2024} conclude, as we do, that the outflows in their sample are likely to be powered by episodic AGN activity. 

\subsubsection{What makes JADES-GS-206183's outflow exceptionally strong?}
In addition, the hypothesis of a fossil AGN-driven outflow may also explain why the Na D outflow in JADES-GS-206183 is so powerful, even comparable with those in local ULIRGs with ongoing AGN activities. If we are seeing an AGN-driven outflow that has been powered by multiple individual AGN episodes during the whole long-term AGN activity phase that has now died out, or where we are between outbursts, then we may have caught this system at a special time. In that case, it would not be surprising that this Na D outflow is the strongest one known at cosmic noon, given the modest numbers observed to date at $z>1$. On the contrary, in the local Universe, \citet{Sun2024} has discovered a much larger number ($\sim100$) of Na D outflows in PSBs without strong ongoing AGN activities (H$\alpha$ EW $<$ 3\AA), though a significant fraction of them locates at the Seyfert region of the [SII]-BPT diagram, which perhaps indicates weak AGN activities. Among those outflows, some do have a comparably high outflow velocity as JADES-GS-206183. Therefore, those local Na D outflows might also be witnessed by us at a similar time point as JADES-GS-206183 -- the end or quiet-phase of the long-term AGN phase. Those outflows observed at such a unique time are essential for understanding the detailed mechanism of episodic-AGN-driven outflow, especially its time evolution.

\subsection{Multiphase outflow}

Galactic outflows are commonly multiphase. In JADES-GS-206183, besides the neutral outflow traced by Na D absorption, we also find  tentative evidence of an ionized-phase outflow. 

In Section~\ref{sec:AGN}, we have shown that the profile of $[\NII]$, H$\alpha$, and $[\SII]$ of JADES-GS-206183 suggest they may share the same broad component, which is consistent with expectations of a warm ionized outflow. However, due to the limited spectral resolution and the complexity of the $[\NII]+$H$\alpha$ profile, the kinematics of ionizing gas derived from multiple component fitting would be uncertain. Therefore, we do not quantify the ionized outflow properties in this work.

Another tentative indication for an ionized outflow is that, from the MUSE IFS data, we mapped the [O II]$\lambda\lambda$3726, 3729 emission of this galaxy.
Even though the seeing of $\sim0.8"$ does not enable us to construct a meaningful velocity and velocity dispersion map of [O II] emission, the whole [O II] flux map provides useful information on galactic-scale ionized gas distribution. 
The left panel of Figure~\ref{fig:MUSE_OII} shows that, this emission is extended to the upper left region up to $\sim1"$ relative to the galaxy stellar distribution. This [O II] extension is consistent with the position of the strong Na D outflow. The MSA shutter was oriented from the upper left to the lower right and the center is slightly shifted to the upper left. Therefore, the Na D outflow we observed from the SMILES NIRSpec MSA spectrum should come from the upper left region of JADES-GS-206183's center. Therefore, 
we speculate that the extended [O II] emission may trace the ionized outflowing gas coupled with the Na D-traced neutral outflow. A future higher spatial-resolution IFU observation of ionized outflow tracers is crucial to confirm the existence of any ionized-phase outflow and its spatial distribution.

\section{Conclusion} \label{sec:concl}

In this work, using the new JWST NIRSpec/MSA spectrum collected by the SMILES survey, we identified an extremely strong neutral outflow traced by Na D $\lambda\lambda5890, 5896$ in the inactive quenching galaxy JADES-GS-206183 at $z=1.317$. Our main findings are the following:
\begin{itemize}

    \item By subtracting the stellar continuum modelled by pPXF from the whole NIRSpec spectrum, we uncovered a significantly deep and blueshifted Na D ISM absorption (Figure~\ref{fig:SMILES_spec}), corresponding to a powerful outflow with a velocity of $v_{\rm out}=828^{+79}_{-49}\,\mathrm{km\,s^{-1}}$, a mass outflow rate of $\log(\dot{M}_{\rm out}/{\rm M_{\odot}\,yr^{-1}})=2.40^{+0.11}_{-0.16}$, and an energy outflow rate of $\log(\dot{E}_{\rm out}/{\rm erg\,s^{-1}}) = 43.73^{+0.11}_{-0.12}$.

    \item Comparing the outflow properties of JADES-GS-206183 with the Na D outflows at cosmic noon identified by the Blue Jay survey from \citet{Davies2024}, as well as the outflows in the local ULIRGs and PSBs with/without AGNs, we found the JADES-GS-206183 Na D outflow has the highest mass outflow rate at $z>1$ (Figure~\ref{fig:SMILES_bluejay_comp_Mout}), also at the top of the Na D outflows in the local galaxies with ongoing intensive SF/AGN activities (Figure~\ref{fig:loal+z2_comp}).

    \item The SFR of JADES-GS-206183 is estimated to be $10.78\pm 0.55\,\mathrm{M_{\odot}\,yr^{-1}}$ based on its Pa$\alpha$ emission, consistent with the value inferred from ALMA CO observations. We performed SED modeling using the panchromatic photometry from HST through NIRCam, together with the VLT/MUSE spectrum that covers key stellar features near the rest-frame 4000 \AA. Depending on the assumed SFH priors, the SED-based SFR lies between approximately 20--40 $\mathrm{M_{\odot}\,yr^{-1}}$. These values are slightly higher but broadly consistent, within the uncertainties, with the SFRs derived from Pa$\alpha$ and CO.
    
    \item Despite the systematic uncertainties associated with different prior assumptions, the reconstructed SFH of JADES-GS-206183 suggests that it experienced an older episode of star formation that peaked between roughly 0.5 and 2 Gyr ago, with possible signs of rejuvenation within the past 10 Myr.
    
    \item In addition, JADES-GS-206183 exhibits several peculiar features regarding its low current star formation. Its morphology shows clear bar and spiral features (Figure~\ref{fig:images}), and it has a large dust attenuation with $A_V=2.27$ mag, indicative of a substantial interstellar medium.
    
    \item We measured the fluxes of the four BPT emission lines H$\beta$, H$\alpha$, $[\OIII]$, and $[\NII]$ from the SMILES NIRspec spectrum and determined the locus of JADES-GS-206183 in the BPT and VO87 diagrams. JADES-GS-206183 has just a tentative detection of $[\OIII]$ (Figure~\ref{fig:SMILES_spec}), which only allows us to derive an upper limit on the $[\OIII]$ flux. JADES-GS-206183 is located in the composite area (Figure~\ref{fig: BPT}), and also in the non-AGN region in VO87 diagrams. Moreover, the detected broad component of H$\alpha$ is more likely to originate from an ionized outflow, rather than an AGN BLR (Figure~\ref{fig:smiles_Ha}).
    Along with its non-AGN classification from \citet{Lyu2022,Lyu2024}, JADES-GS-206183 seems not to host a strong AGN. 

    \item The current weak star formation and AGN activity ($\log(L_{bol, [\OIII]}/\mathrm{erg\,s^{-1}})<43.97$) in JADES-GS-206183 are not sufficient to power the observed strong Na D outflow. In Figure~\ref{fig:SMILES_bluejay_comp_Eout}, this result was confirmed by comparing the observed energy rate of JADES-GS-206183 with the theoretical values considering the energy injection from current star formation and AGN (as a function of SFR and $L_{bol}$). Therefore, we speculate that the strong Na D outflow we observed in JADES-GS-206183 might be an AGN-driven fossil outflow. As a consequence, the fact that this outflow is extremely strong could then be explained by the scenario that we might be witnessing this outflow at a unique time -- the end of a long-term AGN activity phase, in which the outflow has been episodically powered and speeded up by multiple AGN activities during the whole phase.

    \item From the MUSE IFS observations of JADES-GS-206183, we built the [O II] emission map (Figure~\ref{fig:MUSE_OII}) and found it seems to extend to the same direction as the Na D outflow (constrained by the position of the MSA shutter). Along with the outflow interpretation of the broad component detected in H$\alpha$, the results suggest the outflow in JADES-GS-206183 might be multiphase.
    
\end{itemize}

To fully confirm the exact wind-driving mechanism, and understand the timescale of energy source and outflow evolution, it is crucial to obtain the spatially-resolved spectroscopic (e.g. NIRSpec/IFU) observation, which will be essential to map the outflow extent, as well as the spatial distribution of the physical properties of the outflowing gas, such as velocity, mass, temperature. Meanwhile, IFU observations will enable us to map the host galaxy properties, which will reveal the AGN and outflow impact on the host star formation activities, helping us link the outflow to the quenching process in JADES-GS-206183. Finally, along with other multi-wavelength data, e.g., ALMA high-resolution observations are necessary to build a more comprehensive picture of the multiphase outflow of this galaxy.

\section{Acknowledgments}

YS, ZJ, YZ, and CNAW acknowledge support from the NIRCam Science Team contract to the University of Arizona, NAS5-02015. 
GHR acknowledges support from the JWST Mid-Infrared Instrument (MIRI) Science Team Lead, grant 80NSSC18K0555, from NASA Goddard Space Flight Center to the University of Arizona.
FDE acknowledges support by the Science and Technology Facilities Council (STFC), by the ERC through Advanced Grant 695671 ``QUENCH'', and by the UKRI Frontier Research grant RISEandFALL.
AJB acknowledges funding from the "FirstGalaxies" Advanced Grant from the European Research Council (ERC) under the European Union's Horizon 2020 research and innovation programme (Grant agreement No. 789056).

This work is based on
observations made with the VLT/
MUSE and the NASA/ESA/CSA James Webb
Space Telescope. MUSE observations
used in this work are taken by the
MUSE Hubble UDF Survey as parts of the MUSE Consortium, for which the data can be accessed via the AMUSED website \dataset[https://amused.univ-lyon1.fr/project/UDF/]{https://amused.univ-lyon1.fr/project/UDF/}. The JWST data were obtained from the Mikulski
Archive for Space Telescopes at the Space Telescope Science
Institute, which is operated by the Association of Universities for Research in Astronomy, Inc., under NASA contract
NAS 5-03127 for JWST. These observations are associated with programs 1180, 1207, and 1895, for which the data can be accessed via \dataset[https://doi.org/10.17909/8tdj-8n28]{https://doi.org/10.17909/8tdj-8n28} \citep{https://doi.org/10.17909/8tdj-8n28}, \dataset[https://doi.org/10.17909/et3f-zd57]{https://doi.org/10.17909/et3f-zd57}\citep{rieke_george_systematic_2024}, and \dataset[https://doi.org/10.17909/gdyc-7g80]{https://doi.org/10.17909/gdyc-7g80} \citep{https://doi.org/10.17909/gdyc-7g80}.

We respectfully acknowledge the University of Arizona is on the land and territories of Indigenous peoples. Today, Arizona is home to 22 federally recognized tribes, with Tucson being home to the O'odham and the Yaqui. The University strives to build sustainable relationships with sovereign Native Nations and Indigenous communities through education offerings, partnerships, and community service.

\facilities{JWST, VLT/MUSE}

\software{\texttt{AstroPy}\citep{astropy2013,astropy2018,astropy2022},  
\texttt{emcee}\citep{2013PASP..125..306F},
\texttt{lmfit}\citep{Newville2014}, 
\texttt{mpdaf}\citep{Bacon2016},
\texttt{Prospector}\citep{Johnson2021}, \texttt{SciPy}\citep{Virtanen2020}}

\bibliography{reference}{}

\begin{thebibliography}{}
\expandafter\ifx\csname natexlab\endcsname\relax\def\natexlab#1{#1}\fi
\providecommand{\url}[1]{\href{#1}{#1}}
\providecommand{\dodoi}[1]{doi:~\href{http://doi.org/#1}{\nolinkurl{#1}}}
\providecommand{\doeprint}[1]{\href{http://ascl.net/#1}{\nolinkurl{http://ascl.net/#1}}}
\providecommand{\doarXiv}[1]{\href{https://arxiv.org/abs/#1}{\nolinkurl{https://arxiv.org/abs/#1}}}

\bibitem[{{Alberts} {et~al.}(2024){Alberts}, {Lyu}, {Shivaei}, {Rieke}, {P{\'e}rez-Gonz{\'a}lez}, {Bonaventura}, {Zhu}, {Helton}, {Ji}, {Morrison}, {Robertson}, {Stone}, {Sun}, {Williams}, \& {Willmer}}]{Alberts2024}
{Alberts}, S., {Lyu}, J., {Shivaei}, I., {et~al.} 2024, \apj, 976, 224, \dodoi{10.3847/1538-4357/ad7396}

\bibitem[{{Aravena} {et~al.}(2020){Aravena}, {Boogaard}, {G{\'o}nzalez-L{\'o}pez}, {Decarli}, {Walter}, {Carilli}, {Smail}, {Weiss}, {Assef}, {Bauer}, {Bouwens}, {Cortes}, {Cox}, {da Cunha}, {Daddi}, {D{\'\i}az-Santos}, {Inami}, {Ivison}, {Novak}, {Popping}, {Riechers}, {van der Werf}, \& {Wagg}}]{Aravena2020}
{Aravena}, M., {Boogaard}, L., {G{\'o}nzalez-L{\'o}pez}, J., {et~al.} 2020, \apj, 901, 79, \dodoi{10.3847/1538-4357/ab99a2}

\bibitem[{{Astropy Collaboration} {et~al.}(2013){Astropy Collaboration}, {Robitaille}, {Tollerud}, {Greenfield}, {Droettboom}, {Bray}, {Aldcroft}, {Davis}, {Ginsburg}, {Price-Whelan}, {Kerzendorf}, {Conley}, {Crighton}, {Barbary}, {Muna}, {Ferguson}, {Grollier}, {Parikh}, {Nair}, {Unther}, {Deil}, {Woillez}, {Conseil}, {Kramer}, {Turner}, {Singer}, {Fox}, {Weaver}, {Zabalza}, {Edwards}, {Azalee Bostroem}, {Burke}, {Casey}, {Crawford}, {Dencheva}, {Ely}, {Jenness}, {Labrie}, {Lim}, {Pierfederici}, {Pontzen}, {Ptak}, {Refsdal}, {Servillat}, \& {Streicher}}]{astropy2013}
{Astropy Collaboration}, {Robitaille}, T.~P., {Tollerud}, E.~J., {et~al.} 2013, \aap, 558, A33, \dodoi{10.1051/0004-6361/201322068}

\bibitem[{{Astropy Collaboration} {et~al.}(2018){Astropy Collaboration}, {Price-Whelan}, {Sip{\H{o}}cz}, {G{\"u}nther}, {Lim}, {Crawford}, {Conseil}, {Shupe}, {Craig}, {Dencheva}, {Ginsburg}, {VanderPlas}, {Bradley}, {P{\'e}rez-Su{\'a}rez}, {de Val-Borro}, {Aldcroft}, {Cruz}, {Robitaille}, {Tollerud}, {Ardelean}, {Babej}, {Bach}, {Bachetti}, {Bakanov}, {Bamford}, {Barentsen}, {Barmby}, {Baumbach}, {Berry}, {Biscani}, {Boquien}, {Bostroem}, {Bouma}, {Brammer}, {Bray}, {Breytenbach}, {Buddelmeijer}, {Burke}, {Calderone}, {Cano Rodr{\'\i}guez}, {Cara}, {Cardoso}, {Cheedella}, {Copin}, {Corrales}, {Crichton}, {D'Avella}, {Deil}, {Depagne}, {Dietrich}, {Donath}, {Droettboom}, {Earl}, {Erben}, {Fabbro}, {Ferreira}, {Finethy}, {Fox}, {Garrison}, {Gibbons}, {Goldstein}, {Gommers}, {Greco}, {Greenfield}, {Groener}, {Grollier}, {Hagen}, {Hirst}, {Homeier}, {Horton}, {Hosseinzadeh}, {Hu}, {Hunkeler}, {Ivezi{\'c}}, {Jain}, {Jenness}, {Kanarek}, {Kendrew}, {Kern}, {Kerzendorf}, {Khvalko}, {King}, {Kirkby}, {Kulkarni},
  {Kumar}, {Lee}, {Lenz}, {Littlefair}, {Ma}, {Macleod}, {Mastropietro}, {McCully}, {Montagnac}, {Morris}, {Mueller}, {Mumford}, {Muna}, {Murphy}, {Nelson}, {Nguyen}, {Ninan}, {N{\"o}the}, {Ogaz}, {Oh}, {Parejko}, {Parley}, {Pascual}, {Patil}, {Patil}, {Plunkett}, {Prochaska}, {Rastogi}, {Reddy Janga}, {Sabater}, {Sakurikar}, {Seifert}, {Sherbert}, {Sherwood-Taylor}, {Shih}, {Sick}, {Silbiger}, {Singanamalla}, {Singer}, {Sladen}, {Sooley}, {Sornarajah}, {Streicher}, {Teuben}, {Thomas}, {Tremblay}, {Turner}, {Terr{\'o}n}, {van Kerkwijk}, {de la Vega}, {Watkins}, {Weaver}, {Whitmore}, {Woillez}, {Zabalza}, \& {Astropy Contributors}}]{astropy2018}
{Astropy Collaboration}, {Price-Whelan}, A.~M., {Sip{\H{o}}cz}, B.~M., {et~al.} 2018, \aj, 156, 123, \dodoi{10.3847/1538-3881/aabc4f}

\bibitem[{{Astropy Collaboration} {et~al.}(2022){Astropy Collaboration}, {Price-Whelan}, {Lim}, {Earl}, {Starkman}, {Bradley}, {Shupe}, {Patil}, {Corrales}, {Brasseur}, {N{\"o}the}, {Donath}, {Tollerud}, {Morris}, {Ginsburg}, {Vaher}, {Weaver}, {Tocknell}, {Jamieson}, {van Kerkwijk}, {Robitaille}, {Merry}, {Bachetti}, {G{\"u}nther}, {Aldcroft}, {Alvarado-Montes}, {Archibald}, {B{\'o}di}, {Bapat}, {Barentsen}, {Baz{\'a}n}, {Biswas}, {Boquien}, {Burke}, {Cara}, {Cara}, {Conroy}, {Conseil}, {Craig}, {Cross}, {Cruz}, {D'Eugenio}, {Dencheva}, {Devillepoix}, {Dietrich}, {Eigenbrot}, {Erben}, {Ferreira}, {Foreman-Mackey}, {Fox}, {Freij}, {Garg}, {Geda}, {Glattly}, {Gondhalekar}, {Gordon}, {Grant}, {Greenfield}, {Groener}, {Guest}, {Gurovich}, {Handberg}, {Hart}, {Hatfield-Dodds}, {Homeier}, {Hosseinzadeh}, {Jenness}, {Jones}, {Joseph}, {Kalmbach}, {Karamehmetoglu}, {Ka{\l}uszy{\'n}ski}, {Kelley}, {Kern}, {Kerzendorf}, {Koch}, {Kulumani}, {Lee}, {Ly}, {Ma}, {MacBride}, {Maljaars}, {Muna}, {Murphy}, {Norman},
  {O'Steen}, {Oman}, {Pacifici}, {Pascual}, {Pascual-Granado}, {Patil}, {Perren}, {Pickering}, {Rastogi}, {Roulston}, {Ryan}, {Rykoff}, {Sabater}, {Sakurikar}, {Salgado}, {Sanghi}, {Saunders}, {Savchenko}, {Schwardt}, {Seifert-Eckert}, {Shih}, {Jain}, {Shukla}, {Sick}, {Simpson}, {Singanamalla}, {Singer}, {Singhal}, {Sinha}, {Sip{\H{o}}cz}, {Spitler}, {Stansby}, {Streicher}, {{\v{S}}umak}, {Swinbank}, {Taranu}, {Tewary}, {Tremblay}, {de Val-Borro}, {Van Kooten}, {Vasovi{\'c}}, {Verma}, {de Miranda Cardoso}, {Williams}, {Wilson}, {Winkel}, {Wood-Vasey}, {Xue}, {Yoachim}, {Zhang}, {Zonca}, \& {Astropy Project Contributors}}]{astropy2022}
{Astropy Collaboration}, {Price-Whelan}, A.~M., {Lim}, P.~L., {et~al.} 2022, \apj, 935, 167, \dodoi{10.3847/1538-4357/ac7c74}

\bibitem[{{Bacon} {et~al.}(2016){Bacon}, {Piqueras}, {Conseil}, {Richard}, \& {Shepherd}}]{Bacon2016}
{Bacon}, R., {Piqueras}, L., {Conseil}, S., {Richard}, J., \& {Shepherd}, M. 2016, {MPDAF: MUSE Python Data Analysis Framework}, Astrophysics Source Code Library, record ascl:1611.003

\bibitem[{{Bacon} {et~al.}(2023){Bacon}, {Brinchmann}, {Conseil}, {Maseda}, {Nanayakkara}, {Wendt}, {Bacher}, {Mary}, {Weilbacher}, {Krajnovi{\'c}}, {Boogaard}, {Bouch{\'e}}, {Contini}, {Epinat}, {Feltre}, {Guo}, {Herenz}, {Kollatschny}, {Kusakabe}, {Leclercq}, {Michel-Dansac}, {Pello}, {Richard}, {Roth}, {Salvignol}, {Schaye}, {Steinmetz}, {Tresse}, {Urrutia}, {Verhamme}, {Vitte}, {Wisotzki}, \& {Zoutendijk}}]{Bacon2023}
{Bacon}, R., {Brinchmann}, J., {Conseil}, S., {et~al.} 2023, \aap, 670, A4, \dodoi{10.1051/0004-6361/202244187}

\bibitem[{{Baldwin} {et~al.}(1981){Baldwin}, {Phillips}, \& {Terlevich}}]{Baldwin1981}
{Baldwin}, J.~A., {Phillips}, M.~M., \& {Terlevich}, R. 1981, \pasp, 93, 5, \dodoi{10.1086/130766}

\bibitem[{{Baron} \& {Netzer}(2019)}]{Baron2019}
{Baron}, D., \& {Netzer}, H. 2019, \mnras, 486, 4290, \dodoi{10.1093/mnras/stz1070}

\bibitem[{{Baron} {et~al.}(2020){Baron}, {Netzer}, {Davies}, \& {Xavier Prochaska}}]{Baron2020}
{Baron}, D., {Netzer}, H., {Davies}, R.~I., \& {Xavier Prochaska}, J. 2020, \mnras, 494, 5396, \dodoi{10.1093/mnras/staa1018}

\bibitem[{{Baron} {et~al.}(2022){Baron}, {Netzer}, {Lutz}, {Prochaska}, \& {Davies}}]{Baron2022}
{Baron}, D., {Netzer}, H., {Lutz}, D., {Prochaska}, J.~X., \& {Davies}, R.~I. 2022, \mnras, 509, 4457, \dodoi{10.1093/mnras/stab3232}

\bibitem[{{Belli} {et~al.}(2024){Belli}, {Park}, {Davies}, {Mendel}, {Johnson}, {Conroy}, {Benton}, {Bugiani}, {Emami}, {Leja}, {Li}, {Maheson}, {Mathews}, {Naidu}, {Nelson}, {Tacchella}, {Terrazas}, \& {Weinberger}}]{Belli2024}
{Belli}, S., {Park}, M., {Davies}, R.~L., {et~al.} 2024, \nat, 630, 54, \dodoi{10.1038/s41586-024-07412-1}

\bibitem[{{Boogaard} {et~al.}(2019){Boogaard}, {Decarli}, {Gonz{\'a}lez-L{\'o}pez}, {van der Werf}, {Walter}, {Bouwens}, {Aravena}, {Carilli}, {Bauer}, {Brinchmann}, {Contini}, {Cox}, {da Cunha}, {Daddi}, {D{\'\i}az-Santos}, {Hodge}, {Inami}, {Ivison}, {Maseda}, {Matthee}, {Oesch}, {Popping}, {Riechers}, {Schaye}, {Schouws}, {Smail}, {Weiss}, {Wisotzki}, {Bacon}, {Cortes}, {Rix}, {Somerville}, {Swinbank}, \& {Wagg}}]{Boogaard2019}
{Boogaard}, L.~A., {Decarli}, R., {Gonz{\'a}lez-L{\'o}pez}, J., {et~al.} 2019, \apj, 882, 140, \dodoi{10.3847/1538-4357/ab3102}

\bibitem[{{Boogaard} {et~al.}(2020){Boogaard}, {van der Werf}, {Weiss}, {Popping}, {Decarli}, {Walter}, {Aravena}, {Bouwens}, {Riechers}, {Gonz{\'a}lez-L{\'o}pez}, {Smail}, {Carilli}, {Kaasinen}, {Daddi}, {Cox}, {D{\'\i}az-Santos}, {Inami}, {Cortes}, \& {Wagg}}]{2020ApJ...902..109B}
{Boogaard}, L.~A., {van der Werf}, P., {Weiss}, A., {et~al.} 2020, \apj, 902, 109, \dodoi{10.3847/1538-4357/abb82f}

\bibitem[{{Bushouse} {et~al.}(2022){Bushouse}, {Eisenhamer}, {Dencheva}, {Davies}, {Greenfield}, {Morrison}, {Hodge}, {Simon}, {Grumm}, {Droettboom}, {Slavich}, {Sosey}, {Pauly}, {Miller}, {Jedrzejewski}, {Hack}, {Davis}, {Crawford}, {Law}, {Gordon}, {Regan}, {Cara}, {MacDonald}, {Bradley}, {Shanahan}, {Jamieson}, {Teodoro}, \& {Williams}}]{Bushouse2022}
{Bushouse}, H., {Eisenhamer}, J., {Dencheva}, N., {et~al.} 2022, {JWST Calibration Pipeline}, 1.6.2,  Zenodo, \dodoi{10.5281/zenodo.7041998}

\bibitem[{{Bushouse} {et~al.}(2024){Bushouse}, Eisenhamer, Dencheva, Davies, Greenfield, Morrison, Hodge, Simon, Grumm, Droettboom, Slavich, Sosey, Pauly, Miller, Jedrzejewski, Hack, Davis, Crawford, Law, Gordon, Regan, Cara, MacDonald, Bradley, Shanahan, Jamieson, Teodoro, Williams, Pena-Guerrero, Graham, Molter, Brandt, Hayes, Cooper, \& Clarke}]{https://doi.org/10.5281/zenodo.6984365}
{Bushouse}, H., Eisenhamer, J., Dencheva, N., {et~al.} 2024, JWST Calibration Pipeline,  Zenodo, \dodoi{10.5281/ZENODO.6984365}

\bibitem[{{Byler} {et~al.}(2017){Byler}, {Dalcanton}, {Conroy}, \& {Johnson}}]{Byler2017}
{Byler}, N., {Dalcanton}, J.~J., {Conroy}, C., \& {Johnson}, B.~D. 2017, \apj, 840, 44, \dodoi{10.3847/1538-4357/aa6c66}

\bibitem[{{Calzetti} {et~al.}(2000){Calzetti}, {Armus}, {Bohlin}, {Kinney}, {Koornneef}, \& {Storchi-Bergmann}}]{Calzetti2000}
{Calzetti}, D., {Armus}, L., {Bohlin}, R.~C., {et~al.} 2000, \apj, 533, 682, \dodoi{10.1086/308692}

\bibitem[{Cappellari(2017)}]{cappellari2017}
Cappellari, M. 2017, \mnras, 466, 798, \dodoi{10.1093/mnras/stw3020}

\bibitem[{{Cappellari}(2023)}]{Cappellari2023}
{Cappellari}, M. 2023, \mnras, 526, 3273, \dodoi{10.1093/mnras/stad2597}

\bibitem[{Cappellari \& Emsellem(2004)}]{cappellari2004}
Cappellari, M., \& Emsellem, E. 2004, \pasp, 116, 138, \dodoi{10.1086/381875}

\bibitem[{{Carnall} {et~al.}(2023){Carnall}, {McLure}, {Dunlop}, {McLeod}, {Wild}, {Cullen}, {Magee}, {Begley}, {Cimatti}, {Donnan}, {Hamadouche}, {Jewell}, \& {Walker}}]{Carnall2023}
{Carnall}, A.~C., {McLure}, R.~J., {Dunlop}, J.~S., {et~al.} 2023, \nat, 619, 716, \dodoi{10.1038/s41586-023-06158-6}

\bibitem[{{Carniani} {et~al.}(2024){Carniani}, {Venturi}, {Parlanti}, {de Graaff}, {Maiolino}, {Arribas}, {Bonaventura}, {Boyett}, {Bunker}, {Cameron}, {Charlot}, {Chevallard}, {Curti}, {Curtis-Lake}, {Eisenstein}, {Giardino}, {Hausen}, {Kumari}, {Maseda}, {Nelson}, {Perna}, {Rix}, {Robertson}, {Del Pino}, {Sandles}, {Scholtz}, {Simmonds}, {Smit}, {Tacchella}, {{\"U}bler}, {Williams}, {Willott}, \& {Witstok}}]{Carniani2024}
{Carniani}, S., {Venturi}, G., {Parlanti}, E., {et~al.} 2024, \aap, 685, A99, \dodoi{10.1051/0004-6361/202347230}

\bibitem[{{Cazzoli} {et~al.}(2016){Cazzoli}, {Arribas}, {Maiolino}, \& {Colina}}]{Cazzoli2016}
{Cazzoli}, S., {Arribas}, S., {Maiolino}, R., \& {Colina}, L. 2016, \aap, 590, A125, \dodoi{10.1051/0004-6361/201526788}

\bibitem[{{Chabrier}(2003)}]{Chabrier2003}
{Chabrier}, G. 2003, \pasp, 115, 763, \dodoi{10.1086/376392}

\bibitem[{{Chen} {et~al.}(2010){Chen}, {Tremonti}, {Heckman}, {Kauffmann}, {Weiner}, {Brinchmann}, \& {Wang}}]{Chen2010}
{Chen}, Y.-M., {Tremonti}, C.~A., {Heckman}, T.~M., {et~al.} 2010, \aj, 140, 445, \dodoi{10.1088/0004-6256/140/2/445}

\bibitem[{{Choi} {et~al.}(2016){Choi}, {Dotter}, {Conroy}, {Cantiello}, {Paxton}, \& {Johnson}}]{Choi2016}
{Choi}, J., {Dotter}, A., {Conroy}, C., {et~al.} 2016, \apj, 823, 102, \dodoi{10.3847/0004-637X/823/2/102}

\bibitem[{{Cicone} {et~al.}(2014){Cicone}, {Maiolino}, {Sturm}, {Graci{\'a}-Carpio}, {Feruglio}, {Neri}, {Aalto}, {Davies}, {Fiore}, {Fischer}, {Garc{\'\i}a-Burillo}, {Gonz{\'a}lez-Alfonso}, {Hailey-Dunsheath}, {Piconcelli}, \& {Veilleux}}]{Cicone2014}
{Cicone}, C., {Maiolino}, R., {Sturm}, E., {et~al.} 2014, \aap, 562, A21, \dodoi{10.1051/0004-6361/201322464}

\bibitem[{{Coil} {et~al.}(2011){Coil}, {Weiner}, {Holz}, {Cooper}, {Yan}, \& {Aird}}]{Coil2011}
{Coil}, A.~L., {Weiner}, B.~J., {Holz}, D.~E., {et~al.} 2011, \apj, 743, 46, \dodoi{10.1088/0004-637X/743/1/46}

\bibitem[{{Concas} {et~al.}(2019){Concas}, {Popesso}, {Brusa}, {Mainieri}, \& {Thomas}}]{Concas2019}
{Concas}, A., {Popesso}, P., {Brusa}, M., {Mainieri}, V., \& {Thomas}, D. 2019, \aap, 622, A188, \dodoi{10.1051/0004-6361/201732152}

\bibitem[{{Conroy} \& {Gunn}(2010)}]{Conroy2010}
{Conroy}, C., \& {Gunn}, J.~E. 2010, \apj, 712, 833, \dodoi{10.1088/0004-637X/712/2/833}

\bibitem[{{Conroy} {et~al.}(2009){Conroy}, {Gunn}, \& {White}}]{Conroy2009}
{Conroy}, C., {Gunn}, J.~E., \& {White}, M. 2009, \apj, 699, 486, \dodoi{10.1088/0004-637X/699/1/486}

\bibitem[{{Conroy} {et~al.}(2018){Conroy}, {Villaume}, {van Dokkum}, \& {Lind}}]{Conroy2018}
{Conroy}, C., {Villaume}, A., {van Dokkum}, P.~G., \& {Lind}, K. 2018, \apj, 854, 139, \dodoi{10.3847/1538-4357/aaab49}

\bibitem[{{Couch} \& {Sharples}(1987)}]{Couch1987}
{Couch}, W.~J., \& {Sharples}, R.~M. 1987, \mnras, 229, 423, \dodoi{10.1093/mnras/229.3.423}

\bibitem[{{Davies} {et~al.}(2020){Davies}, {Baron}, {Shimizu}, {Netzer}, {Burtscher}, {de Zeeuw}, {Genzel}, {Hicks}, {Koss}, {Lin}, {Lutz}, {Maciejewski}, {M{\"u}ller-S{\'a}nchez}, {Orban de Xivry}, {Ricci}, {Riffel}, {Riffel}, {Rosario}, {Schartmann}, {Schnorr-M{\"u}ller}, {Shangguan}, {Sternberg}, {Sturm}, {Storchi-Bergmann}, {Tacconi}, \& {Veilleux}}]{Davies2020}
{Davies}, R., {Baron}, D., {Shimizu}, T., {et~al.} 2020, \mnras, 498, 4150, \dodoi{10.1093/mnras/staa2413}

\bibitem[{{Davies} {et~al.}(2024){Davies}, {Belli}, {Park}, {Mendel}, {Johnson}, {Conroy}, {Benton}, {Bugiani}, {Emami}, {Leja}, {Li}, {Maheson}, {Mathews}, {Naidu}, {Nelson}, {Tacchella}, {Terrazas}, \& {Weinberger}}]{Davies2024}
{Davies}, R.~L., {Belli}, S., {Park}, M., {et~al.} 2024, \mnras, 528, 4976, \dodoi{10.1093/mnras/stae327}

\bibitem[{{de Graaff} {et~al.}(2024){de Graaff}, {Setton}, {Brammer}, {Cutler}, {Suess}, {Labb{\'e}}, {Leja}, {Weibel}, {Maseda}, {Whitaker}, {Bezanson}, {Boogaard}, {Cleri}, {De Lucia}, {Franx}, {Greene}, {Hirschmann}, {Matthee}, {McConachie}, {Naidu}, {Oesch}, {Price}, {Rix}, {Valentino}, {Wang}, \& {Williams}}]{deGraaff2024}
{de Graaff}, A., {Setton}, D.~J., {Brammer}, G., {et~al.} 2024, Nature Astronomy, \dodoi{10.1038/s41550-024-02424-3}

\bibitem[{{D'Eugenio} {et~al.}(2024){D'Eugenio}, {P{\'e}rez-Gonz{\'a}lez}, {Maiolino}, {Scholtz}, {Perna}, {Circosta}, {{\"U}bler}, {Arribas}, {B{\"o}ker}, {Bunker}, {Carniani}, {Charlot}, {Chevallard}, {Cresci}, {Curtis-Lake}, {Jones}, {Kumari}, {Lamperti}, {Looser}, {Parlanti}, {Rix}, {Robertson}, {Rodr{\'\i}guez Del Pino}, {Tacchella}, {Venturi}, \& {Willott}}]{D'Eugenio2024}
{D'Eugenio}, F., {P{\'e}rez-Gonz{\'a}lez}, P.~G., {Maiolino}, R., {et~al.} 2024, Nature Astronomy, 8, 1443, \dodoi{10.1038/s41550-024-02345-1}

\bibitem[{{D'Eugenio} {et~al.}(2025){D'Eugenio}, {Cameron}, {Scholtz}, {Carniani}, {Willott}, {Curtis-Lake}, {Bunker}, {Parlanti}, {Maiolino}, {Willmer}, {Jakobsen}, {Robertson}, {Johnson}, {Tacchella}, {Cargile}, {Rawle}, {Arribas}, {Chevallard}, {Curti}, {Egami}, {Eisenstein}, {Kumari}, {Looser}, {Rieke}, {Rodr{\'\i}guez Del Pino}, {Saxena}, {{\"U}bler}, {Venturi}, {Witstok}, {Baker}, {Bhatawdekar}, {Bonaventura}, {Boyett}, {Charlot}, {Danhaive}, {Hainline}, {Hausen}, {Helton}, {Ji}, {Ji}, {Jones}, {Juod{\v{z}}balis}, {Maseda}, {P{\'e}rez-Gonz{\'a}lez}, {Perna}, {Pusk{\'a}s}, {Shivaei}, {Silcock}, {Simmonds}, {Smit}, {Sun}, {Villanueva}, {Williams}, \& {Zhu}}]{DEugenio2025}
{D'Eugenio}, F., {Cameron}, A.~J., {Scholtz}, J., {et~al.} 2025, \apjs, 277, 4, \dodoi{10.3847/1538-4365/ada148}

\bibitem[{{Diamond-Stanic} {et~al.}(2009){Diamond-Stanic}, {Rieke}, \& {Rigby}}]{Diamond-Stanic2009}
{Diamond-Stanic}, A.~M., {Rieke}, G.~H., \& {Rigby}, J.~R. 2009, \apj, 698, 623, \dodoi{10.1088/0004-637X/698/1/623}

\bibitem[{{Dressler} \& {Gunn}(1983)}]{Dressler1983}
{Dressler}, A., \& {Gunn}, J.~E. 1983, \apj, 270, 7, \dodoi{10.1086/161093}

\bibitem[{{Duras} {et~al.}(2020){Duras}, {Bongiorno}, {Ricci}, {Piconcelli}, {Shankar}, {Lusso}, {Bianchi}, {Fiore}, {Maiolino}, {Marconi}, {Onori}, {Sani}, {Schneider}, {Vignali}, \& {La Franca}}]{Duras2020}
{Duras}, F., {Bongiorno}, A., {Ricci}, F., {et~al.} 2020, \aap, 636, A73, \dodoi{10.1051/0004-6361/201936817}

\bibitem[{{Eisenstein} {et~al.}(2023{\natexlab{a}}){Eisenstein}, {Willott}, {Alberts}, {Arribas}, {Bonaventura}, {Bunker}, {Cameron}, {Carniani}, {Charlot}, {Curtis-Lake}, {D'Eugenio}, {Endsley}, {Ferruit}, {Giardino}, {Hainline}, {Hausen}, {Jakobsen}, {Johnson}, {Maiolino}, {Rieke}, {Rieke}, {Rix}, {Robertson}, {Stark}, {Tacchella}, {Williams}, {Willmer}, {Baker}, {Baum}, {Bhatawdekar}, {Boyett}, {Chen}, {Chevallard}, {Circosta}, {Curti}, {Danhaive}, {DeCoursey}, {de Graaff}, {Dressler}, {Egami}, {Helton}, {Hviding}, {Ji}, {Jones}, {Kumari}, {L{\"u}tzgendorf}, {Laseter}, {Looser}, {Lyu}, {Maseda}, {Nelson}, {Parlanti}, {Perna}, {Pusk{\'a}s}, {Rawle}, {Rodr{\'\i}guez Del Pino}, {Sandles}, {Saxena}, {Scholtz}, {Sharpe}, {Shivaei}, {Silcock}, {Simmonds}, {Skarbinski}, {Smit}, {Stone}, {Suess}, {Sun}, {Tang}, {Topping}, {{\"U}bler}, {Villanueva}, {Wallace}, {Whitler}, {Witstok}, \& {Woodrum}}]{Eisenstein2023}
{Eisenstein}, D.~J., {Willott}, C., {Alberts}, S., {et~al.} 2023{\natexlab{a}}, arXiv e-prints, arXiv:2306.02465, \dodoi{10.48550/arXiv.2306.02465}

\bibitem[{{Eisenstein} {et~al.}(2023{\natexlab{b}}){Eisenstein}, {Johnson}, {Robertson}, {Tacchella}, {Hainline}, {Jakobsen}, {Maiolino}, {Bonaventura}, {Bunker}, {Cameron}, {Cargile}, {Curtis-Lake}, {Hausen}, {Pusk{\'a}s}, {Rieke}, {Sun}, {Willmer}, {Willott}, {Alberts}, {Arribas}, {Baker}, {Baum}, {Bhatawdekar}, {Carniani}, {Charlot}, {Chen}, {Chevallard}, {Curti}, {DeCoursey}, {D'Eugenio}, {de Graaff}, {Egami}, {Helton}, {Ji}, {Jones}, {Kumari}, {L{\"u}tzgendorf}, {Laseter}, {Looser}, {Lyu}, {Maseda}, {Nelson}, {Parlanti}, {Rauscher}, {Rawle}, {Rieke}, {Rix}, {Rujopakarn}, {Sandles}, {Saxena}, {Scholtz}, {Sharpe}, {Shivaei}, {Simmonds}, {Smit}, {Topping}, {{\"U}bler}, {Venturi}, {Williams}, {Witstok}, \& {Woodrum}}]{Eisenstein2023b}
{Eisenstein}, D.~J., {Johnson}, B.~D., {Robertson}, B., {et~al.} 2023{\natexlab{b}}, arXiv e-prints, arXiv:2310.12340, \dodoi{10.48550/arXiv.2310.12340}

\bibitem[{{Falc{\'o}n-Barroso} {et~al.}(2011){Falc{\'o}n-Barroso}, {S{\'a}nchez-Bl{\'a}zquez}, {Vazdekis}, {Ricciardelli}, {Cardiel}, {Cenarro}, {Gorgas}, \& {Peletier}}]{Falcon-Barroso2011}
{Falc{\'o}n-Barroso}, J., {S{\'a}nchez-Bl{\'a}zquez}, P., {Vazdekis}, A., {et~al.} 2011, \aap, 532, A95, \dodoi{10.1051/0004-6361/201116842}

\bibitem[{{Ferruit} {et~al.}(2022){Ferruit}, {Jakobsen}, {Giardino}, {Rawle}, {Alves de Oliveira}, {Arribas}, {Beck}, {Birkmann}, {B{\"o}ker}, {Bunker}, {Charlot}, {de Marchi}, {Franx}, {Henry}, {Karakla}, {Kassin}, {Kumari}, {L{\'o}pez-Caniego}, {L{\"u}tzgendorf}, {Maiolino}, {Manjavacas}, {Marston}, {Moseley}, {Muzerolle}, {Pirzkal}, {Rauscher}, {Rix}, {Sabbi}, {Sirianni}, {te Plate}, {Valenti}, {Willott}, \& {Zeidler}}]{Ferruit2022}
{Ferruit}, P., {Jakobsen}, P., {Giardino}, G., {et~al.} 2022, \aap, 661, A81, \dodoi{10.1051/0004-6361/202142673}

\bibitem[{{Feruglio} {et~al.}(2015){Feruglio}, {Fiore}, {Carniani}, {Piconcelli}, {Zappacosta}, {Bongiorno}, {Cicone}, {Maiolino}, {Marconi}, {Menci}, {Puccetti}, \& {Veilleux}}]{Feruglio2015}
{Feruglio}, C., {Fiore}, F., {Carniani}, S., {et~al.} 2015, \aap, 583, A99, \dodoi{10.1051/0004-6361/201526020}

\bibitem[{{Fluetsch} {et~al.}(2021){Fluetsch}, {Maiolino}, {Carniani}, {Arribas}, {Belfiore}, {Bellocchi}, {Cazzoli}, {Cicone}, {Cresci}, {Fabian}, {Gallagher}, {Ishibashi}, {Mannucci}, {Marconi}, {Perna}, {Sturm}, \& {Venturi}}]{Fluetsch2021}
{Fluetsch}, A., {Maiolino}, R., {Carniani}, S., {et~al.} 2021, \mnras, 505, 5753, \dodoi{10.1093/mnras/stab1666}

\bibitem[{{Foreman-Mackey} {et~al.}(2013{\natexlab{a}}){Foreman-Mackey}, {Hogg}, {Lang}, \& {Goodman}}]{Foreman-Mackey2013}
{Foreman-Mackey}, D., {Hogg}, D.~W., {Lang}, D., \& {Goodman}, J. 2013{\natexlab{a}}, \pasp, 125, 306, \dodoi{10.1086/670067}

\bibitem[{{Foreman-Mackey} {et~al.}(2013{\natexlab{b}}){Foreman-Mackey}, {Hogg}, {Lang}, \& {Goodman}}]{2013PASP..125..306F}
---. 2013{\natexlab{b}}, \pasp, 125, 306, \dodoi{10.1086/670067}

\bibitem[{{Gaia Collaboration} {et~al.}(2023){Gaia Collaboration}, {Vallenari}, {Brown}, {Prusti}, {de Bruijne}, {Arenou}, {Babusiaux}, {Biermann}, {Creevey}, {Ducourant}, {Evans}, {Eyer}, {Guerra}, {Hutton}, {Jordi}, {Klioner}, {Lammers}, {Lindegren}, {Luri}, {Mignard}, {Panem}, {Pourbaix}, {Randich}, {Sartoretti}, {Soubiran}, {Tanga}, {Walton}, {Bailer-Jones}, {Bastian}, {Drimmel}, {Jansen}, {Katz}, {Lattanzi}, {van Leeuwen}, {Bakker}, {Cacciari}, {Casta{\~n}eda}, {De Angeli}, {Fabricius}, {Fouesneau}, {Fr{\'e}mat}, {Galluccio}, {Guerrier}, {Heiter}, {Masana}, {Messineo}, {Mowlavi}, {Nicolas}, {Nienartowicz}, {Pailler}, {Panuzzo}, {Riclet}, {Roux}, {Seabroke}, {Sordo}, {Th{\'e}venin}, {Gracia-Abril}, {Portell}, {Teyssier}, {Altmann}, {Andrae}, {Audard}, {Bellas-Velidis}, {Benson}, {Berthier}, {Blomme}, {Burgess}, {Busonero}, {Busso}, {C{\'a}novas}, {Carry}, {Cellino}, {Cheek}, {Clementini}, {Damerdji}, {Davidson}, {de Teodoro}, {Nu{\~n}ez Campos}, {Delchambre}, {Dell'Oro}, {Esquej},
  {Fern{\'a}ndez-Hern{\'a}ndez}, {Fraile}, {Garabato}, {Garc{\'\i}a-Lario}, {Gosset}, {Haigron}, {Halbwachs}, {Hambly}, {Harrison}, {Hern{\'a}ndez}, {Hestroffer}, {Hodgkin}, {Holl}, {Jan{\ss}en}, {Jevardat de Fombelle}, {Jordan}, {Krone-Martins}, {Lanzafame}, {L{\"o}ffler}, {Marchal}, {Marrese}, {Moitinho}, {Muinonen}, {Osborne}, {Pancino}, {Pauwels}, {Recio-Blanco}, {Reyl{\'e}}, {Riello}, {Rimoldini}, {Roegiers}, {Rybizki}, {Sarro}, {Siopis}, {Smith}, {Sozzetti}, {Utrilla}, {van Leeuwen}, {Abbas}, {{\'A}brah{\'a}m}, {Abreu Aramburu}, {Aerts}, {Aguado}, {Ajaj}, {Aldea-Montero}, {Altavilla}, {{\'A}lvarez}, {Alves}, {Anders}, {Anderson}, {Anglada Varela}, {Antoja}, {Baines}, {Baker}, {Balaguer-N{\'u}{\~n}ez}, {Balbinot}, {Balog}, {Barache}, {Barbato}, {Barros}, {Barstow}, {Bartolom{\'e}}, {Bassilana}, {Bauchet}, {Becciani}, {Bellazzini}, {Berihuete}, {Bernet}, {Bertone}, {Bianchi}, {Binnenfeld}, {Blanco-Cuaresma}, {Blazere}, {Boch}, {Bombrun}, {Bossini}, {Bouquillon}, {Bragaglia}, {Bramante}, {Breedt},
  {Bressan}, {Brouillet}, {Brugaletta}, {Bucciarelli}, {Burlacu}, {Butkevich}, {Buzzi}, {Caffau}, {Cancelliere}, {Cantat-Gaudin}, {Carballo}, {Carlucci}, {Carnerero}, {Carrasco}, {Casamiquela}, {Castellani}, {Castro-Ginard}, {Chaoul}, {Charlot}, {Chemin}, {Chiaramida}, {Chiavassa}, {Chornay}, {Comoretto}, {Contursi}, {Cooper}, {Cornez}, {Cowell}, {Crifo}, {Cropper}, {Crosta}, {Crowley}, {Dafonte}, {Dapergolas}, {David}, {David}, {de Laverny}, {De Luise}, {De March}, {De Ridder}, {de Souza}, {de Torres}, {del Peloso}, {del Pozo}, {Delbo}, {Delgado}, {Delisle}, {Demouchy}, {Dharmawardena}, {Di Matteo}, {Diakite}, {Diener}, {Distefano}, {Dolding}, {Edvardsson}, {Enke}, {Fabre}, {Fabrizio}, {Faigler}, {Fedorets}, {Fernique}, {Fienga}, {Figueras}, {Fournier}, {Fouron}, {Fragkoudi}, {Gai}, {Garcia-Gutierrez}, {Garcia-Reinaldos}, {Garc{\'\i}a-Torres}, {Garofalo}, {Gavel}, {Gavras}, {Gerlach}, {Geyer}, {Giacobbe}, {Gilmore}, {Girona}, {Giuffrida}, {Gomel}, {Gomez}, {Gonz{\'a}lez-N{\'u}{\~n}ez},
  {Gonz{\'a}lez-Santamar{\'\i}a}, {Gonz{\'a}lez-Vidal}, {Granvik}, {Guillout}, {Guiraud}, {Guti{\'e}rrez-S{\'a}nchez}, {Guy}, {Hatzidimitriou}, {Hauser}, {Haywood}, {Helmer}, {Helmi}, {Sarmiento}, {Hidalgo}, {Hilger}, {H{\l}adczuk}, {Hobbs}, {Holland}, {Huckle}, {Jardine}, {Jasniewicz}, {Jean-Antoine Piccolo}, {Jim{\'e}nez-Arranz}, {Jorissen}, {Juaristi Campillo}, {Julbe}, {Karbevska}, {Kervella}, {Khanna}, {Kontizas}, {Kordopatis}, {Korn}, {K{\'o}sp{\'a}l}, {Kostrzewa-Rutkowska}, {Kruszy{\'n}ska}, {Kun}, {Laizeau}, {Lambert}, {Lanza}, {Lasne}, {Le Campion}, {Lebreton}, {Lebzelter}, {Leccia}, {Leclerc}, {Lecoeur-Taibi}, {Liao}, {Licata}, {Lindstr{\o}m}, {Lister}, {Livanou}, {Lobel}, {Lorca}, {Loup}, {Madrero Pardo}, {Magdaleno Romeo}, {Managau}, {Mann}, {Manteiga}, {Marchant}, {Marconi}, {Marcos}, {Marcos Santos}, {Mar{\'\i}n Pina}, {Marinoni}, {Marocco}, {Marshall}, {Martin Polo}, {Mart{\'\i}n-Fleitas}, {Marton}, {Mary}, {Masip}, {Massari}, {Mastrobuono-Battisti}, {Mazeh}, {McMillan}, {Messina}, {Michalik},
  {Millar}, {Mints}, {Molina}, {Molinaro}, {Moln{\'a}r}, {Monari}, {Mongui{\'o}}, {Montegriffo}, {Montero}, {Mor}, {Mora}, {Morbidelli}, {Morel}, {Morris}, {Muraveva}, {Murphy}, {Musella}, {Nagy}, {Noval}, {Oca{\~n}a}, {Ogden}, {Ordenovic}, {Osinde}, {Pagani}, {Pagano}, {Palaversa}, {Palicio}, {Pallas-Quintela}, {Panahi}, {Payne-Wardenaar}, {Pe{\~n}alosa Esteller}, {Penttil{\"a}}, {Pichon}, {Piersimoni}, {Pineau}, {Plachy}, {Plum}, {Poggio}, {Pr{\v{s}}a}, {Pulone}, {Racero}, {Ragaini}, {Rainer}, {Raiteri}, {Rambaux}, {Ramos}, {Ramos-Lerate}, {Re Fiorentin}, {Regibo}, {Richards}, {Rios Diaz}, {Ripepi}, {Riva}, {Rix}, {Rixon}, {Robichon}, {Robin}, {Robin}, {Roelens}, {Rogues}, {Rohrbasser}, {Romero-G{\'o}mez}, {Rowell}, {Royer}, {Ruz Mieres}, {Rybicki}, {Sadowski}, {S{\'a}ez N{\'u}{\~n}ez}, {Sagrist{\`a} Sell{\'e}s}, {Sahlmann}, {Salguero}, {Samaras}, {Sanchez Gimenez}, {Sanna}, {Santove{\~n}a}, {Sarasso}, {Schultheis}, {Sciacca}, {Segol}, {Segovia}, {S{\'e}gransan}, {Semeux}, {Shahaf}, {Siddiqui}, {Siebert},
  {Siltala}, {Silvelo}, {Slezak}, {Slezak}, {Smart}, {Snaith}, {Solano}, {Solitro}, {Souami}, {Souchay}, {Spagna}, {Spina}, {Spoto}, {Steele}, {Steidelm{\"u}ller}, {Stephenson}, {S{\"u}veges}, {Surdej}, {Szabados}, {Szegedi-Elek}, {Taris}, {Taylor}, {Teixeira}, {Tolomei}, {Tonello}, {Torra}, {Torra}, {Torralba Elipe}, {Trabucchi}, {Tsounis}, {Turon}, {Ulla}, {Unger}, {Vaillant}, {van Dillen}, {van Reeven}, {Vanel}, {Vecchiato}, {Viala}, {Vicente}, {Voutsinas}, {Weiler}, {Wevers}, {Wyrzykowski}, {Yoldas}, {Yvard}, {Zhao}, {Zorec}, {Zucker}, \& {Zwitter}}]{Gaia2023}
{Gaia Collaboration}, {Vallenari}, A., {Brown}, A.~G.~A., {et~al.} 2023, \aap, 674, A1, \dodoi{10.1051/0004-6361/202243940}

\bibitem[{{Greene} \& {Ho}(2005)}]{Greene2005}
{Greene}, J.~E., \& {Ho}, L.~C. 2005, \apj, 630, 122, \dodoi{10.1086/431897}

\bibitem[{{Heckman} {et~al.}(2000){Heckman}, {Lehnert}, {Strickland}, \& {Armus}}]{Heckman2000}
{Heckman}, T.~M., {Lehnert}, M.~D., {Strickland}, D.~K., \& {Armus}, L. 2000, \apjs, 129, 493, \dodoi{10.1086/313421}

\bibitem[{{Hickox} {et~al.}(2014){Hickox}, {Mullaney}, {Alexander}, {Chen}, {Civano}, {Goulding}, \& {Hainline}}]{Hickox2014}
{Hickox}, R.~C., {Mullaney}, J.~R., {Alexander}, D.~M., {et~al.} 2014, \apj, 782, 9, \dodoi{10.1088/0004-637X/782/1/9}

\bibitem[{{Hogg} {et~al.}(2010){Hogg}, {Bovy}, \& {Lang}}]{Hogg2010}
{Hogg}, D.~W., {Bovy}, J., \& {Lang}, D. 2010, arXiv e-prints, arXiv:1008.4686, \dodoi{10.48550/arXiv.1008.4686}

\bibitem[{{Hopkins} {et~al.}(2005){Hopkins}, {Hernquist}, {Martini}, {Cox}, {Robertson}, {Di Matteo}, \& {Springel}}]{Hopkins2005}
{Hopkins}, P.~F., {Hernquist}, L., {Martini}, P., {et~al.} 2005, \apjl, 625, L71, \dodoi{10.1086/431146}

\bibitem[{{Hopkins} {et~al.}(2014){Hopkins}, {Kere{\v{s}}}, {O{\~n}orbe}, {Faucher-Gigu{\`e}re}, {Quataert}, {Murray}, \& {Bullock}}]{Hopkins2014}
{Hopkins}, P.~F., {Kere{\v{s}}}, D., {O{\~n}orbe}, J., {et~al.} 2014, \mnras, 445, 581, \dodoi{10.1093/mnras/stu1738}

\bibitem[{{Horne}(1986)}]{Horne1986}
{Horne}, K. 1986, \pasp, 98, 609, \dodoi{10.1086/131801}

\bibitem[{{Inami} {et~al.}(2017){Inami}, {Bacon}, {Brinchmann}, {Richard}, {Contini}, {Conseil}, {Hamer}, {Akhlaghi}, {Bouch{\'e}}, {Cl{\'e}ment}, {Desprez}, {Drake}, {Hashimoto}, {Leclercq}, {Maseda}, {Michel-Dansac}, {Paalvast}, {Tresse}, {Ventou}, {Kollatschny}, {Boogaard}, {Finley}, {Marino}, {Schaye}, \& {Wisotzki}}]{Inami2017}
{Inami}, H., {Bacon}, R., {Brinchmann}, J., {et~al.} 2017, \aap, 608, A2, \dodoi{10.1051/0004-6361/201731195}

\bibitem[{{Jakobsen} {et~al.}(2022){Jakobsen}, {Ferruit}, {Alves de Oliveira}, {Arribas}, {Bagnasco}, {Barho}, {Beck}, {Birkmann}, {B{\"o}ker}, {Bunker}, {Charlot}, {de Jong}, {de Marchi}, {Ehrenwinkler}, {Falcolini}, {Fels}, {Franx}, {Franz}, {Funke}, {Giardino}, {Gnata}, {Holota}, {Honnen}, {Jensen}, {Jentsch}, {Johnson}, {Jollet}, {Karl}, {Kling}, {K{\"o}hler}, {Kolm}, {Kumari}, {Lander}, {Lemke}, {L{\'o}pez-Caniego}, {L{\"u}tzgendorf}, {Maiolino}, {Manjavacas}, {Marston}, {Maschmann}, {Maurer}, {Messerschmidt}, {Moseley}, {Mosner}, {Mott}, {Muzerolle}, {Pirzkal}, {Pittet}, {Plitzke}, {Posselt}, {Rapp}, {Rauscher}, {Rawle}, {Rix}, {R{\"o}del}, {Rumler}, {Sabbi}, {Salvignol}, {Schmid}, {Sirianni}, {Smith}, {Strada}, {te Plate}, {Valenti}, {Wettemann}, {Wiehe}, {Wiesmayer}, {Willott}, {Wright}, {Zeidler}, \& {Zincke}}]{Jakobsen2022}
{Jakobsen}, P., {Ferruit}, P., {Alves de Oliveira}, C., {et~al.} 2022, \aap, 661, A80, \dodoi{10.1051/0004-6361/202142663}

\bibitem[{{Ji} \& {Giavalisco}(2022)}]{Ji2022}
{Ji}, Z., \& {Giavalisco}, M. 2022, \apj, 935, 120, \dodoi{10.3847/1538-4357/ac7f43}

\bibitem[{{Ji} {et~al.}(2024){Ji}, {Williams}, {Tacchella}, {Suess}, {Baker}, {Alberts}, {Bunker}, {Johnson}, {Robertson}, {Sun}, {Eisenstein}, {Rieke}, {Maseda}, {Hainline}, {Hausen}, {Rieke}, {Willmer}, {Egami}, {Shivaei}, {Carniani}, {Charlot}, {Chevallard}, {Curtis-Lake}, {Looser}, {Maiolino}, {Willott}, {Chen}, {Helton}, {Lyu}, {Nelson}, {Bhatawdekar}, {Boyett}, \& {Sandles}}]{Ji2024}
{Ji}, Z., {Williams}, C.~C., {Tacchella}, S., {et~al.} 2024, \apj, 974, 135, \dodoi{10.3847/1538-4357/ad6e7f}

\bibitem[{{Johnson} {et~al.}(2021){Johnson}, {Leja}, {Conroy}, \& {Speagle}}]{Johnson2021}
{Johnson}, B.~D., {Leja}, J., {Conroy}, C., \& {Speagle}, J.~S. 2021, \apjs, 254, 22, \dodoi{10.3847/1538-4365/abef67}

\bibitem[{{Kauffmann} {et~al.}(2003){Kauffmann}, {Heckman}, {Tremonti}, {Brinchmann}, {Charlot}, {White}, {Ridgway}, {Brinkmann}, {Fukugita}, {Hall}, {Ivezi{\'c}}, {Richards}, \& {Schneider}}]{Kauffmann2003}
{Kauffmann}, G., {Heckman}, T.~M., {Tremonti}, C., {et~al.} 2003, \mnras, 346, 1055, \dodoi{10.1111/j.1365-2966.2003.07154.x}

\bibitem[{{Kelson} {et~al.}(2000){Kelson}, {Illingworth}, {van Dokkum}, \& {Franx}}]{Kelson2000}
{Kelson}, D.~D., {Illingworth}, G.~D., {van Dokkum}, P.~G., \& {Franx}, M. 2000, \apj, 531, 159, \dodoi{10.1086/308445}

\bibitem[{{Kere{\v{s}}} {et~al.}(2009){Kere{\v{s}}}, {Katz}, {Fardal}, {Dav{\'e}}, \& {Weinberg}}]{Keres2009}
{Kere{\v{s}}}, D., {Katz}, N., {Fardal}, M., {Dav{\'e}}, R., \& {Weinberg}, D.~H. 2009, \mnras, 395, 160, \dodoi{10.1111/j.1365-2966.2009.14541.x}

\bibitem[{{Kewley} {et~al.}(2001){Kewley}, {Dopita}, {Sutherland}, {Heisler}, \& {Trevena}}]{Kewley2001}
{Kewley}, L.~J., {Dopita}, M.~A., {Sutherland}, R.~S., {Heisler}, C.~A., \& {Trevena}, J. 2001, \apj, 556, 121, \dodoi{10.1086/321545}

\bibitem[{{Kewley} {et~al.}(2013){Kewley}, {Maier}, {Yabe}, {Ohta}, {Akiyama}, {Dopita}, \& {Yuan}}]{Kewley2013}
{Kewley}, L.~J., {Maier}, C., {Yabe}, K., {et~al.} 2013, \apjl, 774, L10, \dodoi{10.1088/2041-8205/774/1/L10}

\bibitem[{{King} \& {Pounds}(2015)}]{King2015}
{King}, A., \& {Pounds}, K. 2015, \araa, 53, 115, \dodoi{10.1146/annurev-astro-082214-122316}

\bibitem[{{Kroupa}(2001)}]{Kroupa2001}
{Kroupa}, P. 2001, \mnras, 322, 231, \dodoi{10.1046/j.1365-8711.2001.04022.x}

\bibitem[{{Leitherer} {et~al.}(1999){Leitherer}, {Schaerer}, {Goldader}, {Delgado}, {Robert}, {Kune}, {de Mello}, {Devost}, \& {Heckman}}]{Leitherer1999}
{Leitherer}, C., {Schaerer}, D., {Goldader}, J.~D., {et~al.} 1999, \apjs, 123, 3, \dodoi{10.1086/313233}

\bibitem[{{Leja} {et~al.}(2019){Leja}, {Carnall}, {Johnson}, {Conroy}, \& {Speagle}}]{Leja2019}
{Leja}, J., {Carnall}, A.~C., {Johnson}, B.~D., {Conroy}, C., \& {Speagle}, J.~S. 2019, \apj, 876, 3, \dodoi{10.3847/1538-4357/ab133c}

\bibitem[{{Leja} {et~al.}(2017){Leja}, {Johnson}, {Conroy}, {van Dokkum}, \& {Byler}}]{Leja2017}
{Leja}, J., {Johnson}, B.~D., {Conroy}, C., {van Dokkum}, P.~G., \& {Byler}, N. 2017, \apj, 837, 170, \dodoi{10.3847/1538-4357/aa5ffe}

\bibitem[{{Leja} {et~al.}(2022){Leja}, {Speagle}, {Ting}, {Johnson}, {Conroy}, {Whitaker}, {Nelson}, {van Dokkum}, \& {Franx}}]{Leja2022}
{Leja}, J., {Speagle}, J.~S., {Ting}, Y.-S., {et~al.} 2022, \apj, 936, 165, \dodoi{10.3847/1538-4357/ac887d}

\bibitem[{{Liddle}(2007)}]{Liddle2007}
{Liddle}, A.~R. 2007, \mnras, 377, L74, \dodoi{10.1111/j.1745-3933.2007.00306.x}

\bibitem[{{Liu} {et~al.}(2017){Liu}, {Tozzi}, {Wang}, {Brandt}, {Vignali}, {Xue}, {Schneider}, {Comastri}, {Yang}, {Bauer}, {Paolillo}, {Luo}, {Gilli}, {Wang}, {Giavalisco}, {Ji}, {Alexander}, {Mainieri}, {Shemmer}, {Koekemoer}, \& {Risaliti}}]{Liu2017}
{Liu}, T., {Tozzi}, P., {Wang}, J.-X., {et~al.} 2017, \apjs, 232, 8, \dodoi{10.3847/1538-4365/aa7847}

\bibitem[{{Looser} {et~al.}(2024){Looser}, {D'Eugenio}, {Maiolino}, {Witstok}, {Sandles}, {Curtis-Lake}, {Chevallard}, {Tacchella}, {Johnson}, {Baker}, {Suess}, {Carniani}, {Ferruit}, {Arribas}, {Bonaventura}, {Bunker}, {Cameron}, {Charlot}, {Curti}, {de Graaff}, {Maseda}, {Rawle}, {Rix}, {Del Pino}, {Smit}, {{\"U}bler}, {Willott}, {Alberts}, {Egami}, {Eisenstein}, {Endsley}, {Hausen}, {Rieke}, {Robertson}, {Shivaei}, {Williams}, {Boyett}, {Chen}, {Ji}, {Jones}, {Kumari}, {Nelson}, {Perna}, {Saxena}, \& {Scholtz}}]{Looser2024}
{Looser}, T.~J., {D'Eugenio}, F., {Maiolino}, R., {et~al.} 2024, \nat, 629, 53, \dodoi{10.1038/s41586-024-07227-0}

\bibitem[{{Luo} {et~al.}(2017){Luo}, {Brandt}, {Xue}, {Lehmer}, {Alexander}, {Bauer}, {Vito}, {Yang}, {Basu-Zych}, {Comastri}, {Gilli}, {Gu}, {Hornschemeier}, {Koekemoer}, {Liu}, {Mainieri}, {Paolillo}, {Ranalli}, {Rosati}, {Schneider}, {Shemmer}, {Smail}, {Sun}, {Tozzi}, {Vignali}, \& {Wang}}]{Luo2017}
{Luo}, B., {Brandt}, W.~N., {Xue}, Y.~Q., {et~al.} 2017, \apjs, 228, 2, \dodoi{10.3847/1538-4365/228/1/2}

\bibitem[{{Luo} {et~al.}(2022){Luo}, {Rowlands}, {Alatalo}, {Sazonova}, {Abdurro'uf}, {Heckman}, {Medling}, {Deustua}, {Nyland}, {Lanz}, {Petric}, {Otter}, {Aalto}, {Dimassimo}, {French}, {Gallagher}, {Roediger}, \& {Stepanoff}}]{Luo2022}
{Luo}, Y., {Rowlands}, K., {Alatalo}, K., {et~al.} 2022, \apj, 938, 63, \dodoi{10.3847/1538-4357/ac8b7d}

\bibitem[{{Lyu} {et~al.}(2022){Lyu}, {Alberts}, {Rieke}, \& {Rujopakarn}}]{Lyu2022}
{Lyu}, J., {Alberts}, S., {Rieke}, G.~H., \& {Rujopakarn}, W. 2022, \apj, 941, 191, \dodoi{10.3847/1538-4357/ac9e5d}

\bibitem[{{Lyu} {et~al.}(2024){Lyu}, {Alberts}, {Rieke}, {Shivaei}, {P{\'e}rez-Gonz{\'a}lez}, {Sun}, {Hainline}, {Baum}, {Bonaventura}, {Bunker}, {Egami}, {Eisenstein}, {Florian}, {Ji}, {Johnson}, {Morrison}, {Rieke}, {Robertson}, {Rujopakarn}, {Tacchella}, {Scholtz}, \& {Willmer}}]{Lyu2024}
{Lyu}, J., {Alberts}, S., {Rieke}, G.~H., {et~al.} 2024, \apj, 966, 229, \dodoi{10.3847/1538-4357/ad3643}

\bibitem[{{Madau}(1995)}]{Madau1995}
{Madau}, P. 1995, \apj, 441, 18, \dodoi{10.1086/175332}

\bibitem[{{Maltby} {et~al.}(2019){Maltby}, {Almaini}, {McLure}, {Wild}, {Dunlop}, {Rowlands}, {Hartley}, {Hatch}, {Socolovsky}, {Wilkinson}, {Amorin}, {Bradshaw}, {Carnall}, {Castellano}, {Cimatti}, {Cresci}, {Cullen}, {De Barros}, {Fontanot}, {Garilli}, {Koekemoer}, {McLeod}, {Pentericci}, \& {Talia}}]{Maltby2019}
{Maltby}, D.~T., {Almaini}, O., {McLure}, R.~J., {et~al.} 2019, \mnras, 489, 1139, \dodoi{10.1093/mnras/stz2211}

\bibitem[{{Naab} \& {Ostriker}(2017)}]{Naab2017}
{Naab}, T., \& {Ostriker}, J.~P. 2017, \araa, 55, 59, \dodoi{10.1146/annurev-astro-081913-040019}

\bibitem[{{Netzer}(2009)}]{Netzer2009}
{Netzer}, H. 2009, \mnras, 399, 1907, \dodoi{10.1111/j.1365-2966.2009.15434.x}

\bibitem[{{Netzer}(2019)}]{Netzer2019}
---. 2019, \mnras, 488, 5185, \dodoi{10.1093/mnras/stz2016}

\bibitem[{{Neufeld} {et~al.}(2024){Neufeld}, {van Dokkum}, {Asali}, {Covelo-Paz}, {Leja}, {Lin}, {Matthee}, {Oesch}, {Reddy}, {Shivaei}, {Whitaker}, {Wuyts}, {Brammer}, {Marchesini}, {Maseda}, {Naidu}, {Nelson}, {Velichko}, {Weibel}, \& {Xiao}}]{Neufeld2024}
{Neufeld}, C., {van Dokkum}, P., {Asali}, Y., {et~al.} 2024, \apj, 972, 156, \dodoi{10.3847/1538-4357/ad6158}

\bibitem[{{Newville} {et~al.}(2014){Newville}, {Stensitzki}, {Allen}, \& {Ingargiola}}]{Newville2014}
{Newville}, M., {Stensitzki}, T., {Allen}, D.~B., \& {Ingargiola}, A. 2014, {LMFIT: Non-Linear Least-Square Minimization and Curve-Fitting for Python}, 0.8.0,  Zenodo, \dodoi{10.5281/zenodo.11813}

\bibitem[{{Noll} {et~al.}(2009){Noll}, {Burgarella}, {Giovannoli}, {Buat}, {Marcillac}, \& {Mu{\~n}oz-Mateos}}]{Noll2009}
{Noll}, S., {Burgarella}, D., {Giovannoli}, E., {et~al.} 2009, \aap, 507, 1793, \dodoi{10.1051/0004-6361/200912497}

\bibitem[{{Oesch} \& {Magee}(2023)}]{https://doi.org/10.17909/gdyc-7g80}
{Oesch}, P., \& {Magee}, D. 2023, The JWST FRESCO Survey,  STScI/MAST, \dodoi{10.17909/GDYC-7G80}

\bibitem[{{Oesch} {et~al.}(2023){Oesch}, {Brammer}, {Naidu}, {Bouwens}, {Chisholm}, {Illingworth}, {Matthee}, {Nelson}, {Qin}, {Reddy}, {Shapley}, {Shivaei}, {van Dokkum}, {Weibel}, {Whitaker}, {Wuyts}, {Covelo-Paz}, {Endsley}, {Fudamoto}, {Giovinazzo}, {Herard-Demanche}, {Kerutt}, {Kramarenko}, {Labbe}, {Leonova}, {Lin}, {Magee}, {Marchesini}, {Maseda}, {Mason}, {Matharu}, {Meyer}, {Neufeld}, {Prieto Lyon}, {Schaerer}, {Sharma}, {Shuntov}, {Smit}, {Stefanon}, {Wyithe}, \& {Xiao}}]{Oesch2023}
{Oesch}, P.~A., {Brammer}, G., {Naidu}, R.~P., {et~al.} 2023, \mnras, 525, 2864, \dodoi{10.1093/mnras/stad2411}

\bibitem[{{Osterbrock} \& {Ferland}(2006)}]{Osterbrock2006}
{Osterbrock}, D.~E., \& {Ferland}, G.~J. 2006, {Astrophysics of gaseous nebulae and active galactic nuclei}

\bibitem[{{Park} {et~al.}(2024){Park}, {Belli}, {Conroy}, {Johnson}, {Davies}, {Leja}, {Tacchella}, {Mendel}, {Benton}, {Bugiani}, {Emami}, {Khoram}, {Li}, {Maheson}, {Mathews}, {Naidu}, {Nelson}, {Terrazas}, \& {Weinberger}}]{Park2024}
{Park}, M., {Belli}, S., {Conroy}, C., {et~al.} 2024, \apj, 976, 72, \dodoi{10.3847/1538-4357/ad7e15}

\bibitem[{{Pennell} {et~al.}(2017){Pennell}, {Runnoe}, \& {Brotherton}}]{Pennell2017}
{Pennell}, A., {Runnoe}, J.~C., \& {Brotherton}, M.~S. 2017, \mnras, 468, 1433, \dodoi{10.1093/mnras/stx556}

\bibitem[{{Perrotta} {et~al.}(2021){Perrotta}, {George}, {Coil}, {Tremonti}, {Rupke}, {Davis}, {Diamond-Stanic}, {Geach}, {Hickox}, {Moustakas}, {Petter}, {Rudnick}, {Sell}, {Swiggum}, \& {Whalen}}]{Perrotta2021}
{Perrotta}, S., {George}, E.~R., {Coil}, A.~L., {et~al.} 2021, \apj, 923, 275, \dodoi{10.3847/1538-4357/ac2fa4}

\bibitem[{{Rafelski} {et~al.}(2015){Rafelski}, {Teplitz}, {Gardner}, {Coe}, {Bond}, {Koekemoer}, {Grogin}, {Kurczynski}, {McGrath}, {Bourque}, {Atek}, {Brown}, {Colbert}, {Codoreanu}, {Ferguson}, {Finkelstein}, {Gawiser}, {Giavalisco}, {Gronwall}, {Hanish}, {Lee}, {Mehta}, {de Mello}, {Ravindranath}, {Ryan}, {Scarlata}, {Siana}, {Soto}, \& {Voyer}}]{Rafelski2015}
{Rafelski}, M., {Teplitz}, H.~I., {Gardner}, J.~P., {et~al.} 2015, \aj, 150, 31, \dodoi{10.1088/0004-6256/150/1/31}

\bibitem[{{Rieke} {et~al.}(2024){Rieke}, {Alberts}, {Shivaei}, {Lyu}, {Willmer}, {P{\'e}rez-Gonz{\'a}lez}, \& {Williams}}]{Rieke2024}
{Rieke}, G.~H., {Alberts}, S., {Shivaei}, I., {et~al.} 2024, \apj, 975, 83, \dodoi{10.3847/1538-4357/ad6cd2}

\bibitem[{{Rieke} {et~al.}(2009){Rieke}, {Alonso-Herrero}, {Weiner}, {P{\'e}rez-Gonz{\'a}lez}, {Blaylock}, {Donley}, \& {Marcillac}}]{Rieke2009}
{Rieke}, G.~H., {Alonso-Herrero}, A., {Weiner}, B.~J., {et~al.} 2009, \apj, 692, 556, \dodoi{10.1088/0004-637X/692/1/556}

\bibitem[{{Rieke} {et~al.}(2023{\natexlab{a}}){Rieke}, {Robertson}, {Tacchella}, {Willmer}, {Johnson}, {Carniani}, {Bunker}, \& {Willott}}]{https://doi.org/10.17909/8tdj-8n28}
{Rieke}, M., {Robertson}, B., {Tacchella}, S., {et~al.} 2023{\natexlab{a}}, Data from the JWST Advanced Deep Extragalactic Survey (JADES),  STScI/MAST, \dodoi{10.17909/8TDJ-8N28}

\bibitem[{{Rieke} {et~al.}(2023{\natexlab{b}}){Rieke}, {Robertson}, {Tacchella}, {Hainline}, {Johnson}, {Hausen}, {Ji}, {Willmer}, {Eisenstein}, {Pusk{\'a}s}, {Alberts}, {Arribas}, {Baker}, {Baum}, {Bhatawdekar}, {Bonaventura}, {Boyett}, {Bunker}, {Cameron}, {Carniani}, {Charlot}, {Chevallard}, {Chen}, {Curti}, {Curtis-Lake}, {Danhaive}, {DeCoursey}, {Dressler}, {Egami}, {Endsley}, {Helton}, {Hviding}, {Kumari}, {Looser}, {Lyu}, {Maiolino}, {Maseda}, {Nelson}, {Rieke}, {Rix}, {Sandles}, {Saxena}, {Sharpe}, {Shivaei}, {Skarbinski}, {Smit}, {Stark}, {Stone}, {Suess}, {Sun}, {Topping}, {{\"U}bler}, {Villanueva}, {Wallace}, {Williams}, {Willott}, {Whitler}, {Witstok}, \& {Woodrum}}]{Rieke2023}
{Rieke}, M.~J., {Robertson}, B., {Tacchella}, S., {et~al.} 2023{\natexlab{b}}, \apjs, 269, 16, \dodoi{10.3847/1538-4365/acf44d}

\bibitem[{{Rieke, George} {et~al.}(2024){Rieke, George}, {Alberts, Stacey}, {Lyu, Jianwei}, \& {Shivaei, Irene}}]{rieke_george_systematic_2024}
{Rieke, George}, {Alberts, Stacey}, {Lyu, Jianwei}, \& {Shivaei, Irene}. 2024, Systematic {Mid}-infrared {Instrument} {Legacy} {Extragalactic} {Survey} ({SMILES}),  STScI/MAST, \dodoi{10.17909/ET3F-ZD57}

\bibitem[{{Rupke} {et~al.}(2005{\natexlab{a}}){Rupke}, {Veilleux}, \& {Sanders}}]{Rupke2005a}
{Rupke}, D.~S., {Veilleux}, S., \& {Sanders}, D.~B. 2005{\natexlab{a}}, \apjs, 160, 87, \dodoi{10.1086/432886}

\bibitem[{{Rupke} {et~al.}(2005{\natexlab{b}}){Rupke}, {Veilleux}, \& {Sanders}}]{Rupke2005c}
---. 2005{\natexlab{b}}, \apj, 632, 751, \dodoi{10.1086/444451}

\bibitem[{{Rupke} {et~al.}(2005{\natexlab{c}}){Rupke}, {Veilleux}, \& {Sanders}}]{Rupke2005b}
---. 2005{\natexlab{c}}, \apjs, 160, 115, \dodoi{10.1086/432889}

\bibitem[{{Rupke}(2018)}]{Rupke2018}
{Rupke}, D. S.~N. 2018, Galaxies, 6, 138, \dodoi{10.3390/galaxies6040138}

\bibitem[{{Rupke} \& {Veilleux}(2015)}]{Rupke2015}
{Rupke}, D. S.~N., \& {Veilleux}, S. 2015, \apj, 801, 126, \dodoi{10.1088/0004-637X/801/2/126}

\bibitem[{{Salpeter}(1955)}]{Salpeter1955}
{Salpeter}, E.~E. 1955, \apj, 121, 161, \dodoi{10.1086/145971}

\bibitem[{{S{\'a}nchez-Bl{\'a}zquez} {et~al.}(2006){S{\'a}nchez-Bl{\'a}zquez}, {Peletier}, {Jim{\'e}nez-Vicente}, {Cardiel}, {Cenarro}, {Falc{\'o}n-Barroso}, {Gorgas}, {Selam}, \& {Vazdekis}}]{Sanchez-Blazquez2006}
{S{\'a}nchez-Bl{\'a}zquez}, P., {Peletier}, R.~F., {Jim{\'e}nez-Vicente}, J., {et~al.} 2006, \mnras, 371, 703, \dodoi{10.1111/j.1365-2966.2006.10699.x}

\bibitem[{{Sanders} {et~al.}(1988){Sanders}, {Soifer}, {Elias}, {Madore}, {Matthews}, {Neugebauer}, \& {Scoville}}]{Sanders1988}
{Sanders}, D.~B., {Soifer}, B.~T., {Elias}, J.~H., {et~al.} 1988, \apj, 325, 74, \dodoi{10.1086/165983}

\bibitem[{{Sarzi} {et~al.}(2016){Sarzi}, {Kaviraj}, {Nedelchev}, {Tiffany}, {Shabala}, {Deller}, \& {Middelberg}}]{Sarzi2016}
{Sarzi}, M., {Kaviraj}, S., {Nedelchev}, B., {et~al.} 2016, \mnras, 456, L25, \dodoi{10.1093/mnrasl/slv165}

\bibitem[{{Silva-Lima} {et~al.}(2022){Silva-Lima}, {Martins}, {Coelho}, \& {Gadotti}}]{Silva-Lima2022}
{Silva-Lima}, L.~A., {Martins}, L.~P., {Coelho}, P. R.~T., \& {Gadotti}, D.~A. 2022, \aap, 661, A105, \dodoi{10.1051/0004-6361/202142432}

\bibitem[{{Somerville} \& {Dav{\'e}}(2015)}]{Somerville2015}
{Somerville}, R.~S., \& {Dav{\'e}}, R. 2015, \araa, 53, 51, \dodoi{10.1146/annurev-astro-082812-140951}

\bibitem[{{Steidel} {et~al.}(2010){Steidel}, {Erb}, {Shapley}, {Pettini}, {Reddy}, {Bogosavljevi{\'c}}, {Rudie}, \& {Rakic}}]{Steidel2010}
{Steidel}, C.~C., {Erb}, D.~K., {Shapley}, A.~E., {et~al.} 2010, \apj, 717, 289, \dodoi{10.1088/0004-637X/717/1/289}

\bibitem[{{Sun}(2024)}]{https://doi.org/10.5281/zenodo.14052875}
{Sun}, F. 2024, nircam\_grism,  Zenodo, \dodoi{10.5281/ZENODO.14052875}

\bibitem[{{Sun} {et~al.}(2023){Sun}, {Egami}, {Pirzkal}, {Rieke}, {Baum}, {Boyer}, {Boyett}, {Bunker}, {Cameron}, {Curti}, {Eisenstein}, {Gennaro}, {Greene}, {Jaffe}, {Kelly}, {Koekemoer}, {Kumari}, {Maiolino}, {Maseda}, {Perna}, {Rest}, {Robertson}, {Schlawin}, {Smit}, {Stansberry}, {Sunnquist}, {Tacchella}, {Williams}, \& {Willmer}}]{Sun2023}
{Sun}, F., {Egami}, E., {Pirzkal}, N., {et~al.} 2023, \apj, 953, 53, \dodoi{10.3847/1538-4357/acd53c}

\bibitem[{{Sun} {et~al.}(2024){Sun}, {Lee}, {Zabludoff}, {French}, {Helton}, {Kerrison}, {Tremonti}, \& {Yang}}]{Sun2024}
{Sun}, Y., {Lee}, G.-H., {Zabludoff}, A.~I., {et~al.} 2024, \mnras, 528, 5783, \dodoi{10.1093/mnras/stae366}

\bibitem[{{Sun} {et~al.}(2025){Sun}, {Lyu}, {Rieke}, {Ji}, {Sun}, {Zhu}, {Bunker}, {Cargile}, {Circosta}, {D'Eugenio}, {Egami}, {Hainline}, {Helton}, {Rinaldi}, {Robertson}, {Scholtz}, {Shivaei}, {Stone}, {Tacchella}, {Williams}, {Willmer}, \& {Willott}}]{Sun2025}
{Sun}, Y., {Lyu}, J., {Rieke}, G.~H., {et~al.} 2025, \apj, 978, 98, \dodoi{10.3847/1538-4357/ad973b}

\bibitem[{{Tacchella} {et~al.}(2022){Tacchella}, {Conroy}, {Faber}, {Johnson}, {Leja}, {Barro}, {Cunningham}, {Deason}, {Guhathakurta}, {Guo}, {Hernquist}, {Koo}, {McKinnon}, {Rockosi}, {Speagle}, {van Dokkum}, \& {Yesuf}}]{Tacchella2022}
{Tacchella}, S., {Conroy}, C., {Faber}, S.~M., {et~al.} 2022, \apj, 926, 134, \dodoi{10.3847/1538-4357/ac449b}

\bibitem[{{Taylor} {et~al.}(2024){Taylor}, {Maltby}, {Almaini}, {Merrifield}, {Wild}, {Rowlands}, \& {Harrold}}]{Taylor2024}
{Taylor}, E., {Maltby}, D., {Almaini}, O., {et~al.} 2024, \mnras, 535, 1684, \dodoi{10.1093/mnras/stae2463}

\bibitem[{{Tremonti} {et~al.}(2007){Tremonti}, {Moustakas}, \& {Diamond-Stanic}}]{Tremonti2007}
{Tremonti}, C.~A., {Moustakas}, J., \& {Diamond-Stanic}, A.~M. 2007, \apjl, 663, L77, \dodoi{10.1086/520083}

\bibitem[{{Veilleux} {et~al.}(2005){Veilleux}, {Cecil}, \& {Bland-Hawthorn}}]{Veilleux2005}
{Veilleux}, S., {Cecil}, G., \& {Bland-Hawthorn}, J. 2005, \araa, 43, 769, \dodoi{10.1146/annurev.astro.43.072103.150610}

\bibitem[{{Veilleux} \& {Osterbrock}(1987)}]{Veilleux1987}
{Veilleux}, S., \& {Osterbrock}, D.~E. 1987, in NASA Conference Publication, Vol. 2466, NASA Conference Publication, ed. C.~J. {Lonsdale Persson}, 737--740

\bibitem[{{Virtanen} {et~al.}(2020){Virtanen}, {Gommers}, {Oliphant}, {Haberland}, {Reddy}, {Cournapeau}, {Burovski}, {Peterson}, {Weckesser}, {Bright}, {van der Walt}, {Brett}, {Wilson}, {Millman}, {Mayorov}, {Nelson}, {Jones}, {Kern}, {Larson}, {Carey}, {Polat}, {Feng}, {Moore}, {VanderPlas}, {Laxalde}, {Perktold}, {Cimrman}, {Henriksen}, {Quintero}, {Harris}, {Archibald}, {Ribeiro}, {Pedregosa}, {van Mulbregt}, \& {SciPy 1. 0 Contributors}}]{Virtanen2020}
{Virtanen}, P., {Gommers}, R., {Oliphant}, T.~E., {et~al.} 2020, Nature Methods, 17, 261, \dodoi{10.1038/s41592-019-0686-2}

\bibitem[{{Weiner} {et~al.}(2009){Weiner}, {Coil}, {Prochaska}, {Newman}, {Cooper}, {Bundy}, {Conselice}, {Dutton}, {Faber}, {Koo}, {Lotz}, {Rieke}, \& {Rubin}}]{Weiner2009}
{Weiner}, B.~J., {Coil}, A.~L., {Prochaska}, J.~X., {et~al.} 2009, \apj, 692, 187, \dodoi{10.1088/0004-637X/692/1/187}

\bibitem[{{Wu}(2025)}]{Wu2025}
{Wu}, P.-F. 2025, \apj, 978, 131, \dodoi{10.3847/1538-4357/ad98ef}

\bibitem[{{Zhu} {et~al.}(2024{\natexlab{a}}){Zhu}, {Rieke}, {Ji}, {Simmonds}, {Sun}, {Sun}, {Alberts}, {Bhatawdekar}, {Bunker}, {Cargile}, {Carniani}, {de Graaff}, {Hainline}, {Helton}, {Jones}, {Lyu}, {Rieke}, {Rinaldi}, {Robertson}, {Scholtz}, {{\"U}bler}, {Williams}, \& {Willmer}}]{Zhu2024}
{Zhu}, Y., {Rieke}, M.~J., {Ji}, Z., {et~al.} 2024{\natexlab{a}}, arXiv e-prints, arXiv:2409.11464, \dodoi{10.48550/arXiv.2409.11464}

\bibitem[{{Zhu} {et~al.}(2024{\natexlab{b}}){Zhu}, {Alberts}, {Lyu}, {Morrison}, {Rieke}, {Sun}, {Helton}, {Ji}, {Bhatawdekar}, {Bonaventura}, {Bunker}, {Lin}, {Rieke}, {Rinaldi}, {Shivaei}, {Willmer}, \& {Zhang}}]{Zhu2024arXiv}
{Zhu}, Y., {Alberts}, S., {Lyu}, J., {et~al.} 2024{\natexlab{b}}, arXiv e-prints, arXiv:2410.14804, \dodoi{10.48550/arXiv.2410.14804}

\bibitem[{{Zhu} {et~al.}(2025){Zhu}, {Bonaventura}, {Sun}, {Rieke}, {Alberts}, {Lyu}, {Morrison}, {Ji}, {Egami}, {Helton}, {Rieke}, {Rinaldi}, {Sun}, \& {Willmer}}]{Zhu2025}
{Zhu}, Y., {Bonaventura}, N., {Sun}, Y., {et~al.} 2025, arXiv e-prints, arXiv:2508.12599, \dodoi{10.48550/arXiv.2508.12599}

\bibitem[{{Zubovas} {et~al.}(2022){Zubovas}, {Bialopetravi{\v{c}}ius}, \& {Kazlauskait{\.{e}}}}]{Zubovas2022}
{Zubovas}, K., {Bialopetravi{\v{c}}ius}, J., \& {Kazlauskait{\.{e}}}, M. 2022, \mnras, 515, 1705, \dodoi{10.1093/mnras/stac1887}

\bibitem[{{Zubovas} \& {Nardini}(2020)}]{Zubovas2020}
{Zubovas}, K., \& {Nardini}, E. 2020, \mnras, 498, 3633, \dodoi{10.1093/mnras/staa2652}

\end{thebibliography}
\bibliographystyle{aasjournal}

\end{document}